\newcommand{\kc}{$\cdots$}
\newcommand{\kaa}{$^{\ast}$}
\newcommand{\kam}{m$_{\rm 1}$}
\newcommand{\kac}{c$_{\rm 1}$}

\newcommand{\kgf}{$\log gf$}
\newcommand{\kba}{$^{\rm 1}$}
\newcommand{\kbb}{$^{\rm 2}$}
\newcommand{\kbc}{$^{\rm 3}$}
\newcommand{\kbd}{$^{\rm 4}$}
\newcommand{\kbe}{$^{\rm 5}$}
\newcommand{\kbf}{$^{\rm 6}$}

\newcommand{\ka}{$^{a}$}
\newcommand{\kw}{$^{b}$}

\newcommand{\kt}{$T_{\rm eff}$}
\newcommand{\klt}{$\log T_{\rm eff}$}
\newcommand{\kj}{$^{c}$}
\newcommand{\kp}{$^{d}$}
\newcommand{\kpe}{$^{e}$}
\newcommand{\kpf}{$^{f}$}

\newcommand{\keb}{$E(B-V)$}
\newcommand{\kev}{$E(V-K)$}
\newcommand{\kx}{$(b-y)$} 
\newcommand{\kxo}{$(b-y)_{\rm 0}$} 
\newcommand{\kl}{$-$} 
 
\newcommand{\km}{$\xi$} 
\newcommand{\kbx}{$^{\ddagger}$}           
\newcommand{\kb}{Source$^{\ddagger}$}           
\newcommand{\kq}{$^{\dagger}$}           

\newcommand{\kg}{$\log g$}
\newcommand{\kd}{$(u-B)_{\rm K}$}
\newcommand{\kub}{$(uBV)_{\rm K}$}
\newcommand{\kf}{($B-V$)}
\newcommand{\kv}{($V-K$)}
\newcommand{\kfo}{$(B-V)_{\rm 0}$}
\newcommand{\kvo}{$(V-K)_{\rm 0}$}

\newcommand{\kr}{$\beta$}
\newcommand{\ks}{km~s$^{\rm -1}$}
\newcommand{\ksi}{$v \sin i$}
\newcommand{\ksa}{$\overline{v \sin i}$}
\newcommand{\ksb}{$\surd~(\overline{v-\bar{v})^{\rm 2}}$}

\newcommand{\kmg}{Mg\,{\sc ii}}     
\newcommand{\kti}{Ti\,{\sc ii}}     
\newcommand{\kff}{Fe\,{\sc ii}}     
\newcommand{\kfe}{Fe\,{\sc i}}     
\newcommand{\kca}{Ca\,{\sc i}}     
\newcommand{\kcr}{Cr\,{\sc i}}     
\newcommand{\kcc}{Cr\,{\sc ii}}     
\newcommand{\kbt}{Ba\,{\sc ii}}     
\newcommand{\ksc}{Sc\,{\sc ii}}     
\documentclass{aa}
\usepackage{graphics}
\begin{document}

   \thesaurus{06     
              (08.01.1;  
               08.06.3;  
               08.08.2;  
               08.16.4;  
               08.22.3   
               10.08.1; } 
   \title{A spectroscopic study of field BHB star candidates\thanks{
 Based on observations obtained at KPNO, operated by the Association of
 Universities  for Research in Astronomy, Inc., under contract with the National
 Science Foundation, and the European Southern Observatory, Chile.}
 \fnmsep \thanks{
 Tables 4 and 5 are only available in electronic form at the CDS via  
 anonymous ftp to cdsarc.u-strasbg.fr}}

   \author{T. Kinman\inst{1}
           \and
           F. Castelli\inst{2}
           \and
           C.Cacciari\inst{3}
           \and
           A. Bragaglia\inst{3}
           \and
           D. Harmer\inst{1}
           \and
           F. Valdes\inst{1}
          }

   \offprints{T. Kinman}

   \institute{Kitt Peak National Observatory, National Optical Astronomy
              Observatories, Box 26732, Tucson, AZ 85726-6732, U.S.A. \\
              email: kinman@noao.edu
     \and
      CNR-Gruppo Nazionale Astronomia and Osservatorio Astronomico di 
      Trieste, v. Tiepolo 11, I-34131 Trieste, Italy.\\
      email: castelli@ts.astro.it
     \and
      Osservatorio Astronomico di Bologna, v. Ranzani 1,
      I-40127 Bologna, Italy.\\
      email: cacciari@bo.astro.it, angela@bo.astro.it
      }

   \date{Received 23 February 2000 / Accepted 7 June 2000 }

  \titlerunning{The chemical composition of field BHB star candidates}

  \authorrunning{Kinman et al.}

   \maketitle

   \begin{abstract}

   New spectroscopic observations are presented for a sample
   of thirty-one blue horizontal branch (BHB) star candidates 
    that are sufficiently nearby to have reliable proper motions.
    Comments are given on a further 
   twenty-five stars that have previously been suggested as BHB star candidates
   but which were not included in our sample.
   Moderately high-resolution spectra ($\lambda$/$\Delta \lambda \approx$
   15\,000) of twenty five of our program stars were taken with the coud\'{e} 
   feed spectrograph at Kitt Peak. Twelve of the program stars were also 
   observed with the 
   CAT spectrograph at ESO. Six of these program stars were observed from both 
   hemispheres. IUE low-resolution 
   spectra are available for most of our candidates and were used, in 
   addition to other methods, in the determination of their  \kt~and 
   reddening. A compilation of the visual photometry for these stars 
   (including new photometry obtained at Kitt Peak) is also given.
   Abundances were obtained from these spectra using models computed by
   Castelli with an updated version of the ATLAS9 code (Kurucz 1993a).

   All thirty one candidates are halo stars.
  Of these, twenty eight are classified as BHB stars because: 
\begin{description}
\item[(1)]they lie close to the ZAHB (in a similar position to the BHB stars in
globular clusters) in the \kt~versus \kg~plot. For all but one of these
stars,  far-UV data were available which were consistent with
other data (Str\"omgren photometry, energy distributions, H$\gamma$
profiles) for deriving \kt~and \kg. 
\item[(2)]they have a distribution of \ksi~($\leq$40~\ks) that is similar  to
that found for the BHB in globular clusters. Peterson et al. (1995)
and Cohen \& McCarthy (1997) have shown that the BHB stars in the globular clusters
M13 and M92  have a higher  \ksi~($\leq$ 40~\ks) than those in M3 and NGC 288
($\leq$20~\ks). The mean deprojected rotational velocity ($\overline{v}$) was
calculated for both the two globular clusters and the nearby BHB star samples.
A comparison of these suggests that both globular cluster \ksi~ types are
present in our nearby sample. No obvious trend is seen between \ksi~ and
either \kfo~ or [Fe/H].
\item[(3)]they have $-$0.99$\geq$[Fe/H]$\geq$$-$2.95 (mean [Fe/H] $-$1.67;
dispersion 0.42 dex), which is similar to that found for field halo RR Lyrae
and red HB stars.  These local halo field stars appear  (on average) to be more
metal-poor  than the halo globular clusters. The local sample of red giant
stars given by Chiba \& Yoshii (1998) contains a greater fraction of
metal-poor stars than either our halo  samples or the halo globular clusters.
The stars in our sample that have a \kt~that exceeds about
 8\,500 K show the He\,{\sc i}  ($\lambda$
4471) line with a strength that corresponds to the solar helium abundance.
\item[(4)]they show a similar enhancement of the $\alpha$-elements ($<$[Mg/Fe]$>$
= +0.43$\pm$0.04 and also $<$[Ti/Fe]$>$ = +0.44$\pm$0.02) to that
found for other halo field stars of similar metallicity.
\end{description}

      \keywords{Stars: abundances --
                Stars: fundamental parameters --
                Stars: horizontal branch --
                Stars: AGB and post-AGB  --
                Stars: variables: RR Lyrae --
                Galaxy: halo
               }
   \end{abstract}


\section{Introduction}

The {\it field} blue horizontal branch (BHB) stars have often been used to trace
 the galactic halo. Recent surveys of {\it distant BHB stars} include those of 
Pier (1983),
Sommer-Larsen \& Christensen (1986),
Flynn \& Sommer-Larsen (1988), 
Sommer-Larsen et al.  (1989),
Preston et al. (1991),
Arnold \& Gilmore (1992),
Kinman et al. (1994, hereafter KSK),
Beers et al. (1996) 
and Sluis \& Arnold (1998).

 The {\it nearby BHB stars} have been discovered sporadically over the past 
sixty years $-$ the majority by Str\"{o}mgren 4-colour photometry. Pre-eminent
 among the discoverers have been A. G. Davis Philip (Philip
1994) and Stetson (1991).  The only attempt, however,  to obtain a complete
sample of the nearby  BHB stars (and hence a local space density) appears to
be that by Green \& Morrison (1993).  Following  Philip et al. 
(1990), they showed that  a BHB star must not only have  the appropriate
Str\"{o}mgren \kx~ and \kac~ indices, but must also show  little or no
rotational broadening in high-resolution spectra.  This criterion must now be 
 somewhat modified since Peterson et al.  (1995) found BHB stars
with \ksi~as large as 40 \ks~in the   globular cluster M13.  Philip et al.
also considered that a BHB star must have an  appropriate location in
the  C(19$-$V)$_{\rm 0}$  vs. (b$-$y)$_{\rm 0}$  diagram\footnote{The
C(19$-$V)  index is derived from the magnitude at 1\,900\AA~ in IUE spectra and
the V magnitude of the star.}.  In the solar neighbourhood, disk stars greatly
outnumber halo stars and there  is a relatively high probability of finding
disk objects whose Str\"{o}mgren  indices are close to those of BHB stars. To
emphasize this, we give, in the  Appendix~A, a non-exhaustive list of stars
whose colours resemble those of BHB  stars but which most probably do not
belong to this category.  {\it The use of high-resolution 
spectra is mandatory for the selection of BHB stars in the solar 
neighbourhood since both accurate abundances and \ksi~are needed as criteria.}

 High resolution studies of nearby RR Lyrae stars have been made by  
Clementini et al. (1995) and by  Lambert et al. (1996). Both the RR Lyrae and
BHB   stars may be expected to have similar galactic kinematics.    There are,
however,   {\it disk} RR Lyrae stars in the nearby field, but there are (as
far as we know) no corresponding nearby field BHB stars that have disk 
kinematics\footnote{
One BHB star is known in the old metal-rich galactic cluster NGC  
6791~(Green et al. 1996) and extended blue   horizontal branches
have been found in the two disk metal-rich globular   clusters NGC 6388 and
NGC 6441 by Rich et al.   (1997). A search for   metal-rich BHB stars in the
galactic bulge has been started by Terndrup   et al. (1999).}.

While it is known that the field BHB stars generally show the low metal 
abundances that characterize halo stars, 
early determinations of these abundances show a rather wide
scatter (see Table A27 in KSK).    The first reliable determination was
probably that based on the co-added photographic spectra of HD 161817 by
Adelman et al. (1987). The metallic lines in the visible spectra of
BHB stars are relatively weak and early photographic spectra did not have
adequate signal-to-noise to measure these lines with sufficient accuracy.
Also, until relatively recently, it was not known with certainty whether or
not the evolution from the tip of the giant branch to the blue end of the
horizontal branch (with significant mass-loss) would change the composition in
the stellar atmospheres and whether diffusion effects would be present.
Glaspey et al. (1989) observed two HB stars in the globular cluster NGC 6752.
The hotter (16\,000 K) showed low rotation, a strong overabundance of iron and
a helium deficiency. The cooler (10\,000 K) showed a higher rotation and no
abundance anomalies compared to the red giants in the cluster. An example of a
hot (16\,430 K) field HB star is Feige 86 which was analyzed by Bonifacio et al.
 (1995). They found overabundances of the heavy elements and
other pecularities which might be attributed to diffusion. Lambert et al.
 (1992) used an echelle spectrograph with a                
 CCD detector to obtain  moderately high-resolution
spectra ($\lambda$/$\Delta \lambda \approx$ 18\,000) of two BHB stars (with
\kt~$\leq$10\,000 K) in the globular clusters M4 and NGC 6397. They found that
their metallicities agreed well with those found previously for the red giants
in these clusters.  Caloi (1999) proposed that the gap observed in the HB
sequence in many globular clusters at a \kf~of about zero is a surface
phenomenon and that stars with \kt~$>$10\,000 K will show peculiar chemical
compositions. Grundahl et al. (1999) have noted that a jump in both
Str\"{o}mgren $u$ and \kg~ occurs for stars hotter than \kt~=~11\,500 K in the
EHB of globular clusters and suggest that this marks the onset of radiative
levitation. This would explain the results of Glaspey et al. (1989) and the
more recent discoveries of strong overabundances of Fe in these hotter stars
in the globular clusters NGC 6752 (Moehler et al. 1999) and M13 (Behr et al.
1999). Behr et al. (2000) find that the HB stars in M13 that are cooler than
\kt~=~11\,000 K have high rotation (\ksi~$\sim$ 40~\ks) while the hotter stars
have a low rotation as might be expected if radiative levitation is operating.

All stars in our sample are cooler than 11\,000 K because hotter stars cannot
easily be identified as BHB stars by their Str\"{o}mgren indices. We should
therefore expect them to have chemical abundances that are similar to those
of other halo field stars such as halo RR Lyrae stars and halo red giants.
We should not expect abundance anomalies to be present,  and none have been
reported,  in the cooler field BHB stars that have previously been observed. In
addition to HD~161817 (Adelman \& Hill 1987), abundances have been derived
from CCD spectra for ten other nearby field BHB stars (Adelman \& Philip
1990, 1992, 1994, 1996a, 1996c). We consider that the abundances of BHB stars
based on photographic spectra by Klochkova \& Panchuk (1990) are  less
accurate because of the poor agreement of their equivalent widths with those
obtained from CCD spectra.

The aim of the present study is to provide data for a reliable sample of the
BHB stars in the solar neighbourhood. This includes the colour distribution,
the reddenings, the stellar parameters, the projected stellar rotations (\ksi~) 
and the abundance 
ratios. These data can be compared with data for BHB in globular clusters and
in other parts of the galaxy. 
 The galactic orbits of about half the stars in our present sample have
recently been calculated and analyzed by Altmann \& de Boer (2000).
It is intended, in a future paper, to derive the galactic orbits not only for
this BHB sample but also for other local samples of halo stars so that these
may be compared. These samples can help us to determine a better overall
definition of the local halo and determine to what extent it may be
distinguished from the disk populations\footnote{
For instance Majewski (1999) questions whether there is any  difference
between populations that have been identified as ``Intermediate Population II"
and as the ``Flat halo".  If the concept of stellar populations is to be
useful, there is a continual need to refine  the definitions of each
population so that misunderstandings are less likely to occur.}.

\section{The Selection of Candidates: notes on the individual objects}

Green \& Morrison (1993) found 10 BHB stars among the 23 candidates that they
studied and considered  that their sample was incomplete for BHB stars with V
$>$ 8.5. Many {\it nearby} BHB stars have been identified among high proper
motion stars and so any sample of them will have kinematic bias. Thus Stetson 
(1991) made 4-colour observations of high proper motion early-type stars taken
from the SAO Catalogue. More bright BHB stars might well be discovered by
using a more recent source of proper motions such as the PPM Star Catalogue
(R\"{o}ser \& Bastian 1991) in which a larger fraction of the stars have
spectral types. To do this, even for a part of the sky, would be a large
undertaking and we have therefore chosen to limit our observations to
previously identified BHB star candidates. Str\"{o}mgren photometry can only
be used to distinguish BHB stars that are redder than \kx~$\sim$$-$0.01~mag,
so that the hotter stars (belonging to the extended horizontal branch) are
excluded. This paper enlarges the local sample of definite BHB stars, but does
not affect our knowledge of the local BHB space density because this depends
only on the number of the very brightest of these stars. Our sample is limited
to stars that are brighter than $V$ = 10.9; these stars are near enough to have
significant proper motions and bright enough for their high-resolution spectra
to be obtainable with the Kitt Peak coud\'{e} feed spectrograph and with the
ESO-CAT spectrograph.

Thirty-one nearby BHB candidates were selected from the literature. 
BD~+00~0145 was listed by
              Huenemoerder et al. (1984). 
Twelve candidates were described by Philip (1984) who
gave finding charts and some references to their original identification.
These same stars and a (FHB) numbering system are also given in a more recent 
compilation and discussion of BHB star candidates by  Gray et al. 
(1996). The remaining eighteen candidates 
were identified as possible HB stars by Stetson (1991) as a result of
his 4-colour photometry of high proper motion A stars;  some of these had 
been identified earlier as BHB stars as noted below. 
\begin{description}
\item[\object{HD~2857}]   (FHB No. 61) Noted by Oke et al. 
 (1966).
\item[\object{HD~4850}]   Stetson (1991).  
\item[\object{BD~+00~0145}] Noted by Cowley (1958).  Kilkenny (1984)
 classified it as A0 from his 4-colour photometry.  Our colours (\kf~=~+0.04,
 \kd~=+~1.83) would not make it a BHB star if the  reddening given by the
 maps of 
 Schlegel et al. (1998, hereafter SFD) (\keb~=~0.028) is correct.
 Huenemoerder et al.   (1984)  included it in their list of HB stars
 but derived a high gravity (\kg~ = 3.9)  for this star. We find a similar
 gravity that is too high for it to be a  BHB star.
\item[\object{HD~8376}]   Stetson (1991).  
\item[\object{HD~13780}]   Stetson (1991).  
\item[\object{HD~14829}]   (FHB No. 23)~Philip (1969). 
\item[\object{HD~16456}]   Stetson (1991). This is the type c 
  RR Lyrae star CS Eri which was discovered by Przybylski (1970).
\item[\object{HD~31943}]   Stetson (1991).    
\item[\object{HD~252940}]   Stetson (1991).  
\item[\object{HD~60778}]   (FHB No. 47). Noted by Roman (1955a).
\item[\object{HD~74721}]   (FHB No. 48). Noted by Roman (1955a).
  Adelman \& Philip (1996a) give [Fe/H]~=~\kl~1.40 (see Sect.10.1). 
\item[\object{HD~78913}]   Stetson (1991).               
\item[\object{HD~86986}]   (FHB No. 66). Noted by Oke et al. 
 (1966).  Adelman \& Philip (1996a)  give [Fe/H]~=~\kl~1.80 (see Sect. 10.1).
\item[\object{HD~87047}]   Stetson (1991).    
\item[\object{HD~87112}]   Stetson (1991).              
\item[\object{HD~93329}]   Stetson (1991). Adelman \& Philip (1996a) give
 [Fe/H]~=~\kl~1.40 (see Sect. 10.1).
\item[\object{BD~+32~2188}] (FHB No. 1) Originally noted by Slettebak \& Stock 
 (1959), Gray et al. (1996)  described  this star as ``UV bright"
 or ``above horizontal branch". Mitchell et al. (1998) refer to this star as
 SBS 10. They derive \kt~= 11\,200 K and \kg~= 2.2 from \kac ~and the
 equivalent widths of H$\gamma$ \& H$\delta$ and \kt~= 10\,700 K and \kg~
= 2.28 from a fit of the observed high-resolution  spectrum to a grid of
 synthetic spectra. They classify it as a post-AGB star in their \kt~{\it vs}
 \kg~diagram in which the star lies close to the track for a 0.546 M$_{\sun}$
 post-AGB star (Sch\"{o}nberner 1983). Our results  agree with this
 classification.
\item[\object{HD~106304}]   Stetson (1991). This star had 
 previously been classified as a metal-poor HB star by Przybylski (1971).
\item[\object{BD~+42~2309}] (FHB No. 03). Identified as a BHB star by
  Philip (1967).
\item[\object{HD~109995}]   (FHB No. 67). Originally noted by Slettebak,
 Bahner \& Stock (1961), an early abundance analysis was made by
 Wallerstein \& Hunziker (1964). 
 Adelman \& Philip (1994) give [Fe/H]~=~\kl~1.89 (see Sect. 10.1).
\item[\object{BD~+25~2602}] Stetson (1991). Identified as an HB star by
 Hill et al. (1982).
\item[\object{HD~117880}]   (FHB No. 49). Noted by Roman (1955a, 1955b) 
 whose radial velocity ($-$44.6 \ks) differs completely from that given by 
 Greenstein \& Sargent (1974: +141 \ks) and by Kilkenny \& Muller 
 (1989) with which our velocity agrees. 
 Adelman \& Philip (1992) observed this star but only give an 
 abundance from two Si\,{\sc ii} lines. 
\item[\object{HD~128801}]   Stetson (1991).                  
 Adelman \& Philip (1996a) give [Fe/H]~=$-$1.26 (see Sect. 10.1). The [Ca/Fe]
 ratio which they derive ($-$1.03) is very low. 
\item[\object{HD~130095}]   (FHB No. 68). First suggested to be a halo star by
 Luyten (1957) (as CoD $-$26~10505) and later by 
 Greenstein \& Eggen (1966). Found to be a velocity variable by
 Przybylski \& Kennedy (1965b) and also Hill (1971). It does not,
 however, show light variations (ESA Hipparcos Catalogue 1997) nor was it
 found to be a photometric binary by Carney (1983).
 Adelman \& Philip (1996a) give [Fe/H] = $-$2.03 (see Sect.10.1).
\item[\object{HD~130201}]   Stetson (1991).          
\item[\object{HD~139961}]   Stetson (1991). 
 This star was first noted as an HB star by Graham \& Doremus (1968). 
 It is NSV~7204 in the New Catalogue of Suspected Variable stars (1982). 
 Corben et al. (1972) found a  range of 0.08 magnitudes in V over six
 observations. It does not appear to be variable according to the
 ESA Hipparcos Catalogue (1997). 
\item[\object{HD~161817}]   (FHB No. 69). Albitzky (1933) took the 
 first spectrum of HD~161817 and noted its large radial velocity.  Slettebak
 (1952) gives a referenced account of the early spectroscopic observations of
 this star. Burbidge \& Burbidge (1956) were the first to show that it was a
 metal-weak Population II star. Other early abundance determinations are
 mentioned by Takeda \& Sadakane (1997) who made a non-LTE study of its C, N, O
 and S abundances.  They adopted \kt~= 7\,500 K and \kg~= 3.0 and found \ksi~
 from between 14.3 \ks~and 15.9 \ks. Their re-analysis of Adelman \&
 Philips (1994) data leads to [Fe/H]~$\simeq$\kl~1.5;  Adelman \& Philip (1994)
 and Adelman \& Philip (1996a) derived  [Fe/H] = $-$1.74 and [Fe/H]=$-$1.66
 respectively (see also Sect. 10.1)
\item[\object{HD~167105}]   Stetson (1991).                  
 Adelman \& Philip (1996a) give [Fe/H]~=~$-$1.80 (see Sect. 10.1). 
\item[\object{HD~180903}]   Stetson (1991).               
\item[\object{HD~202759}]   (FHB No. 70). Noted by MacConnell et al.  
 (1971) as a probable HB star. It was shown by Przybylski 
 \& Bessell (1974) to have a very low $V$ amplitude (0.075 mag) with a
 period of 11.5 hours; it is  classified as a type c RR Lyrae star
 (AW Mic). Przybylski \& Bessell deduced from its colour that this star 
 must be very close to the blue edge of the instability strip; they 
 derived a \kt~of 7400 K and \kg~= 3.1 in good agreement with the
 values derived by us. It was confirmed spectroscopically as an HB~star by
 Kodaira \& Philip (1984). Adelman \& Philip (1990) give 
 [Fe/H]~=~$-$2.36 (see Sect. 10.1) 
\item[\object{HD~213468}]   Stetson (1991). The large radial velocity
 was discovered by Przybylski \& Kennedy (1965a) and it was noted as a
 probable HB star by MacConnell et al. (1971). 
\end{description} 


\begin{table*}
{{\bf Table~1.} Summary of Photometric Data for Horizontal Branch Star Candidates}
\begin{flushleft}
\begin{tabular}{clccccccccl} 
\hline\noalign{\smallskip}
  No.&~~Object&  $V$   &\kf &\kd &\kx &\kr &\kam &\kac &\kv &\kb \\
        &      &       &       &       &     &   &  &  &       &   \\
  (1)&~~~~~(2)  &(3)&(4)&(5)~~~~ &(6)&(7) &(8)&(9) &~~~(10) &(11) \\
\noalign{\smallskip}
 \hline 
\noalign{\smallskip}
 1 &HD~2857& 9.990 &+0.180 &2.094& \kc &2.787& \kc &\kc  &\kc&K(19,27,4)\\
 & & {\bf~9.990}&{\bf~+0.180}&{\bf~2.094}&{\bf~+0.135}&{\bf~2.787}&{\bf~0.113}& {\bf~1.212}  & {\bf~0.67}& (1) \\
              &      &   &  &     &  &  &     &      &            \\
2 &HD~4850 & {\bf~9.619}&{\bf~+0.066}&\kc        &{\bf~0.048}&{\bf~2.846}&{\bf~0.132}&{\bf~1.282}&\kc & (2)(3)\\
              &      &   &  &     &  &  &     &      &            \\
 3 &BD +00 0145    &10.58 &+0.040 &1.831& \kc &2.897& \kc &\kc  & \kc&K(7,8,3)\\
 & & {\bf~10.58}&{\bf~+0.040}&{\bf~1.831}&{\bf~+0.023}&{\bf~2.897}&{\bf~0.154}&{\bf~1.032}& \kc & (4)(5) \\
              &      &   &  &     &  &  &     &      &            \\
 4 &HD~8376   & 9.640 &+0.126 &2.123&\kc  &2.820& \kc &\kc & \kc&K(11,15,7)\\
 &  &{\bf~9.655}&{\bf~+0.126}&{\bf~2.123}&{\bf~+0.092}&{\bf~2.835}&{\bf~0.104}&  {\bf~1.273}& \kc &  (3)  \\
              &      &   &  &     &  &  &     &      &            \\
5&HD~13780  &{\bf~9.811}&\kc        &\kc        &{\bf~+0.088}&{\bf~2.816}&{\bf~0.119}&  {\bf~1.285} &  \kc & (3)   \\
              &      &   &  &     &  &  &     &      &            \\
 6 &HD~14829&10.29 &+0.033&2.004&\kc  &2.858&\kc  &\kc  &\kc&K(11,11,6)\\
 & & {\bf~10.306}&{\bf~+0.033}&{\bf~2.004}&{\bf~+0.036}&{\bf~2.858}&{\bf~0.135}&  {\bf~1.241}&{\bf 0.15}& (2) (6)\\
            &      &      &      &     &     &     &     &     &            \\
7 &HD~31943 & {\bf~8.262}&\kc        &\kc        &{\bf~+0.083}&{\bf~2.814}&{\bf~0.142}&   {\bf~1.226}  & \kc & (1) (3)   \\
              &      &   &  &     &  &  &     &      &            \\
 8 &HD~252940&9.090  &+0.211 &2.115&\kc  &2.769&\kc  &\kc &\kc &K(14,17,8)\\
 & & {\bf~9.098}&{\bf~+0.211}&{\bf~2.115}&{\bf~+0.159}&{\bf~2.768}&{\bf~0.091}&  {\bf~1.215} &  \kc  & (1) (3)   \\
              &      &   &  &     &  &  &     &      &            \\
 9 &HD~60778& 9.090&+0.104 &2.104& \kc &2.833& \kc &\kc&\kc  &K(8,8,6) \\
 & & {\bf~9.103}&{\bf~+0.105}&{\bf~2.104}&{\bf~+0.078}&{\bf~2.834}&{\bf~0.118}&{\bf~1.294}&{\bf 0.41} & (1) (3)  \\
              &      &   &  &     &  &  &     &      &            \\
10 &HD~74721& 8.700&+0.033 &2.028& \kc &2.857& \kc &\kc &\kc &K(12,12,6)\\
 & & {\bf~8.713}&{\bf~+0.033}&{\bf~2.028}&{\bf~+0.028}&{\bf~2.859}&{\bf~0.127}&{\bf~1.273}&{\bf 0.20}& (1) (3)   \\
              &      &   &  &     &  &  &     &      &         \\ 
11 &HD~78913 & {\bf~9.285}&{\bf~+0.089}&\kc    &{\bf~+0.066}&{\bf~2.842}&{\bf~0.118}&{\bf~1.281}& \kc & (1) (3)\\
              &      &   &  &     &  &  &     &      &            \\
12 &HD~86986& 8.000&+0.121 &2.103&\kc  &2.809& \kc &\kc&\kc  &K(11,5,4)     \\
 & & {\bf~8.000}&{\bf~+0.122}&{\bf~2.103}&{\bf~+0.092}&{\bf~2.825}&{\bf~0.109}&  {\bf~1.278}&{\bf~0.48}&(1) (3)     \\
              &      &   &  &     &  &  &     &      &            \\
13 &HD~87047      & 9.740&+0.112 &2.084&\kc  &2.796& \kc &\kc &\kc &K(6,12,8)    \\
 & & {\bf~9.752}&{\bf~+0.115}&{\bf~2.084}&{\bf~+0.091}&{\bf~2.797}&{\bf~0.105}&  {\bf~1.273} &  \kc &(3)  \\
              &      &   &  &     &  &  &     &      &            \\
14 &HD~87112      &9.710 &$-$0.023&1.860&\kc  &2.839&\kc  &\kc &\kc  &K(7,10,6)\\
 & & {\bf~9.717}&{\bf~$-$0.023}&{\bf~1.860}&{\bf~+0.001}&{\bf~2.840}&{\bf~0.115}&  {\bf~1.161}& \kc & (3)    \\
              &      &   &  &     &  &  &     &      &            \\
15 &HD~93329& 8.780&+0.080 &\kc  &\kc  &2.814&\kc  &\kc &\kc &K(3,9,5)  \\
 & & {\bf~8.790}&{\bf~+0.080}& \kc       &{\bf~+0.060}&{\bf~2.825}&{\bf~0.123}&   {\bf~1.315}  &  \kc & (1)(3)(8)  \\
              &      &   &  &     &  &  &     &      &            \\
16 &BD +32 2188&10.750&$-$0.050 &\kc &\kc  &\kc  &\kc  &\kc&\kc  &K(1,2)\\
 & & {\bf~10.756}&{\bf~$-$0.060}&\kc  &{\bf~+0.012}&{\bf~2.633\kaa\kaa}&{\bf~0.069}&  {\bf~0.921} &   \kc  & (1)   \\
              &      &   &  &     &  &  &     &      &            \\
17 &HD~106304 & {\bf~9.069}&{\bf~+0.022}&\kc    &{\bf~+0.025}&{\bf~2.845}&{\bf~0.114}&{\bf~1.162}&  \kc & (1) (3) \\
              &      &   &  &     &  &  &     &      &            \\
18 &BD +42 2309 & {\bf~10.771}&{\bf~+0.030}&\kc    &{\bf~+0.037}&{\bf~2.844\kaa}&{\bf~0.134}&{\bf~1.259}& \kc & (1)(3) \\
              &      &   &  &     &  &  &     &      &            \\
19 &HD~109995&7.630 &+0.052 &2.083&\kc &\kc  & \kc &\kc &\kc &K(1,4) \\
 & & {\bf~7.602}&{\bf~+0.055}&{\bf~2.083}&{\bf~+0.049}&{\bf~2.848}&{\bf~0.117}&{\bf~1.305}&{\bf~0.30} & (1)(3)    \\
              &      &   &  &     &  &  &     &      &            \\
\noalign{\smallskip} 
\hline
\end{tabular}
\end{flushleft}
\end{table*}

\begin{table*}
\addtocounter{table}{-1}  
{{\bf Table~1.} Summary of Photometric Data for Horizontal Branch Star Candidates (continued)}
\begin{flushleft}
\begin{tabular}{clccccccccl} 
\hline\noalign{\smallskip}
  No.&~~Object &  $V$   &\kf &\kd &\kx &\kr &\kam &\kac &\kv& \kb \\
        &       &       &       &       &     &   &  &  &      \\
  (1)&~~~~~(2)  & (3)&(4)&(5)&(6)~~~~ &(7)&(8) &(9)&(10) &~~~(11) \\
\noalign{\smallskip}
 \hline 
\noalign{\smallskip}
              &      &   &  &     &  &  &     &      &            \\
20 &BD +25 2602 &10.120&+0.070&2.056&\kc  &\kc  & \kc &\kc &\kc &K(1,1)      \\
 & & {\bf~10.120}&{\bf~+0.060}&{\bf~2.056}&{\bf~+0.048}&{\bf~2.850}&{\bf~0.128}&  {\bf~1.298}& \kc  & (1)(3) \\
              &      &   &  &     &  &  &     &      &            \\
21 &HD~117880 & {\bf~9.064}&{\bf~+0.075}&\kc   &{\bf~+0.056}&{\bf~2.855}&{\bf~0.125}&  {\bf~1.207} &{\bf 0.25} & (1) (3)        \\
              &      &   &  &     &  &  &     &      &            \\
        &      &      &   &  &     &  &  &     &      &            \\
22 &HD~128801&8.730 &$-$0.036&1.745&\kc  &\kc  &\kc  &\kc &\kc &K(5,11)\\
 & & {\bf~8.738}&{\bf~$-$0.036}&{\bf~1.745}&{\bf~$-$0.005}&{\bf~2.816}&{\bf~0.109}&  {\bf~1.056}& \kc & (1)(3)    \\
              &      &   &  &     &  &  &     &      &            \\
23 &HD~130095 & {\bf~8.128}&{\bf~+0.085}& \kc       &{\bf~+0.065}&{\bf~2.855}&{\bf~0.108}&   {\bf~1.256}  &{\bf 0.31} & (1)(3)       \\
              &      &   &  &     &  &  &     &      &            \\
24 &HD~130201 & {\bf~10.110}&{\bf~+0.075}&\kc  &{\bf~+0.061}&{\bf~2.860}&{\bf~0.109}&  {\bf~1.245} &    \kc  & (1)(3) \\
              &      &   &  &     &  &  &     &      &            \\
25 &HD~139961 & {\bf~ 8.860}&{\bf~+0.100}&\kc   &{\bf~+0.078}&{\bf~2.858}&{\bf~0.115}&  {\bf~1.298}&  \kc  & (1)(3)   \\
              &      &   &  &     &  &  &     &      &            \\
26 &HD~161817 & {\bf~6.976}&{\bf~+0.160}&\kc    &{\bf~+0.127}&{\bf~2.746}&{\bf~0.100}&{\bf~1.197}&{\bf 0.61} & (1)(3)  \\
              &      &   &  &     &  &  &     &      &            \\
27 &HD~167105 & {\bf~8.966}&{\bf~+0.025}&\kc    &{\bf~+0.036}&{\bf~2.849}&{\bf~0.120}&{\bf~1.260}& \kc & (1)(3)(8) \\
              &      &   &  &     &  &  &     &      &            \\
28 &HD~180903 & {\bf~9.568}& \kc       &\kc        &{\bf~+0.174}&{\bf~2.800}&{\bf~0.095}&  {\bf~1.255}& \kc & (1)(3) \\
              &      &   &  &     &  &  &     &      &            \\
29 &HD~202759 & {\bf~09.09v}& \kc       & \kc       &{\bf~+0.178}&{\bf~2.770}&{\bf~0.082}&   {\bf~1.164}  &  \kc & (1)(3)  \\
              &      &   &  &     &  &  &     &      &            \\
30 &HD~212468 & {\bf~10.926}&{\bf~+0.009}& \kc       &{\bf~+0.018}&{\bf~2.849}&{\bf~0.126}&   {\bf~1.246}  &  \kc & (1)(3)  \\
              &      &   &  &     &  &  &     &      &            \\
\noalign{\smallskip}
 \hline
 \end{tabular}
$\ddagger$~~~(K)(m,n,o): new observations by Kinman where m and n are the 
number of nights and the total number of BV \\
~~~~~observations and o is the total number of observations of $\beta$. \\
~~~~~Other sources used to form adopted values (columns 3 to 9):\\
~~~~~(1)~Hauck \& Mermilliod  (1998);
(2)~Alexander \& Carter 1971; (3)~Stetson 1991; (4) Klemola 1962;     
(5)~Kilkenny 1984;\\ ~~~~~(6) Gray et al. 1996; (7) Cousins 1972; (8) Oja 1987; (9) ESA Hipparcos Catalogue, 1997 (for $V$ magnitudes) \\
~~~~~\kv~ (column 10) are from  Arribas \& Martinez Roger  (1987).\\
\kq~~~~The value of $\beta$ is derived from observations by Philip \& Tifft (1971) only. \\
\kaa\kaa~~Mean of Hauck \& Mermilliod catalogue value and single observation by Kinman (2.644$\pm$0.010)\\ 
\kaa~~~~Mean of Hauck \& Mermilliod catalogue value and single observation by Kinman (2.873$\pm$0.015)\\ 
\end{flushleft}
\end{table*}

\begin{table*}
{{\bf Table 2.} Journal of KPNO spectra of BHB star candidates} 
\begin{flushleft}
\begin{tabular}{lcccccccccc}
\noalign{\smallskip}
\hline
\noalign{\smallskip}
 Star &  RA (J2000)&Dec (J2000)&  V &  Date & Start&  T& S/N & Rad.  Vel. &  no. of &$\sigma$ \\
          &            &      &         & (UT) & (UT)&(min.)& &(\ks)&lines&(\ks) \\
   (1)    &     (2)    &   (3)     &(4)   &   (5)&     (6)&  (7)&(8) &(9)& (10)&(11) \\
\noalign{\smallskip}
\hline
\noalign{\smallskip}
\object{HD~2857}&  00:31:53.8&$-$05:15:43& 9.99&1994~Sep~06& 06:41& 60&124&$-$155.7 & 9 &0.9 \\
\object{BD~+00~0145}&  00:56:26.9&+01:43:45&10.58&1994~Sep~06  & 07:47& 60&\kc&$-$261.0 & 1 &\kc \\
\object{HD~8376}&  01:23:27.8&+31:47:13 &9.66&1994~Sep~06 &  09:01& 60&114&+143.8 & 7 &1.1 \\
\object{HD~14829}&  02:23:09.2&$-$10:40:38&10.30&1995~Jan~01 &03:18& 60&\kc&$-$177.0 & 1 &\kc \\
\object{HD~16456}\ka\kw&  02:37:05.8&$-$42:57:48& 9.0v&1995~Jan~09& 02:05& 60&063&$-$139.8 & 7 &0.4 \\
\object{HD~31943}\kw&  04:57:40.7&$-$43:01:58& 8.26&1995~Jan~09& 04:22& 45&092&+088.2 &11 &0.8 \\
                  &            &           &     &1995~Jan~09& 05:08& 45&104&+088.7 &11 &0.7  \\
\object{HD~252940}&  06:11:37.2&+26:27:30 &9.10&1995~Jan~07  & 04:25& 40&100&+160.5 & 8 &0.7  \\
                  &            &          &    &1995~Jan~09  & 03:10& 40&106&+159.4 & 9 &0.8 \\
\object{HD~60778}&  07:36:11.7&$-$00:08:16 &9.10&1995~Jan~07& 07:24& 50&121&+041.1 &11 &0.7 \\
\object{HD~74721}&  08:45:59.2&+13:15:49 &8.71&1995~Jan~07  & 08:23& 22&108&+030.7 & 9 &0.6 \\
\object{HD~86986}&  10:02:29.6&+14:33:26 &8.00&1995~Jan~07  & 08:49& 12&096&+009.3 & 9 &0.7 \\
\object{HD~87047}&  10:03:12.8&+31:03:19 &9.75&1995~Jan~07  & 09:05& 60&105&+137.2 & 7 &0.4 \\
\object{HD~87112}&  10:04:38.8&+57:49:56 &9.71&1995~Jan~07& 10:11& 60&086&$-$171.9 & 7 &0.9 \\
\object{HD~93329}&  10:46:36.7&+11:11:03 &8.79&1995~Jan~09  & 11:20& 60&108&+205.2 & 8 &0.7 \\
\object{BD~+32~2188}&  11:47:00.5&+31:50:09&10.74&1995~May~04& 04:13& 67&072&+091.6 &11 &1.5 \\
\object{BD~+42~2309}&  12:28:22.2&+41:38:53&10.77&1995~May~02  & 04:00& 60&078&$-$145.3 & 8 &1.7 \\
\object{HD~109995}&  12:38:47.6&+39:18:32 &7.60&1995~May~03  & 03:18& 20&136&$-$129.0 & 8 &0.9 \\
                  &            &          &    &1995~May~03  & 03:40& 20&175&$-$130.1 & 7 &1.1 \\
\object{BD~+25~2602}&  13:09:25.6&+24:19:25&10.12&1995~May~03  & 04:06& 60&099&$-$067.0 & 8 &1.4 \\
\object{HD~117880}&  13:33:29.8&$-$18:30:54& 9.06&1995~May~04& 06:25& 60&137&$+$144.7 & 6 &0.5 \\
\object{HD~128801}&  14:38:48.1&+07:54:40& 8.74&1995~May~04& 07:28& 40&171&$-$081.1 & 6 &0.9 \\
\object{HD~130095}\kw&  14:46:51.9&$-$27:14:50& 8.13&1995~May~03& 07:08& 30&130&+066.0 & 5 &0.7 \\
\object{HD~139961}\kw&  15:42:52.9&$-$44:56:41& 8.86v&1995~May~03& 08:24& 50&097&+145.3 & 5 &0.5 \\
                  &            &           &     &1995~May~04& 08:27& 60&100&+143.2 & 9 &2.2 \\
\object{HD~161817}&  17:46:40.6&+25:44:57 &6.97&1994~Sep~06  & 02:30& 10&205&$-$363.8 &10 &0.6 \\
                  &            &          &    &1995~May~03  & 10:22& 20&203&$-$362.7 &10 &0.6 \\
\object{HD~167105}&  18:11:06.4&+50:47:32 &8.96&1995~May~04  & 09:38& 50&183&$-$173.6 &10 &2.0 \\
\object{HD~180903}\kw&  19:19:16.3&$-$24:23:11 &9.57&1995~May~04& 10:37& 60&100&+103.7 &11 &0.8 \\
\object{HD~202759}\ka\kw&  21:19:05.9&$-$33:55:08 &9.09v&1994~Sep~06& 05:08& 60&108&+021.3 & 8 &0.6 \\
\noalign{\smallskip}
\hline
\end{tabular}
\ka~~~RR Lyrae variable. \\
\kw~~~Also observed with ESO-CAT.\\
\end{flushleft}
\end{table*}
\begin{table*}
{{\bf Table 3.} Journal of ESO-CAT spectra of BHB star candidates.}
\begin{flushleft}
\begin{tabular}{lcccccccccc}
\noalign{\smallskip}
\hline
\noalign{\smallskip}
 Star  & RA(J2000) &Dec(J2000) &V   & Date  &Start & T    &S/N & Rad. Vel.&no. of &$\sigma$ \\
           &           &           &    & (UT)  &(UT)  &(min) &    &(\ks)&lines      &(\ks)   \\
 (1)       & (2)       & (3)       &(4) &(5)    &(6)   &(7)   &(8) &(9)    &(10)       &(11)     \\
\noalign{\smallskip}
\hline
\noalign{\smallskip}
\object{HD~4850}   &00:49:59.7  &$-$47:17:34 & 9.62 &1995~Sep~09 &04:38 &70 & 60 &$-$041.7 &10  &0.6 \\
           & & &                             &1995~Sep~09 &07:39 &60 & 60 &$-$041.8 &8   &0.6 \\
\object{HD~13780}  &02:12:51.4  &$-$49:03:17    & 9.80 &1995~Sep~09 &05:52 &45 & 50 &+025.4 &9   &1.0 \\
           & & &                             &1995~Sep~09 &06:40 &45 & 40 &+025.4 &9    &1.0 \\
\object{HD~16456}\ka\kw  &02:37:05.8  &$-$42:57:48    & 9.0v &1995~Sep~09 &09:43 &30 & 65 &$-$158.9 &9    &0.8 \\
\object{HD~31943}\kw  &04:57:40.7  &$-$43:01:58    & 8.26 &1995~Sep~09 &08:50 &40 & 80 &+084.9 &12   &3.5 \\
\object{HD~78913}  &09:06:55.0  &$-$68:29:22    & 9.28 &1995~Apr~29 &00:26 &70 &105 &+316.8 &7    &0.7 \\
\object{HD~106304} &12:13:53.6 &$-$40:52:25    & 9.07 &1995~Apr~29 &01:48 &60 & 85 &+115.4  &3   &1.0 \\
\object{HD~130095}\kw &14:46:51.9  &$-$27:14:50    & 8.13 &1995~Apr~29 &02:58 &20 & 65 &+065.6 &3    &0.8 \\
\object{HD~130201} &14:48:19.9  &$-$45:40:12    &10.11 &1995~Apr~29 &03:46 &60 & 45 &+069.7 &2    &1.0 \\
           & & &                             &1995~Apr~29 &04:49 &60 & 40 &+069.3 &2    &1.0 \\
\object{HD~139961}\kw &15:42:52.9  &$-$44:56:41    & 8.86v &1995~Apr~29 &06:12 &40 & 65 &+142.6 &4    &0.8 \\
\object{HD~180903}\kw &19:19:16.3  &$-$24:23:11    & 9.57 &1995~Apr~29 &09:19 &45 & 45 &+105.6 &9    &1.1 \\
\object{HD~202759}\ka\kw &21:19:05.9  &$-$33:55:08    & 9.09v &1995~Apr~29 &07:03 &40 & 55 &+027.5 &5    &1.0 \\
           & & &                             &1995~Sep~09 &00:08 &60 & 65 &+018.5 &10   &0.7 \\
\object{HD~213468} &22:32:17.3  &$-$42:34:49    &10.92 &1995~Sep~09 &02:12 &60 & 25 &$-$174.3 &3    &0.8 \\
           & & &                             &1995~Sep~09 &03:15 &60 & 35 &$-$173.6 &5    &0.5 \\
\noalign{\smallskip}
\hline
\noalign{\smallskip}
\end{tabular}

\ka~~~RR Lyrae variable. \\
\kw~~~Also observed at Kitt Peak.\\
\end{flushleft}
\end{table*}

\begin{figure}
\resizebox{8.0cm}{!}{\includegraphics{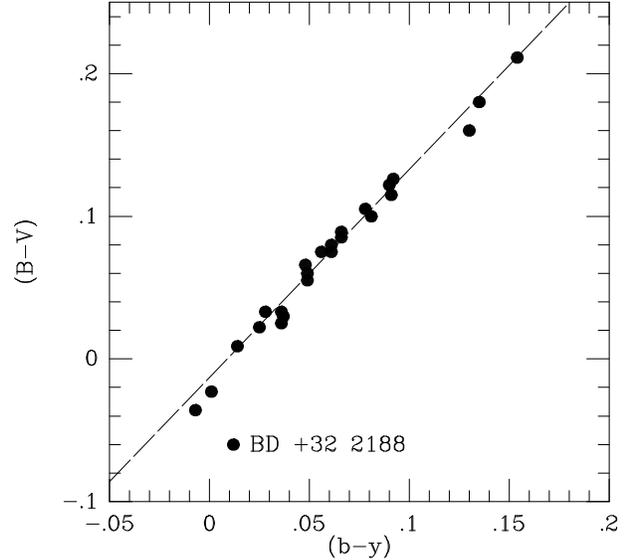}}
\hfill
\parbox[b]{88mm}{ 
\caption[]{A plot of \kf~ against \kx~ for the program BHB star
candidates. These colours show a close linear relationship except
for the PAGB star BD +32 2188.
}
\label{Fig01}}
\end{figure}

\section{Photometric observations in the visible spectrum}

Table~1 
is a compilation of both existing and new photometry for all our BHB  
candidates except for the RR Lyrae variable HD~016456\footnote{
The amplitude of this RR Lyrae is large enough ($\Delta V \sim$ 0.5 mag) for
its colours to be quite variable. Although Str\"{o}mgren photometry for this
star is given by Gray \& Olsen (1991) and Stetson (1991), there are not enough
individual observations (with phases) to determine the stellar parameters.}.
The final adopted photometry is given in boldface.

The new photometric observations were made by Kinman with the Mk III
photometer on the Kitt Peak 1.3-m telescope (with chopping secondary) on the
\kub~system as described by KSK. Additional \kub~observations were also made
with the Kitt Peak 0.9-m telescope using a 512 $\times$ 512 CCD under control
of the CCDPHOT program; details of this observing system are given by Kinman
(1998). The \kub~photometry  gives \kf~on the Johnson system and a hybrid
\kd~index from the Str\"{o}mgren $u$ filter and the Johnson $B$ filter. The \kd~vs
 \kf~diagram can be used to separate BHB from other stellar types as
described by KSK. A \kd~vs \kf~diagram using the most recent data is shown
in Fig.~3 of Kinman (1998). There is a satisfactory separation of the BHB
stars and  RR Lyrae stars with \kf~$\geq$ 0.00, but bluer than this the
separation becomes rapidly more difficult. The \kd~vs \kf~diagram gives a
satisfactory way of distinguishing fainter BHB stars at high galactic
latitudes because the risk of confusion with other types of early-type stars
is not too severe and the integration times are smaller than for the
Str\"{o}mgren photometry; this is an important consideration for faint stars.
In the solar neighbourhood, however, there is a wide variety of early-type
stars and  these diagrams can only be used to provide BHB candidates.

\begin{figure*}
\resizebox{17.5cm}{!}{\includegraphics{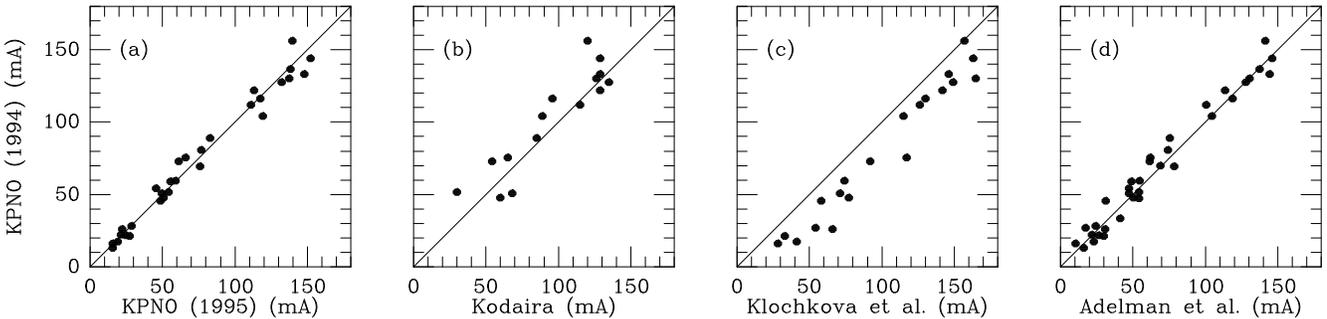}}
\hfill
\parbox[b]{180mm}{
\caption{Comparison of the equivalent widths of the lines in HD 161817 in a 
 KPNO spectrum 
 taken in 1994 with {\bf a} those from a 
KPNO spectrum taken in 1995;  
 {\bf b} those given by Kodaira (1964); 
 {\bf c} those given by Klochkova and Panchuk (1990) and {\bf d} those given
by Adelman et al. (1987). }
\label{Fig02}}
\end{figure*}

 \begin{figure*}
\resizebox{17cm}{!}{\includegraphics{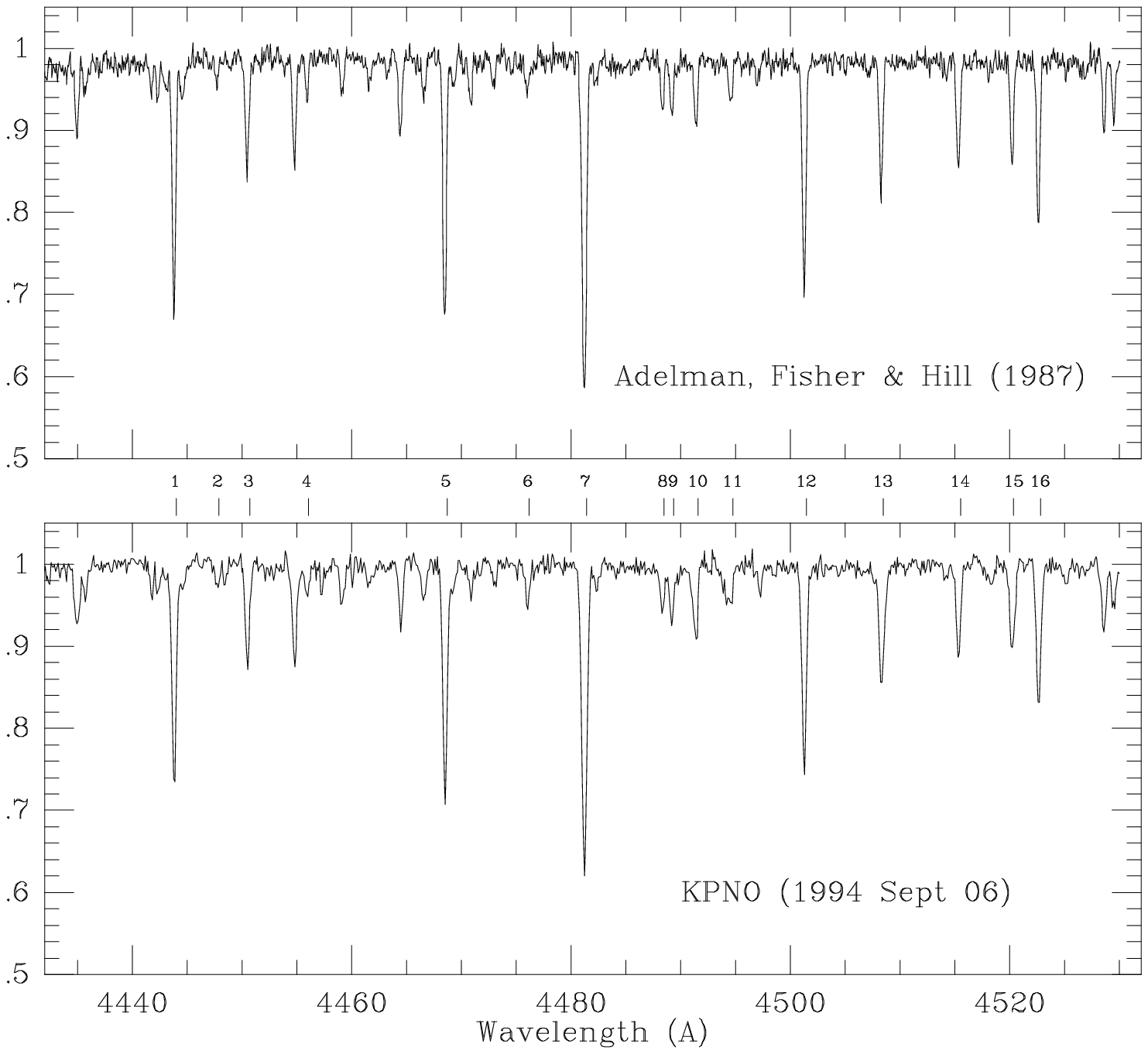}}
\hfill
\parbox[b]{180mm}{
\caption{Comparison of the spectrum of HD 161817 by Adelman et al. 
(1987) (above) with that taken at the Kitt Peak Coud\'e Feed telescope 
on 1999 Sept 06 UT (below). Line identifications:
(1) 4443.8 (Ti\,{\sc ii}),      
(2) 4447.7 (Fe\,{\sc i}),  
(3) 4450.5 (Ti\,{\sc ii}),
(4) 4455.9 (Ca\,{\sc i}),
(5) 4468.5 (Ti\,{\sc ii}),
(6) 4476.0 (Fe\,{\sc i}),  
(7) 4481.2 (Mg\,{\sc ii}),
(8) 4488.3 (Ti\,{\sc ii}),
(9) 4489.2 (Fe\,{\sc ii}),  
(10) 4491.4 (Fe\,{\sc ii}),  
(11) 4494.6 (Fe\,{\sc i}),  
(12) 4501.3 (Ti\,{\sc ii}),
(13) 4508.3 (Fe\,{\sc ii}),  
(14) 4515.3 (Fe\,{\sc ii}),  
(15) 4520.2 (Fe\,{\sc ii}) and
(16) 4522.6 (Fe\,{\sc ii}).
}
\label{Fig03}}
\end{figure*}

Some idea of the accuracy of the adopted photometric data can be appreciated
from the plots of \kx~against \kf~shown in Fig.~1. With the exception of
the Post-AGB star (BD +32 2188), which has a lower gravity than the remaining
stars, the BHB star candidates approximately follow the linear relationship:
\begin{displaymath}
    B-V = 1.459 (b-y) - 0.013
\end{displaymath}
which is shown by the dashed line in Fig~1. None of the BHB stars depart from
this relation by more than 0.01~mag in \kx. This suggests that these
quantities are not likely to be in error by more than one or two hundredths of
a magnitude.

Hipparcos magnitudes (which are of high accuracy and on a very homogeneous
system) are available for twenty one of the thirty stars given in Table~1. It
was found that the difference ($\Delta V$) between the Hipparcos magnitude and 
the mean $V$ magnitude\footnote{The observations of Stetson (1991) were given 
double weight in forming these means.}
 for our BHB star candidates could be expressed as the
following linear function of $(B-V)$:
\begin{displaymath}
    \Delta V = 0.002 + 0.275 (B-V) 
\end{displaymath}
The Hipparcos Catalogue (Vol.~1) (1997) gives values of $\Delta V$ for various
$(V-I)$ in Table 1.3.5 and values of $(V-I)$ for different $(B-V)$ in Table
1.3.7. Thus the catalogue values of $\Delta V$ may be obtained for various
$(B-V)$. These $\Delta V$ agree well with our linear relation at a $(B-V)$ of
0.00 and 0.22 but are up to 0.01 magnitudes larger at intermediate $(B-V)$.
The catalogue $\Delta V$ are for ``early type stars" and we have preferred our
relation because it refers to the specific class of stars that we are
studying. Our linear relation was therefore used to convert the Hipparcos
magnitudes to $V$ magnitudes and these are our adopted magnitudes. If no 
Hipparcos magnitude is available, the weighted mean $V$ magnitude was adopted.

Significant systematic differences exist between values of the
Str\"{o}mgren \kr~-index made by different observers (Joner \&  Taylor 1997).
Fortunately, many of the BHB candidates have been observed by Stetson (1991)
and were therefore on one system. New \kr~observations of a selection of our
candidates were made using BHB (and other stars of similar colour that were
observed by Stetson) as standards so as to be on his system\footnote{
The Str\"{o}mgren \kr~index is very valuable because it is not changed by
interstellar extinction but it does require measuring to a high accuracy to be
useful. The central wavelength of the narrow H$\beta$ filter undoubtedly
shifts with temperature and this means that careful calibration is needed in
order to get onto the standard system.}.
These new $\beta$ values are given in the first line of Table~1, when the
source K(n,m,o) is given. It should be noted that the large radial velocities of
BHB stars can cause their H$\beta$ line to be shifted (in the case of HD
161817  by as much as 6~\AA~) from the rest wavelength. The FWHM of the narrow
H$\beta$ filter is only 30~\AA, so that small inaccuracies may be expected
from this cause. As a check, synthetic \kr~indices were determined by
measuring the ``magnitudes" of the H$\gamma$-line through 30\AA~ and 150\AA~
bandpasses in our spectra (which do not include H$\beta$) using the 
{\it magband}
routine in the CTIO package of IRAF. 
 It was found that these synthetic \kr~indices  (on the photoelectric system) 
could be derived as a linear function of the difference between the broad and
 narrow 
 H$\gamma$ ``magnitudes"; these synthetic indices are given in column 5 of 
Table 15.
  In general, these   synthetic \kr~agree well with the mean
photoelectric values of \kr~taken from Hauck \& Mermilliod (1998) and given
in Table~1 and with our adopted values that are also given in Table 16. The
$rms$ difference between our synthetic \kr~and the adopted photoelectric
values for the BHB stars  is $\pm$0.009 if we omit HD 161817 for which the
difference is 0.031.

Photometric data both from the far ultraviolet and from the infrared can also
be used for the determination of the interstellar reddening and stellar
parameters. These data are discussed in Sects 5.4 and 7.5 respectively.

\section{Spectroscopic observations}

High resolution spectra of the thirty-one candidates  were obtained either
with the Kitt Peak coud\'{e} feed spectrograph or with the ESO-CAT
spectrograph at La Silla, Chile; six stars were observed at both observatories.
The journals of the observations are given in Table~2 and Table~3 for Kitt
Peak and ESO respectively. These tables contain the coordinates (columns 2 and
3) and $V$ magnitude (column 4) of each star. The UT date, starting time and
duration of each integration is given in columns 5, 6 and 7. The S/N of each
spectrum (column 8) was determined by using the IRAF $splot$ task which 
determined 
the ($\sqrt(mean~signal)/rms$) near the $\lambda$4481~\kmg~line in each
spectrum.
The measured heliocentric radial velocities and their $rms$ errors are given
in columns 9 and 11 and the number of lines used is in column 10. The agreement
between the two sets of observations for the stars in common is satisfactory
if we consider the number of lines that were available and also that three of
these stars (HD~16456, HD~202759 and possibly HD~139961) are variable. 
 The spectra of
BD~+00~0145 and  HD~014829  have a significantly poorer quality than the
others and were not used for a complete abundance analysis. We were able,
however, to measure the equivalent width of the  $\lambda$4481 \kmg~line in
these spectra and so derive an approximate [Fe/H] for these stars as explained
in Sect. 8.2.

\subsection{KPNO Observations}

The spectroscopic observations of the northern BHB candidates were made by
Kinman and Harmer using the Kitt Peak 0.9 m coud\'{e} feed spectrograph.  The
long collimator (F/31.2; focal length 10.11 m) and camera 5 (F/3.6; focal
length 108.0 cm) were used with grating A (632 grooves/mm) in the second order
with a Corning 4-96 blocking filter. This gives a 300 \AA~ bandpass covering
$\lambda\lambda$ 4260$-$4560 which includes both H$\gamma$, the 
\kmg~$\lambda$4481-line and a selection of \kfe, \kff~and \kti~lines. The
detector was a Ford 3KB chip (3072$\times$1024  pixels) with a pixel size of
15 microns. This gives a 3-pixel resolution of approximately 0.3\AA. The
nominal resolution at 4\,500\AA~is therefore 15\,000. Biases were taken at
the start of each night and a series of flat field quartz calibration
exposures were taken at the start and end of each night. ThAr arc lamp spectra
for wavelength calibration were made at the start, end and at frequent
intervals during each night. The spectra were reduced using standard IRAF
proceedures of bias subtraction, flat field correction and the extraction of
the [1-d] spectrum. The wavelength calibration was made using the ThAr arc
spectrum that was closest in time to the program spectrum.

The spectra were normalized to the continuum level interactively by using an
updated version of the NORMA code (Bonifacio 1989, Castelli \& Bonifacio 1990).
These normalized spectra were used to derive stellar parameters from the
H$\gamma$-profile and for the comparison with the synthetic spectra.

\subsection{ESO-CAT Observations}

The southern BHB candidates were observed by Bragaglia with the CAT + CES
(Coud\'e Auxiliary Telescope, 1.4 m diameter + Coud\'e Echelle Spectrograph)
combination at La Silla, Chile, during  April  and September 1995. This
equipment gives a single echelle order which was observed with two different
instrumental configurations. In April we used an  RCA CCD (ESO \#9), 1024
pixels long, covering about 40 \AA ~at a resolution of 0.14 \AA ~(or R
$\simeq$ 30\,000), while in September the detector was a Loral CCD (ESO \#38),
2688$\times$512 pixels, covering about 50 \AA ~at a resolution of 0.11 \AA
~(or R $\simeq$ 40\,000). In both cases the spectra were centered on the
$\lambda$4481~\kmg~line. Integration times ranged  from 10 to 70 minutes; the
faintest stars were observed twice.

The ESO-CAT spectra also were reduced with standard IRAF proceedures. The
extraction of the [1-d] spectra from the [2-d] images was performed weighting
the pixels according to the variance and without automatic cleaning from
cosmic rays. The wavelength calibration also was computed from a series of
Thorium arc-spectra and is estimated to be accurate to a few hundredths of an
\AA. IRAF tasks were used to clean the spectra from cosmic rays and defects,
for flattening and for normalization.

\begin{figure*}
\resizebox{17.5cm}{!}{\includegraphics{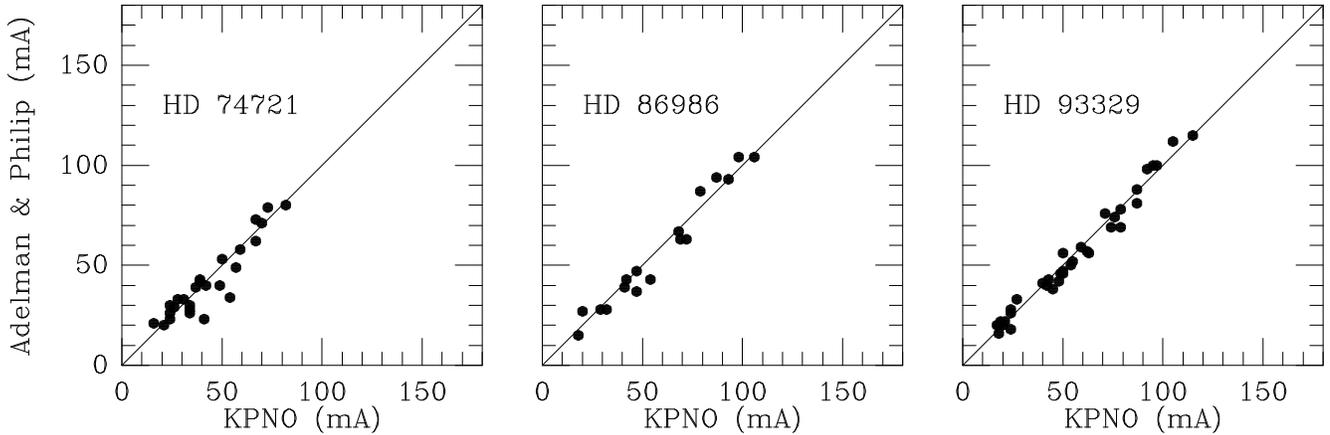}}
\hfill
\parbox[b]{180mm}{
\caption{Comparison of equivalent widths measured by Adelman \& Philip 
 (1994, 1996a) for HD 74721, HD 86986 and HD 93329 with those 
  measured from the KPNO spectra.}
\label{Fig04}}
\end{figure*}

\begin{figure}
\resizebox{6.35cm}{!}{\includegraphics{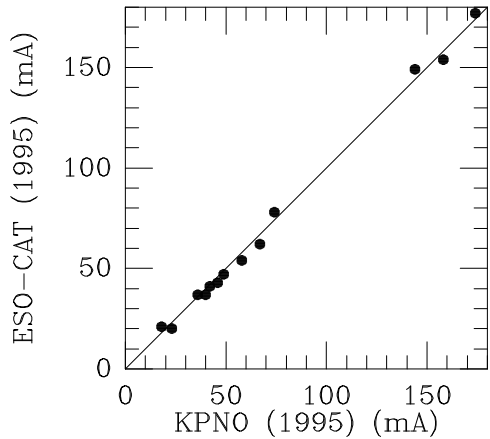}}
\hfill
\parbox[b]{88mm}{ 
\caption[]{A comparison of the equivalent widths from the ESO-CAT and
Kitt Peak spectra of HD~31943 and HD~180903.}
\label{Fig05}}
\end{figure}

 \begin{figure*}
\resizebox{17.0cm}{!}{\includegraphics{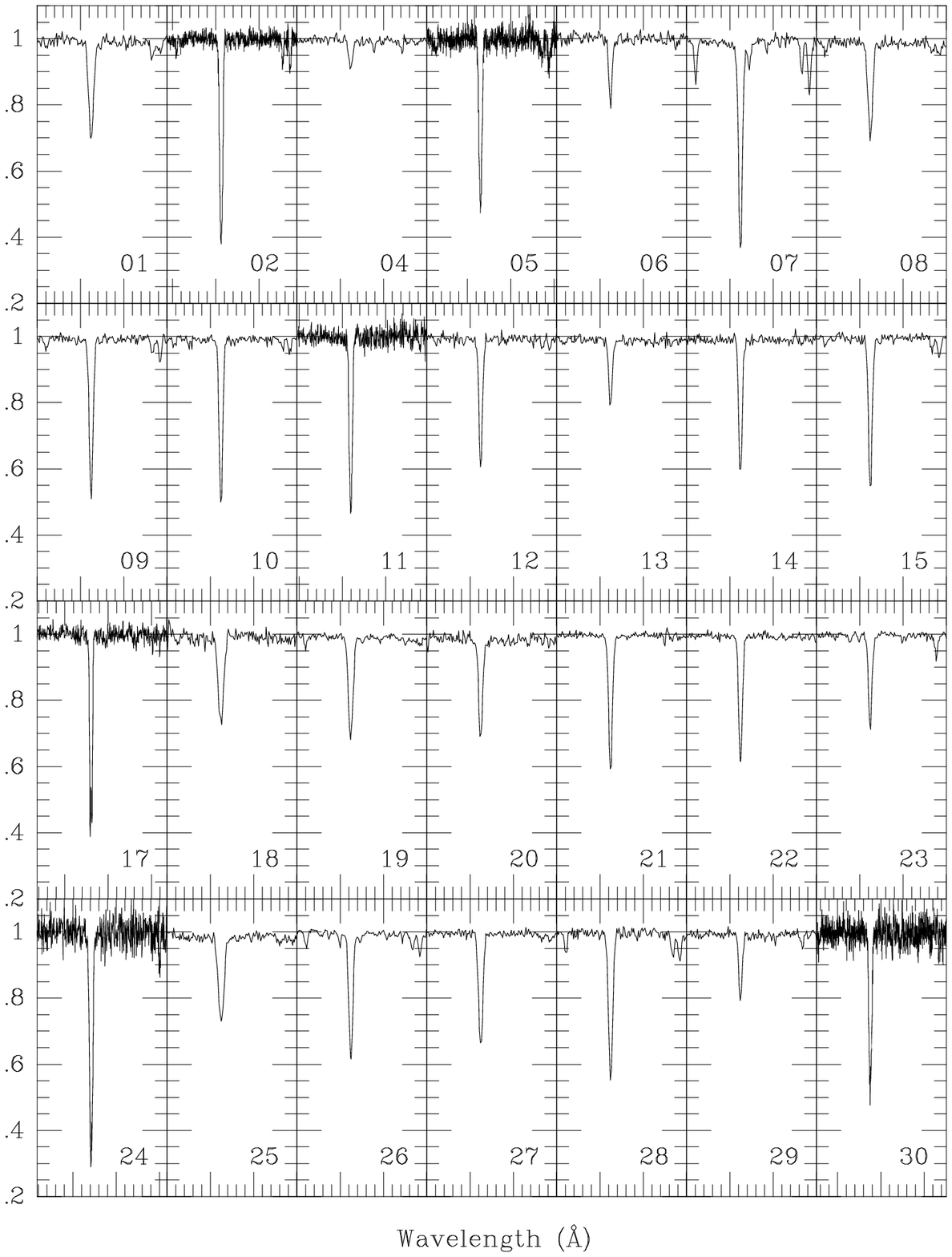}}
\hfill
\parbox[b]{180mm}{
\caption{The spectra of the 28  BHB stars in our sample 
 in the spectral range 4\,475 to 
4\,490 \AA. The stars are numbered as in Table 1.
}
\label{Fig06}}
\end{figure*}

\subsection{The measurement of the Kitt Peak (KPNO) and ESO-CAT spectra}

In order to be able to compare the spectra with the models, they were
transformed to zero velocity using the IRAF $dopcor$ routine. Line positions
and equivalent widths were obtained from the reduced  spectra using the IRAF
$splot$ routine, approximating (or deblending, if necessary) lines with
gaussians functions. When either two KPNO or 
 two ESO-CAT spectra were available for the same star, they were measured 
independently and the values of the equivalent widths were averaged. 
The comparison with the synthetic spectra  was made, however, with the spectrum 
of highest quality in order  
to compare H$\gamma$ profiles, to derive stellar parameters, to test 
abundances (derived from the averaged equivalent widths), to test the
microturbulent velocity and to derive the rotational velocities. 
Our measured equivalent widths, together with the derived abundances (Sect.~8), 
are given in Table~4 and Table~5 for the KPNO and ESO-CAT spectra respectively.
The wavelengths and multiplet numbers in these tables are taken from 
Moore (1945).

Fig.~2 gives a comparison of the equivalent widths that we and other observers
obtained from the spectrum of HD~161817.
Fig.2 (a) compares the equivalent widths obtained from the 1994 Kitt Peak
spectrum of HD~161817 with those obtained from a spectrum that was taken with
the same equipment at the end of the night of 1995 May 03 UT and which has  a
significantly poorer focus than any of our other program spectra; even in this
case, the effect on the equivalent widths appears to be minor.
The comparison in (b) is with the early photographic observations of Kodaira
(1964) which were made with the Palomar coud\'e spectrograph (10\AA/mm) and
shows a fairly large scatter (presumably because of the low S/N of single
photographic exposures) but the systematic differences are small. 
The comparison in (c) with the more recent photographic observations of
Klochkova \& Panchuk (1990), however, shows substantial differences in the
sense that the equivalent widths of these authors are systematically too large
with respect to the present measurements. On the other hand, the systematic
agreement of our data for this star with those of Adelman et al.  
(1987) shown in Fig. 2 (d) is quite good. The Adelman et al. spectrum was
derived from 12 co-added photographic spectra (6.5\AA/mm) and has a resolution
of about 25\,000; a 100\AA-section of this spectrum is shown in Fig.~3 (above)
together with the KPNO spectrum of 1994 Sept 06 UT (below)\footnote{
The first spectra of HD~161817 were taken by Albitzky (1933) who noted ``The
spectrum contains fairly good H lines, strong K  line, and a number of faint
metallic lines. The Mg 4481 is hardly perceptible and was measured on one
plate only." The  \kmg~($\lambda$4481) line is the strongest line (equivalent
width 210 m\AA) in this figure.}.
The noise in the KPNO CCD spectrum is such that some of the fainter
lines (equivalent widths less than about 30 m\AA) can be quite distorted. 
Such lines can generally be recognized and omitted from our analysis. 

Fig.~4 compares the equivalent widths for HD 74721, HD 86986 and HD 93329 from
the KPNO spectra with those measured by Adelman \& Philip (1994, 1996a) using
the same equipment. The agreement is satisfactory except for the Fe\,{\sc ii}
$\lambda$4555 and Ti\,{\sc ii} $\lambda$4563 lines in HD 74721. The KPNO equivalent
widths give abundances that are in agreement with the other lines of these
species and were preferred. Otherwise, these various comparisons give no
evidence for significant systematic differences between our  equivalent widths
and those given by Adelman \&  Philip.

Fig.~5 compares the KPNO equivalent widths of HD~31943 and HD~180903 with
those obtained from the higher resolution ESO-CAT spectra; the agreement is
very satisfactory.  The 28 spectra of the stars in our sample identified by
us as BHB stars 
are shown  for the spectral region 4\,475 to 4\,490 \AA~in Fig. 6; they are 
numbered as in Table~1.

The determination of the chemical composition of the BHB stars requires a
knowledge of the parameters that govern the physical conditions in their
atmospheres such as the effective temperature (\kt), the surface gravity
(\kg), the microturbulent velocity (\km) and also assumptions about
convection. These parameters are determined from both spectroscopic and
photometric data. The latter require correction for interstellar reddening and
this can be determined in several ways. These different methods and the extent
to which they agree are discussed in the next section.

\begin{table*}
{{\bf Table 6.} Galactic coordinates, distances, IUE colour $(18-V)$ and a comparison of the extinctions for the program \\ BHB stars by different methods.}
\begin{flushleft}
\begin{tabular}{lcccccccccc} 
\noalign{\smallskip}
\hline 
Star &l &b &Dist.   & $(18-V)$
  &\multicolumn{5}{c}{\keb}&[M/H]/\kt/\kg\kpf  \\
\cline{6-10 }
      & & &(pc)&  &SFD\ka     & ED\kw&Moon\kj&IUE\kp&IUE\kpe&       \\                               
 (1)&(2)&(3)&(4)&(5)&(6)&(7)&(8)&(9)&(10)&(11)   \\  
\noalign{\smallskip}
\hline 
\noalign{\smallskip}
HD~2857  &110.0&$-$67.6&717&1.085& 0.042&0.005&0.030&0.008&0.010&--1.50/7\,500/2.95    \\
HD~4850  &303.8&$-$69.8&563&0.272& 0.016&\kc  &0.000&0.010&0.010&--1.50/8\,400/3.10      \\
BD~+00~0145&126.8&$-$60.8&600:&$-$0.670&0.028&\kc&0.000&0.024&0.030&--1.50/10\,000/4.00      \\
HD~8376  &130.8&$-$30.6&565&0.554& 0.051&\kc  &0.035 &0.039&0.037&--2.00/8\,100/3.20       \\
HD~13780  &272.8&$-$63.0&640&0.651& 0.018&\kc  &0.008 &0.016&0.015&--1.50/7\,930/3.10        \\
HD~14829  &180.5&$-$62.3&715&$-$0.093& 0.024&0.025&0.000&0.023&0.023&--2.00/8\,950/3.30      \\
HD~16456  &256.3&$-$63.4&308&\kc& 0.021&\kc  &0.007&\kc &\kc & \kc         \\ 
HD~31943  &247.8&$-$38.4&317&0.774& 0.008&\kc  &0.000 &0.011&0.006&--1.00/7\,850/3.00         \\
HD~252940  &185.0&  +03.8&460&1.129& 0.168:&\kc&0.042 &0.051&0.06:&--1.75/7\,600/3.00            \\
HD~60778  &218.2&  +09.9&443&0.517& 0.054&0.040&0.016 &0.027&0.028&--1.50/8\,160/3.10          \\
HD~74721  &213.5&  +31.3&351&$-$0.042& 0.029&0.015&0.000 &0.006&0.005&--1.50/8\,800/3.30        \\
HD~78913  &284.6&$-$14.0&483&0.175& 0.068&\kc  &0.007 &0.060&0.062&--1.50/8\,870/3.25           \\
HD~86986  &221.9&  +48.8&274&0.617& 0.030&0.005&0.023 &0.035&0.025&--1.50/8\,000/3.20           \\
HD~87047  &196.5&  +53.3&633&0.584& 0.019&\kc  &0.000 &0.000&0.000&--2.50/7\,800/3.00           \\
HD~87112  &154.7&+47.7&511&$-$0.602&0.009&\kc&0.000 &0.000&0.000&--1.50/9\,700/3.50             \\
HD~93329  &235.4&  +56.6&386&0.392& 0.029&\kc  &0.000 &0.014&0.014&--1.50/8\,260/3.10           \\
BD~+32~2188&190.5&+75.2&4170&$-$0.872& 0.021&0.000:&0.000&0.023&\kc& \kc      \\
HD~106304  &295.3&  +21.4&369&$-$0.401& 0.082&\kc&0.000 &0.031&0.031&--1.50/9\,600/3.50          \\
BD~+42~2309&139.5&  +74.7&895&0.025& 0.018&0.020&0.000 &0.012&0.013&--1.50/8\,730/3.30          \\
HD~109995  &134.3&  +77.5&211&0.198& 0.017&0.050&0.000 &0.020&0.020&--1.50/8\,558/3.15           \\
BD~+25~2602&359.2&  +85.1&707&\kc & 0.017&\kc  &0.000&\kc  &\kc&\kc           \\
HD~117880  &316.7&  +43.2&358&$-$0.077& 0.087&0.080&0.000 &0.066&0.064&--1.50/9\,300/3.50        \\
HD~128801  &000.8&  +58.1&306&$-$0.791& 0.027&\kc  &0.000 &0.004&0.004&--1.50/10\,135/3.50        \\
HD~130095  &332.3&  +29.0&241&0.103& 0.108:&0.085&0.016 &0.060&0.060&--2.00/8\,925/3.40          \\
HD~130201  &323.4&  +12.5&664&0.089& 0.103&\kc  &0.015 &0.056 &0.055&--1.50/8\,925/3.50          \\
HD~139961  &332.0&  +08.0&370&0.316& 0.149:&\kc&0.042 &0.058&0.060&--1.50/8\,600/3.30            \\
HD~161817  &050.4&  +24.9&185&0.999& 0.073:&0.000&0.000&0.000&0.000&--1.50/7\,525/3.00           \\
HD~167105  &078.7&  +26.9&372&$-$0.091&0.043&\kc  &0.000&0.030 &0.029&--1.50/9\,050/3.25         \\
HD~180903  &013.6&$-$16.6&523&1.220& 0.076&\kc  &0.103 &0.090&0.095&--1.50/7\,700/3.10       \\
HD~202759  &010.8&$-$44.3&447&1.285& 0.098&0.065&0.063 &0.063 &0.06:&--2.00/7\,465/3.00         \\
HD~213468  &355.1&$-$57.9&939&$-$0.228& 0.017&\kc  &0.000 &0.005&0.006&--1.50/9\,100/3.25       \\

\noalign{\smallskip}
\hline 
\noalign{\smallskip}
\end{tabular}           

\ka~~Derived from the whole sky map of Schlegel et al. (1998). \\
\kw~~Derived from the energy distribution. \\
\kj~~Derived from the Str\"{o}mgren colours using the Moon (1985) code. \\
\kp~~Derived from a comparison of the observed with the theoretical (ATLAS9) $(18-V)$ colours ($E(B-V)_{1}$).\\
\kpe~~Derived by comparing the observed $(18-V)$ vs. \kx~ colours with the corresponding 
     theoretical values ($E(B-V)_{2}$).\\
\kpf~~The [M/H]/\kt/\kg~that were used to obtain the reddenings that are 
given in columns (9) and (10). 
\end{flushleft}
\normalsize
\end{table*}

\section{The interstellar reddening for the program stars} 

We have estimated the interstellar reddening for our BHB candidates by two
direct and two indirect methods. The first direct method makes use of the
whole sky map of the dust infrared emission and the second is based on the
empirical calibration of the Str\"omgren colours. The indirect methods use
model atmospheres to compare the observed and computed visible
spectrophotometric data and the observed and computed UV colour indices.

\subsection{The interstellar reddening for the program stars from whole-sky maps}

 The reddening in the direction of our program stars was estimated from the
whole-sky maps of Schlegel et al. (1998, SFD) which give the 
total line-of-sight reddening ($E(B-V)_{total}$) as a function of the 
galactocentric coordinates ($l$, $b$). The reddening between the stars and the
observer (Table 6, column 6) was derived by multiplying $E(B-V)_{total}$ by
\mbox{($1-\exp(-\mid z \mid/h))$} where $z$ is the star's distance above the
galactic plane.  The value that was assumed for the scale height ($h$) was 
was taken to increase linearly from
50 pc for stars at a distance of 200 pc to 120 pc at a distance of 600 pc and
to remain constant thereafter. The stellar distances (Table 6, column 4) were 
computed assuming the $M_V$ {\it vs.} \kf~relation    
given by Preston et al. (1991) with the zero-point modified to give an 
$M_V$ of +0.60 at \kf~= 0.20 (the blue edge of the instability strip). 
The mean difference between the $E(B-V)$ found in this way from the SFD maps
and those given by Harris (1996) for 16 high-latitude globular clusters is
satisfactorily small (+0.004$\pm$0.003). 
At lower latitudes 
($\la$ 30$\degr$), however, the extinction is too patchy for the simple 
exponential model to be reliable and the reddenings found in this way are 
much less certain. The least reliable values (Table 6, column 6) are marked 
with a colon.

\begin{figure}
\resizebox{8.5cm}{!}{\includegraphics{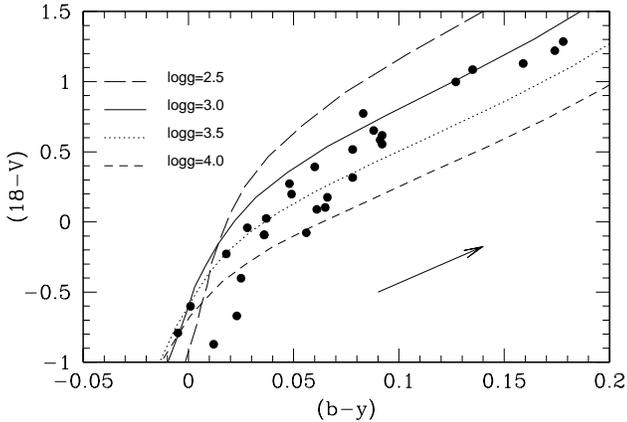}}
\hfill
\parbox[b]{88mm}{ 
\caption[]{The theoretical (ATLAS9) colours $(18-V)$ vs $(b-y)$ for [m/H]=--1.5 
and gravity from 2.5 to 4.0 in steps of 0.5. The arrow indicates the effect of
reddening.}
\label{Fig07}}
\end{figure}
 
\subsection {Reddening from the intrinsic colour calibration}

We used the  UVBYLIST code of Moon (1985) to derive the  intrinsic Str\"omgren
indices from the observed indices of our program BHB stars by means of
empirical calibrations that are taken from the literature.

The stars are divided into  eight photometric groups according to their 
spectral class and a different empirical calibration is used for each group. 
All our program stars except BD+32~2188 have 2.72$\leq \beta \leq$2.88 and 
belong to group 6 (stars of spectral type A3--F0 with luminosity class III--V).
We placed BD+32~2188 in group 4 (B0--A0 bright giants). 

A complete description of the dereddening procedures can be found in Moon
(1985) and also Moon \& Dworetsky (1985).  Here we recall that for group~6,
\kxo,  and hence the reddening, is calculated from the equations given by
Crawford (1979), which relate \kxo~ to the $\beta$, $\delta$c$_{1}$, and
$\delta$m$_{1}$($\beta$) indices. For group~4, a dereddening equation was
derived by coupling linear relations between the c$_{0}$ and $(u-b)_{0}$
colours, determined from Table~IV of Zhang (1983), with the reddening ratios
given by Crawford \& Mandwewala (1976):

\begin{displaymath}
(b-y)_{0} = (b-y) -  E(b-y)  
\end{displaymath}
\begin{displaymath}
m_{0} = m_{1} + 0.32E(b-y) 
\end{displaymath}
\begin{displaymath}
c_{0} = c_{1} - 0.20E(b-y) 
\end{displaymath}

We emphasize, however, that the empirical calibrations used by this method are
based on stars of spectral type B0 to M2 that lie on or near the  {\it main
sequence}.  Hence, for the BHB stars, the reddening, the intrinsic indices,
and results from them,  should be compared with the corresponding quantities
obtained with other methods in order to assess the reliability of this
dereddening procedure.   The reddening values derived from the UVBYLIST
program are given in column 8 of Table 6.

\subsection{Reddening from Spectrophotometric Data}

Spectrophotometric observations are available (Philip \& Hayes 1983, Hayes \&
Philip 1983) for twelve of our candidate BHB stars. An estimate of the
reddening was derived for these stars in the process of obtaining the stellar
parameters (Sect. 7) by fitting the observed energy distributions to a grid of
computed fluxes. For each star, we dereddened the observed  energy
distribution for a set of \keb~ values, sampled at steps of 0.005 mag and
starting from 0.000 mag. The adopted reddening law A($\lambda$) was taken from
Table~1 in Mathis (1990) for $A_{V}/E(B-V)$ = 3.1. For each \keb, the
stellar parameters are those that give the minimum $rms$ (Sect. 7.2).  We
assumed as the most probable \keb, that which gave the minimum $rms$  among
those given by the fitting procedure. These \keb~are listed in column~7 of
Table 6.

\subsection{Reddening from IUE Ultraviolet Data}

 It has been shown (Huenemoerder et al. 1984) that far-UV spectra can be 
useful for classifying BHB stars and for determining their reddening.
All of our candidate BHB stars (except HD 16456
and BD+25 2602) have UV IUE low-resolution (6 \AA) spectra, that have been
previously analysed and discussed (Huenemoerder et al. 1984, Cacciari 1985,
Cacciari et al. 1987 and de Boer et al. 1997 and references
therein ). We felt, however, that we should re-discuss the UV-spectra of these
 stars, especially the short-wavelength spectra (SWP, 1150-1980 \AA), using the
data that is in the IUE Final Archive\footnote{The data in the Final Archive
was reprocessed by the IUE Project using the latest and most accurate flux
calibrations and the most recent image-processing techniques so that no
further processing of this data is needed.}. In this way we could  extract all
the UV-spectra in a {\it homogeneous} way using the final IUE flux calibration
and image-processing techniques (Nichols \& Linsky 1996, Bohlin 1996) and
compare them to the latest model atmospheres for metal-poor stars computed by
Castelli with the ATLAS9 code and the Opacity Distribution Functions (ODFs)
from Kurucz  (Castelli 1999).

 The region of the UV spectrum that is best reproduced by the model 
atmospheres of stars with \kt~between 7\,500 K and 10\,000 K seems to be that 
in the region of 1\,800\AA~(Huenemoerder et al. 1984, Cacciari et al. 1987).
The values of the observed  fluxes at 1\,800\AA~were  obtained 
  from the SWP spectra by
averaging the flux over a rectangular bandpass 150 \AA~wide. The UV-colour
$(18-V)$   (given in Table 6, column 5) is defined as
\begin{displaymath}
(18-V)=-2.5(\log F_{1\,800}-\log F_{V})  
\end{displaymath}
where $logF_V=-0.4V - 8.456$ (Gray 1992). The UV-flux is strongly affected by
interstellar extinction. Consequently, the reddening can be estimated from the
$(18-V)$ colour by comparing it with that predicted by a model atmosphere
assuming that the temperature and/or gravity are known. We used  corrections
for reddening that were based on Seaton's (1979) reddening law, which gives
\begin{displaymath}
 E(18-V)/E(B-V)=4.748
\end{displaymath}
on the assumption that $A_V = 3.1E(B-V)$. The recent reanalysis of the
interstellar extinction by Fitzpatrick (1999) would give a somewhat higher
value (4.85) for this ratio. We also estimated the $(19-V)$-colour, defined
similarly to $(18-V)$, as a check on the consistency of our results.  The
colours $(18-V)$ and $(19-V)$ sample contiguous parts of the energy
distribution and so are highly correlated; $(19-V)$ is closer to the
2\,200\AA~ feature and can be more noisy because it is near the edge of the
energy distribution in the SWP spectra. We verified that both of these colours
gave consistent results but only have used  the more reliable $(18-V)$ colour
so as to avoid duplication.

We started with preliminary values of \kt,~\kg~and abundances that had been
derived from the model atmosphere analysis. When it was available, we
preferred the parameters derived from the H$\gamma$ profile because these
are reddening-independent. These stellar parameters were used to calculate an
intrinsic $(18-V)$ colour and the difference between this and the observed
$(18-V)$ colour gives the reddening $E(B-V)$, called here $E(B-V)_{1}$

The reddening $E(B-V)$ may also be estimated by comparing the observed $(18-V)$
vs. \kx~pairs with those predicted by the models at a given gravity, and is
called here $E(B-V)_{2}$. In Fig~7 we show the program stars and theoretical
relations for [M/H]=--1.5 and gravities 2.5, 3.0, 3.5 and 4.0 in the $(18-V)$
vs. \kx~plane.  If these two estimates of the reddening were consistent
within $\sim$ 0.02 mag, then we assumed that our initial values of \kt~and
\kg~ were reasonably correct. If this was not the case, we repeated the
analysis with different initial values. Our finally adopted values for
$E(B-V)_{1}$ and $E(B-V)_{2}$ are given  in columns 9 and 10 of Table~6,
 where they are compared with the other
reddening determinations. The finally adopted values for [M/H], \kt~and  
\kg~used to obtain these reddenings are given in column 11 of Table 6. They
generally agree with other determinations (Sect.~7). In the second method, the
UV data essentially constrain  the \kg~that is permitted for a given
reddening; in particular this strongly discriminates between FHB stars and
main sequence stars of higher gravity. The {\it internal} accuracies of the
parameters that are found by this way are estimated to be  $\pm$0.1 in 
\kg~and $\pm$100K in \kt.

We estimate that the typical error in these reddenings that comes from
photometric errors and the systematic errors to the absolute visual and UV
photometric calibrations is $\le$ 0.03 mag. Using only the 20 higher latitude
BHB stars where the SFD-derived reddenings  (Table 6, column 6) are 
reliable, the
mean value of the  SFD reddenings {\it minus} the mean of the two reddenings
derived from the IUE data  (Table 6, columns 9 and 10) 
 is +0.017$\pm$0.004. The
mean difference between the reddenings derived by the intrinsic colour
calibration (Sect. 5.2; Table~6, column 11) and the mean of the two reddenings
derived from the IUE data is $-$0.011$\pm$0.004 (28 stars).

A detailed comparison between the different reddening estimates is given in 
Table~6. Clearly systematic differences of the order of a few hundredths of a 
magnitude in \keb~ exist between the reddenings derived by the different 
methods even at high galactic latitudes. At lower latitudes, the differences
are  much larger because of the greater uncertainties in the reddenings derived
from whole sky maps. It does not seem possible to resolve these differences
without additional observations (e.g. mapping the extinction in the direction
of the BHB stars using the Str\"{o}mgren photometry of main sequence field
stars).

\section {The Model Atmosphere analysis}

The high-resolution spectra were analyzed with the model atmosphere technique.
Stellar parameters were estimated from Str\"omgren photometry, from
spectrophotometry, from H$\gamma$ profiles, from IUE ultraviolet  colours
and (for nine stars) from \kv~colours. The estimates obtained by these
different methods and the values of the parameters that we adopted are given
in Table~7.

Having fixed the model atmosphere, we computed abundances  from both the
equivalent widths and line profiles for each star observed at KPNO, except for
BD~+00~0145 and HD~14829. For these two stars  and for the BHB stars that
were observed at ESO (whose spectra only extended over 50 \AA) we estimated
the abundances from the equivalent widths alone.

We used model atmospheres and fluxes that were computed by Castelli with an
updated version of the ATLAS9 code (Kurucz 1993a). We adopted models for stars
with an $\alpha$-element enhancement [$\alpha/\alpha_{\sun}$]=+0.4 (which is
generally appropriate for halo stars). The symbol ``a'' near the metallicity
in column~6 of Table~7 indicates when these models were used. The convective
models (\kt $<$ 8750~K) were calculated with the no-overshooting 
approximation. More details about these models can be found in Castelli et al.
 (1997) and in Castelli (1999). The synthetic grids of
Str\"omgren indices,  Johnson \kv~indices and H$\gamma$  profiles were also
computed by Castelli from the above  models. Grids of models, fluxes and 
colours are available  either at the  Kurucz website (http://kurucz.harvard.edu)
or upon request.

We derived abundances from the equivalent widths using   
the WIDTH code (Kurucz 1993a),  modified so that we could derive the Mg
abundance from the measured equivalent width of the doublet \kmg~4481~\AA.
The SYNTHE code (Kurucz 1993b), together with the atomic line lists from
Kurucz \& Bell (1995), were used to compute the synthetic spectra.

\begin{table*}
{{\bf Table~7.} Comparison of the stellar parameters obtained by different methods.}
\begin{flushleft}
\begin{tabular}{lcccccc} 
\noalign{\smallskip}
\hline
\noalign{\smallskip}
\noalign{\smallskip}
 Star    &Data   &\keb  &\kt (K)    & \kg\kbx &[M/H]&Method \\
\noalign{\smallskip}
\hline 
\noalign{\smallskip}
\\
{\bf HD 2857}
& (c,~\kx ) & 0.042\kba &7\,730$\pm$120&3.15$\pm$0.04 & [--1.5a]  &  interpolation\\
&  (c,~\kx ) & 0.030\kbb &7\,650$\pm$120&3.10$\pm$0.04&     "     &    "\\
&[\kv     & (0.022) \kbc &7420$\pm$95\kaa   & (3.1)  &     "     &   "]\\
& En. Distr. &0.005\kbd &7\,400$\pm$100        &2.8$\pm$0.1 &     "     &fit, RMS(min)=0.0110\\
&  UV     &0.008\kbe,0.010\kbf  &7\,500$\pm$100&2.95$\pm$0.1       & "&     \\
&H$\gamma$&\kc    &7\,550$\pm$150& (3.0)           & " &  fit, RMS(min)=0.0104  \\
Mean&      &{\bf 0.022$\pm$0.009} &{\bf 7566$\pm$58}  &{\bf 3.00$\pm$0.08}  &{\bf  [--1.5a]} &  \\
\\ 

{\bf HD 4850}         
& (a,~r ) & 0.016\kba &8\,610$\pm$275&3.15$\pm$0.04 & [--1.5a]&  interpolation\\
& (c,~\kx )& 0.000\kbb &8\,350$\pm$250&3.35$\pm$0.05&    "    &" \\
&  UV      & 0.010\kbe,0.010\kbf &8\,400$\pm$100 &3.10$\pm$0.1 &    "     & \\
Mean: &      &{\bf 0.009$\pm$0.005}     &{\bf 8453$\pm$80 }  &{\bf 3.20$\pm$0.08}  &{\bf  [--1.5a ]} &   \\
\\ 

{\bf BD +00 145}         
& (a,~r )& 0.028\kba &9\,740$\pm$350&4.05$\pm$0.04 & [--1.5a]& interpolation\\
& (a,~r )& 0.000\kbb &9\,350$\pm$300&4.00$\pm$0.05&     "    &   "\\
&  UV   &0.024\kbe,0.030\kbf  &10\,000$\pm$200 &4.00$\pm$0.2  &       " &   \\
Mean:  &      &{\bf 0.018$\pm$0.009}     &{\bf 9697$\pm$189}  &{\bf 4.02$\pm$0.02}  &{\bf  [--1.5a ]}&    \\
\\

{\bf HD 8376}
& (c,~\kx) & 0.051\kba &8\,270$\pm$250&3.35$\pm$0.05& [--2.5]& interpolation\\
& (c,~\kx )& 0.035\kbb &8\,110$\pm$175&3.25$\pm$0.04 &   "   &  "           \\
&  UV     &0.039\kbe,0.037\kbf &8\,100$\pm$100 &3.20$\pm$0.1 &[--2.0a]&      \\         
&H$\gamma$&\kc    &8\,050$\pm$150 &(3.3)&   [--2.5]& fit, RMS(min)=0.0081\\
Mean: &      &{\bf 0.041$\pm$0.005}     &{\bf 8133$\pm$48} &{\bf 3.27$\pm$0.04}  &{\bf  [--2.5]} &   \\
\\

{\bf HD 13780}
& (c,~\kx) & 0.018\kba &7\,970$\pm$150&3.15$\pm$0.04& [--1.5a]&interpolation\\
& (c,~\kx) & 0.008\kbb &7\,890$\pm$150&3.10$\pm$0.04 &   "     &  "         \\
&  UV     &0.016\kbe,0.015\kbf &7\,930$\pm$100 &3.10$\pm$0.1    &    "  &     \\         
Mean:  &      &{\bf 0.014$\pm$0.003} &{\bf 7930$\pm$23 }  &{\bf 3.12$\pm$0.02 }  &{\bf  [--1.5a ]}  &  \\
\\

{\bf HD 14829}
& (a,r)  &0.024\kba & 9\,010$\pm$275&3.35$\pm$0.04&  [--2.0a]&interpolation\\
& (c,~\kx)  &0.000\kbb & 8\,700$\pm$250&3.30$\pm$0.05&     " &     "        \\
&[ \kv      &(0.018)\kbc & 9\,040$\pm$170\kaa & (3.2)   &     " &      "]  \\   
& En. Distr.&0.025\kbd & 9\,000$\pm$50 &3.0$\pm$0.1 &     "  &fit,RMS(min)=0.0100\\
& UV       &0.023\kbe,0.023\kbf & 8\,950$\pm$100 &3.20$\pm$0.1& "  &        \\
Mean:  &      &{\bf 0.018$\pm$0.006}     &{\bf 8915$\pm$73}  &{\bf 3.21$\pm$0.08}  &{\bf  [--2.0a]}  &  \\
\\
{\bf HD 16456\kq }
&H$\gamma$ &\kc& 6\,750$\pm$150&(2.8)   &    "        &fit,RMS(min)=0.0140 \\
Mean:   &      &\kc &{\bf 6750$\pm$150}  &{\bf (2.8)}  &{\bf  [--1.5a]}  &  \\
\\

{\bf HD 31943 }
&(c, \kx )    &0.008\kba & 8\,020$\pm$150 &3.35$\pm$0.04 &  [--1.0]& interpolation\\
&(c,~\kx)    &0.000\kbb & 7\,950$\pm$150  &3.30$\pm$0.04  &      " &    "     \\
& UV       &0.011\kbe,0.006\kbf & 7\,850$\pm$100 &3.00$\pm$0.1 & "  &\\
&H$\gamma$ & \kc    & 7\,750$\pm$100 &(3.2)  &"&  fit,RMS(min)=0.0061 \\
Mean:   &      &{\bf 0.006$\pm$0.003} &{\bf 7893$\pm$59} &{\bf 3.22$\pm$0.11 }& {\bf  [--1.0]}    \\
\\

{\bf HD 252940 }
&[(c, \kx )    &0.168\kba & 8\,660$\pm$300 &3.65$\pm$0.05 &  [--1.5a]& interpolation]\\
&(c,~\kx)    &0.042\kbb & 7\,490$\pm$120  &2.90$\pm$0.04 & "        & "      \\
& UV       &0.051\kbe ,0.06:\kbf& 7\,600$\pm$100 &3.00$\pm$0.1 &  [--1.75a]     &\\
&H$\gamma$& \kc    & 7\,600$\pm$150 & (2.9) &  [--1.5a]& fit,RMS(min)=0.0104\\
Mean:  &      &{\bf 0.048$\pm$0.006}  &{\bf 7563$\pm$37}  &{\bf 2.95$\pm$0.05}  &{\bf  [--1.5a]}&   \\
\\

\noalign{\smallskip}
\hline 
\noalign{\smallskip}
\end{tabular}           
\end{flushleft}
\end{table*}

\clearpage

\begin{table*}
\addtocounter{table}{-1}
{{\bf Table~7.} Comparison of the stellar parameters obtained by different methods. (continued)}
\begin{flushleft}
\begin{tabular}{lcccccc} 
\noalign{\smallskip}
\hline
\noalign{\smallskip}
\noalign{\smallskip}
 Star    &Data   &\keb  &\kt        & \kg\kbx  &[M/H]&Method \\
\noalign{\smallskip}
\hline 
\noalign{\smallskip}
\\ 

{\bf HD 60778 }
&[ (a,~r )    &0.054\kba & 8\,590$\pm$275 &3.10$\pm$0.05 &  [--1.5a]& interpolation]\\
&(c,~\kx)    &0.016\kbb & 8\,080$\pm$200  &3.20$\pm$0.04 &      "   & "     \\
& [\kv     & (0.028) \kbc &8\,110$\pm$90\kaa  &     (3.1)  &      "   & "]\\
& En. Distr.& 0.040\kbd &8\,100$\pm$100&3.1$\pm$0.1   & "&fit,RMS(min)=0.0105 \\
& UV       &0.027\kbe,0.028\kbf & 8\,160$\pm$100 &3.10$\pm$0.1 &  "&\\
&H$\gamma$& \kc    & 7\,950$\pm$300   &(3.1) &"& fit,RMS(min)=0.0072\\
Mean:  &      &{\bf 0.028$\pm$0.007}     &{\bf 8072$\pm$44}  &{\bf 3.13$\pm$0.03}&{\bf [--1.5a]}    \\
\\

{\bf HD 74721 }
&  (a,~r )    &0.029\kba & 9\,170$\pm$300 &3.30$\pm$0.04 &  [--1.5a]& interpolation\\
&  (a,~r)    &0.000\kbb & 8\,810$\pm$300  &3.25$\pm$0.04 &      "   &   "     \\
& [\kv     & (0.012) \kbc &8\,680$\pm$175\kaa  &  (3.3)     &     "    & "]   \\
& En. Distr.& 0.015\kbd &8\,850$\pm$50 & 3.3$\pm$0.2  &  "&  fit,RMS(min)=0.0093 \\
& UV       &0.006\kbe,0.005\kbf & 8\,800$\pm$100 &3.30$\pm$0.1 &  "&  \\
&H$\gamma$& \kc    & (8\,850) &3.4$\pm$0.1 &"& fit,RMS(min)=0.0080\\
Mean:  &      &{\bf 0.012$\pm$0.006}     &{\bf 8908$\pm$88}  &{\bf 3.31$\pm$0.02}  &{\bf  [--1.5a]}  &  \\
\\

{\bf HD 78913 }
&[  (a,~r )    &0.068\kba & 9\,000$\pm$290 &3.20$\pm$0.04 &  [--1.5a]& interpolation]\\
&(c,~\kx)    &0.007\kbb & 8\,160$\pm$200  &3.25$\pm$0.04 &      "   &    "     \\
& UV       &0.060\kbe,0.062\kbf & 8\,870$\pm$100 &3.25$\pm$0.1 &  "  \\
Mean:  &      &{\bf 0.034$\pm$0.027}     &{\bf 8515$\pm$355}  &{\bf 3.25$\pm$0.00}  &{\bf  [--1.5a]}  &  \\
\\

{\bf HD 86986 }
& (c,~\kx)    &0.030\kba & 8\,050$\pm$170 &3.25$\pm$0.04 &  [--1.5a]& interpolation\\
& (c,~\kx)    &0.023\kbb & 7\,980$\pm$170  &3.20$\pm$0.04 &      "    &    "     \\
& [\kv     &(0.022) \kbc &7\,870$\pm$140\kaa &    (3.2)       &  "    &  "]\\
& En. Distr.& 0.005\kbd &7\,850$\pm$50 &3.1$\pm$0.1  &  "&   fit,RMS(min)=0.0105 \\
& UV       &0.035\kbe,0.025\kbf & 8\,000$\pm$100 &3.20$\pm$0.1  &  " &  \\
&H$\gamma$ & \kc    & 7\,800$\pm$150 &(3.2) &" &    fit,RMS(min)=0.0085\\
Mean:   &      &{\bf 0.022$\pm$0.006}  &{\bf 7936$\pm$47}  &{\bf 3.19$\pm$0.03 }  &{\bf  [--1.5a]}  &  \\
\\

{\bf HD 87047 }
& (c,~\kx)    &0.019\kba & 7\,970$\pm$150 &3.15$\pm$0.04 &  [--2.5]& interpolation\\
& (c,~\kx)    &0.000\kbb & 7\,790$\pm$150  &3.05$\pm$0.04 &  "     &    "     \\
& UV       &0.000\kbe,0.000\kbf & 7\,800$\pm$100 &3.00$\pm$0.1  &        "     &  "     \\
&H$\gamma$& \kc & 7\,750$\pm$100 &(3.1) &        "&    fit,RMS(min)=0.0069\\
Mean:   &      &{\bf 0.006$\pm$0.006} &{\bf7828$\pm$49}  &{\bf 3.07$\pm$0.04}  &{\bf  [--2.5]}  &  \\
\\

{\bf HD 87112 }
& (a,~r)    &0.009\kba & 9\,810$\pm$340 &3.45$\pm$0.04 &  [--1.5a]& interpolation\\
& (a,~r)    &0.000\kbb & 9\,690$\pm$325 &3.45$\pm$0.04 &      "   &    "     \\
& UV       &0.000\kbe,0.000\kbf & 9\,700$\pm$100 &3.5$\pm$0.1 & "  &\\
&H$\gamma$& \kc    & (9\,750) &3.5$\pm$0.1 &      "       &fit,RMS(min)=0.0080\\
Mean:  &      &{\bf 0.003$\pm$0.003}  &{\bf 9733$\pm$38}  &{\bf 3.48$\pm$0.01}  &{\bf  [--1.5a]}  &  \\
\\

{\bf HD 93329 }
& (c,~\kx)    &0.029\kba & 8\,460$\pm$300 &3.25$\pm$0.06 &  [--1.5a]& interpolation\\
& (c,~\kx)    &0.000\kbb & 8\,130$\pm$200  &3.15$\pm$0.05 &    "    &   "    \\
& UV       &0.014\kbe,0.014\kbf & 8\,260$\pm$150 &3.1$\pm$0.1 &          "     &      \\
&H$\gamma$& \kc    & 8\,100$\pm$200        &3.0$\pm$0.1    & "& fit,RMS(min)=0.0046\\
Mean:   &      &{\bf 0.014$\pm$0.008}     &{\bf 8237$\pm$82}  &{\bf 3.12$\pm$0.05}  &{\bf  [--1.5a]}  &  \\
\\

{\bf BD +32 2188 }
& (c,\kr)    &0.021\kba & 10\,420$\pm$100 &2.10$\pm$0.05 &  [--1.0]& interpolation\\
& (c,~\kr)    &0.000\kbb &10\,400$\pm$100  &2.10$\pm$0.04 &      "     &    "     \\
& En. Distr.& 0.000\kbd &10\,500$\pm$500 &2.2$\pm$0.4   &  "  &fit,RMS(min)=0.0396\\
&H$\gamma$& \kc    &(10\,400) &2.1$\pm$0.05 &  "& fit,RMS(min)=0.0085\\
Mean:   &      &{\bf 0.007$\pm$0.007}     &{\bf 10440$\pm$30}  &{\bf 2.12$\pm$0.02}  &{\bf  [--1.0]}  &  \\
\\

\noalign{\smallskip}
\hline 
\noalign{\smallskip}
\end{tabular}           
\end{flushleft}
\end{table*}

\clearpage

\begin{table*}
\addtocounter{table}{-1}
{{\bf Table~7.} Comparison of the stellar parameters obtained by different methods. (continued)}
\begin{flushleft}
\begin{tabular}{lcccccc} 
\noalign{\smallskip}
\hline 
\noalign{\smallskip}
\noalign{\smallskip}
 Star    &Data   &\keb  &\kt        & \kg\kbx &[M/H]&Method \\
\noalign{\smallskip}
\hline 
\noalign{\smallskip}
\\

{\bf HD 106304}
& (a,~r)    &0.082\kba &10\,440$\pm$300 &3.55$\pm$0.04 &  [--1.5a]& interpolation\\
& (a,~r)    &0.000\kbb &9\,200$\pm$300  &3.45$\pm$0.04 & "        &   "     \\
& UV       &0.031\kbe,0.031\kbf &9\,600$\pm$100 &3.5$\pm$0.1 &  "  &\\
Mean:   &      &{\bf 0.038$\pm$0.024}  &{\bf 9747$\pm$365 }  &{\bf 3.50$\pm$0.03 }  &{\bf  [--1.5a]}  &  \\
\\

{\bf BD +42 2309 }
& (a,~r)    &0.018\kba & 8\,870$\pm$300 &3.20$\pm$0.04 &  [--1.5a]& interpolation\\
& (a,~r)    &0.000\kbb & 8\,650$\pm$275  &3.20$\pm$0.04 &      "  &"       \\
& En. Distr.& 0.020\kbd &9\,050$\pm$100&3.0$\pm$0.2    &          "     & fit,RMS(min)=0.0179 \\
& UV       &0.012\kbe,0.013\kbf & 8\,730$\pm$100 &3.3$\pm$0.1 &  "  &\\
&H$\gamma$& \kc    & (8\,850)&3.3$\pm$0.1 &  "&fit,RMS(min)=0.0077 \\
Mean:   &      &{\bf 0.013$\pm$0.004}  &{\bf 8825$\pm$88} &{\bf 3.20$\pm$0.05}  &{\bf  [--1.5a]}  &  \\
\\

{\bf HD 109995 }
& (a,~r)    &0.017\kba & 8\,610$\pm$300 &3.10$\pm$0.04 &  [--1.5a]& interpolation\\
& (c,~\kx)    &0.000\kbb & 8\,300$\pm$250 &3.25$\pm$0.05 &   "    &   "     \\
& [\kv     &(0.022) \kbc &8\,390$\pm$235\kaa      & (3.15)   &   "    &  "]\\
& En. Distr.& 0.050\kbd &8\,900$\pm$100 &2.9$\pm$0.3  &          "   & fit,RMS(min)=0.0190 \\
& UV       &0.020\kbe,0.020\kbf & 8\,560$\pm$200 &3.15$\pm$0.2 & "    &\\
&H$\gamma$& \kc    & 8\,200$\pm$250 & 3.0$\pm$0.1 &"&   fit,RMS(min)=0.0075\\
Mean:   &      &{\bf 0.022$\pm$0.010} &{\bf 8514$\pm$123}  &{\bf 3.08$\pm$0.07 }  &{\bf  [--1.5a]}  &  \\
\\

{\bf BD +25 2602 }
& (a,~r)    &0.017\kba & 8\,610$\pm$275 &3.15$\pm$0.04 &  [--2.0a]& interpolation\\
& (c,~\kx)    &0.000\kbb & 8\,320$\pm$275  &3.25$\pm$0.05 &  "    &"     \\
&H$\gamma$& \kc    & 8\,300$\pm$300 &3.1$\pm$0.1 & "&fit,RMS(min)=0.0074\\
Mean:   &      &{\bf 0.008$\pm$0.008} &{\bf 8410$\pm$100}  &{\bf 3.17$\pm$0.04 }  &{\bf  [--2.0a]}  &  \\
\\

{\bf HD 117880 }
& (a,~r)    &0.087\kba & 9\,590$\pm$325 &3.45$\pm$0.04 &  [--1.5a]& interpolation\\
& [(c,~\kx)    &0.000\kbb &8\,320$\pm$220  &3.55$\pm$0.04 &      " &    "]     \\
& [\kv     & (0.077) \kbc &9\,380$\pm$225\kaa &        (3.3) &     " &   "]\\
& En. Distr.& 0.080\kbd &9\,300$\pm$50&3.3$\pm$0.1              &  "   & fit,RMS(min)=0.0074 \\
& UV       &0.066\kbe,0.0064\kbf& 9\,300$\pm$100 &3.5$\pm$0.1 & " \\
&H$\gamma$& \kc    & 8\,950$\pm$500&3.1$\pm$0.2 &"& fit,RMS(min)=0.0109\\
Mean:   &      &{\bf 0.077$\pm$0.006} &{\bf 9285$\pm$131}  &{\bf 3.34$\pm$0.09}  &{\bf  [--1.5a]}  &  \\
\\

{\bf HD 128801 }
& (a,~r)    &0.027\kba &10\,640$\pm$400 &3.55$\pm$0.04 &  [--1.5a]& interpolation\\
& (a,~r)    &0.000\kbb &10\,160$\pm$400  &3.55$\pm$0.04 & "       & "      \\
& UV       &0.004\kbe,0.004\kbf &10\,140$\pm$200 &3.5$\pm$0.1 &  "\\
&H$\gamma$& \kc    &(10\,300)  &3.6$\pm$0.1  &  "&  fit,RMS(min)=0.0087\\
Mean:   &      &{\bf 0.010$\pm$0.008}     &{\bf 10313$\pm$163}  &{\bf 3.55$\pm$0.02  }  &{\bf  [--1.5a]}  &  \\
\\

{\bf HD 130095 }
&[ (a,~r)    &0.108\kba & 9\,650$\pm$350 &3.35$\pm$0.04 &  [--2.0a]& interpolation]\\
& [(c,~\kx)    &0.016\kbb &8\,300$\pm$250  &3.40$\pm$0.05 &      "   &    "]     \\
& [\kv     & (0.072)\kbc &8\,990$\pm$180\kaa  & (3.3)   &            &"]\\
& En. Distr.& 0.085\kbd &9\,100$\pm$50&3.3$\pm$0.1    &        "     & fit,RMS(min)=0.0073 \\
& UV       &0.060\kbe,0.060\kbf & 8\,920$\pm$100 &3.4$\pm$0.1 & "  &\\
&H$\gamma$& \kc    & (9\,000) &3.2$\pm$0.1 &  "& fit,RMS(min)=0.0082\\
Mean:    &      &{\bf 0.072$\pm$0.012}&{\bf 9010$\pm$90} &{\bf 3.30$\pm$0.03} &{\bf  [--2.0a]}  &  \\
\\

{\bf HD 130201 }
&[ (a,~r)    &0.103\kba & 9\,700$\pm$330 &3.45$\pm$0.04 &  [--1.5a]& interpolation]\\
& (c,~\kx)    &0.015\kbb &8\,370$\pm$250  &3.45$\pm$0.05 &    "    &  "   \\
& UV       &0.056\kbe,0.055\kbf & 8\,920$\pm$100 &3.5$\pm$0.1 &"  &\\
Mean:    &      &{\bf 0.035$\pm$0.020}  &{\bf 8645$\pm$275}&{\bf 3.48$\pm$0.03 }  &{\bf  [--1.5a]}  &  \\
\\

\noalign{\smallskip}
\hline 
\noalign{\smallskip}
\end{tabular}           
\end{flushleft}
\end{table*}

\clearpage

\begin{table*}
\addtocounter{table}{-1}
{{\bf Table~7.} Comparison of the stellar parameters obtained by different methods. (continued)}
\begin{flushleft}
\begin{tabular}{lcccccc} 
\noalign{\smallskip}
\hline 
\noalign{\smallskip}
\noalign{\smallskip}
 Star    &Data   &\keb  &\kt        & \kg\kbx &[M/H]&Method \\
\noalign{\smallskip}
\hline 
\noalign{\smallskip}
\\

{\bf HD 139961 }
& [(a,~r)    &0.149\kba & 9\,840$\pm$350 &3.30$\pm$0.04 &  [--1.5a]& interpolation]\\
& (c,~\kx)    &0.042\kbb &8\,350$\pm$250  &3.30$\pm$0.05 & "      &  "     \\
& UV       &0.058\kbe,0.060\kbf & 8\,600$\pm$100 &3.3$\pm$0.1 & "  &\\
&H$\gamma$& \kc    & 8\,600$\pm$250 &3.1$\pm$0.1& "& fit,RMS(min)=0.0098\\
Mean:  &      &{\bf 0.051$\pm$0.008}     &{\bf 8517$\pm$83}  &{\bf 3.23$\pm$0.07 }  &{\bf  [--1.5a]}  &  \\
\\

{\bf HD 161817 }
&[  (c,~\kx)    &0.073\kba & 8\,120$\pm$150 &3.50$\pm$0.04 &  [--1.5a]& interpolation]\\
& (c,~\kx)    &0.000\kbb &7\,510$\pm$120  &3.00$\pm$0.04 &      "     &"    \\
& [\kv     & (0.000) \kbc &7\,410$\pm$45\kaa &    (3.00)    &      "    &"]\\
& En. Distr.& 0.000\kbd &7\,550$\pm$200 &3.0$\pm$0.2    &        "   & fit,RMS(min)=0.0262 \\
& UV       &0.000\kbe,0.000\kbf & 7\,520$\pm$100   &3.0$\pm$0.1 &  " &   \\
&H$\gamma$ &\kc    & 7\,550$\pm$100      &(3.00) & "& fit,RMS(min)=0.0078\\
Mean: &      &{\bf 0.000$\pm$0.000} &{\bf 7533$\pm$10}  &{\bf 3.00$\pm$0.00 }  &{\bf  [--1.5a]}  &  \\
\\

{\bf HD 167105 }
&  (a,~r)    &0.043\kba & 9\,270$\pm$300 &3.25$\pm$0.04 &  [--1.5a]& interpolation\\
& (a,~r)    &0.000\kbb &8\,730$\pm$300  &3.25$\pm$0.04 &      "    & "       \\
& UV       &0.030\kbe,0.029\kbf & 9\,050$\pm$100 &3.25$\pm$0.1  &  " & \\
&H$\gamma$& \kc    & 9050$\pm$500&3.4$\pm$0.2 &  "& fit,RMS(min)=0.0063\\
Mean:  &      &{\bf 0.024$\pm$0.013}     &{\bf 9025$\pm$111}  &{\bf  3.29$\pm$0.04}  &{\bf  [--1.5a]}  &  \\
\\

{\bf HD 180903}
&[  (c,~\kx)    &0.076\kba & 7\,530$\pm$120 &2.90$\pm$0.04 &  [--1.5a]& interpolation]\\
& (c,~\kx)    &0.103\kbb &7\,750$\pm$120  &3.10$\pm$0.04 &      "       & "     \\
& UV       &0.090\kbe,0.095\kbf & 7\,700$\pm$100 &3.1$\pm$0.1  & "  &\\
&H$\gamma$& \kc   & 7\,600$\pm$100 &(3.1) &  "&fit,RMS(min)=0.0092\\
Mean:   &      &{\bf 0.098$\pm$0.005}     &{\bf 7683$\pm$44}  &{\bf 3.10$\pm$0.00 }  &{\bf  [--1.5a]}  &  \\
\\

{\bf HD 202759 }
&  (c,~\kx)    &0.098\kba &7\,790$\pm$120 &3.35$\pm$0.04 &  [--2.0a]& interpolation\\
& (c,~\kx)    &0.063\kbb &7\,510$\pm$120  &3.05$\pm$0.04 & "        &       "     \\
& En. Distr.& 0.065\kbd &7\,300$\pm$100 &2.8$\pm$0.1 &        "&fit,RMS(min)=0.0141 \\
& UV       &0.063\kbe ,0.06:\kbf& 7\,460$\pm$100 &3.0$\pm$0.1 & " &\\
&H$\gamma$& \kc    & 7\,550$\pm$150        &(3.0) & "&fit,RMS(min)=0.0128 \\
Mean:   &      &{\bf 0.072$\pm$0.009}     &{\bf 7522$\pm$79}  &{\bf 3.05$\pm$0.11 }  &{\bf  [--2.0a]}  &  \\
\\

{\bf HD 213468}
&  (a,~r)    &0.017\kba & 9\,280$\pm$320 &3.30$\pm$0.04 &  [--1.5a]& interpolation\\
& (a,~r)    &0.000\kbb &9\,060$\pm$300  &3.30$\pm$0.04 &      "        &    "     \\
& UV       &0.005\kbe,0.006\kbf & 9\,100$\pm$100 &3.25$\pm$0.1   &  " & \\
Mean:   &      &{\bf 0.008$\pm$0.005}     &{\bf 9147$\pm$68 }  &{\bf 3.28$\pm$0.02}  &{\bf  [--1.5a ]}  &  \\
\\

\noalign{\smallskip}
\hline 
\noalign{\smallskip}
\end{tabular}           
\\

\kbx~~If \kt~$<$ 8000 K, H$\gamma$ is almost independent of the gravity
  and \kg~is then given in parentheses. The gravities
  used to derive \kt~from the \kt,~\kg~\&~\kv~grid are also given in 
  parentheses and were not used to obtain the mean gravity. \\
\kq~~RR Lyrae variable at phase 0.42 (Sect. 10.7). \\
\kaa~~The quoted errors in these \kt~correspond to the range in 
\keb~between columns 6 \& 8 in Table 6.\\
\kba~~\keb~taken from SFD map (Table 6, column 6). \\
\kbb~~\keb~derived from Str\"{o}mgren colours using the Moon (1985) code 
 (Table 6, column 8). \\ 
\kbc~~\keb~ is the mean of the values given in columns 6 \& 8 of Table 6. \\
\kbd~~\keb~ was adjusted to obtain the best fit between the model and the 
  observed Energy Distribution. \\
\kbe~~\keb~ was derived by comparing the observed and theoretical ($18-V$)
 colours for the \kt,~\kg~\& [M/H] shown. \\
\kbf~~\keb~ was derived by comparing the observed and theoretical ($18-V$)
 vs. \kx~ for the \kt,~\kg~\& [M/H] shown. \\
\end{flushleft}
\end{table*}

\section{Stellar parameters}

\subsection{Stellar parameters from Str\"omgren photometry}
 
The stellar parameters \kt~and \kg~were found from the observed Str\"omgren
indices after de-reddening (as discussed in Sect. 5) by interpolation  in the
$uvby\beta$ synthetic grids. The adopted indices are those listed in boldface
in Table~1. Dereddened indices were obtained  both from \keb~values derived
from the SFD whole sky map (Table~6, column 6) and from the \keb~derived from
the UVBYLIST program of Moon (1985) (Table~6, column 8). The reddening relations
given in Sect. 5.2 were used in both cases.

When \kt$>$8\,500~K and \kg$<3.5$, the (c,~\kx) grid does not give an
unambigous determination of the parameters and the (a,~r) grid                
(Str\"{o}mgren 1966) is to be preferred. It should be noted that different 
values for the reddening may be derived for a star by the two methods, so that
it may lie in the (a,~r) plane according to one reddening determination, and
in the (c,~\kx) plane according to the other.

For each star, we started by selecting, from among the available grids of
colour indices, the one which had the metallicity closest to that given in the
literature or from a preliminary estimate based on the strength of the
$\lambda$4481 \kmg~line (KSK). After a new metallicity was found from the
model atmosphere analysis, it was  used to determine, by interpolation, the
colour grid which corresponded to this new metallicity. New parameters were
then redetermined. We found that the stellar parameters were, in practice,
relatively  insensitive to the value used for the metallicity. For this
reason, the stellar parameters found from the Str\"omgren indices and listed
in the first two (or three) lines of Table~7  are those relative to the
approximate  metallicity listed in column 6. At this stage, we also adopted a
microturbulent velocity of $\xi$= 2.0 \ks~for all the stars. In Table~7, the
data on the first line for each star correspond to the \keb~derived from the
SFD whole sky-map (Table~6, column 6), while the data on the second line
correspond to the \keb~derived using Moon's program (Table~6, column 8). The
reddening  $E(B-V)$=$E(b-y)$/0.73 is  given in column~3 of Table~7. The specific
Str\"omgren indices that we used to obtain \kt~and \kg~for each star are
given in the second column of Table~7. The errors in the parameters were
calculated by assuming an uncertainty of $\pm$ 0.015 mag for all Str\"omgren
indices except $\beta$ for which $\pm$ 0.005 mag was adopted. The actual error
in $\beta$ may well be larger than this for some stars as noted in Table~1 and
in Sect. 3.

 \subsection{Stellar parameters from spectrophotometry in the visible}

Spectrophotometric observations are available (Philip \& Hayes 1983, Hayes \&
Philip 1983) for some of our candidate BHB stars. Stellar parameters were
derived for these stars by fitting the observed energy distribution to the
fluxes of that grid, among those available to us, which had the closest
metallicity either to that given in the literature,  or to that obtained
from a preliminary 
estimate based on the strength of the $\lambda$4481 \kmg~line, or to that given
in a preliminary abundance analysis. The observed energy distribution was
dereddened as described in Sect. 5.3. The fitting procedure is that described
by Lane \& Lester (1984) in which the entire energy distribution is fitted to
the model which yields the minimum $rms$ difference. The search for the
minimum $rms$ difference is made by interpolating in the grid of computed
fluxes. The computed fluxes are sampled in steps of 50~K or 100~K in 
\kt~depending
 whether \kt$\le$ 10000~K or \kt$>$10000~K, and in steps of 0.1 dex in
\kg. The fluxes are actually given in steps of 250~K or 500~K in \kt~and in
steps of 0.5 dex in \kg, so the finer sampling was obtained by linear
interpolation.

The parameters  derived from the energy distributions are given on the ``En.
Distr." line in Table~7 and the adopted metallicity is that listed in column
6. The errors in the parameters were estimated from the ranges in \kt~and
\kg~for which $rms$=$rms$(min)+50\% $rms$(min). Lane \& Lester note that the
point-to-point scatter that determines the value of $rms$ may be less
important in their data than the calibration errors over large ranges of
wavelength. In our data the main uncertainty in deriving \kt~and \kg~from
the energy distribution probably comes from the spectrophotometric
observations being available at relatively few wavelengths. This makes it
difficult to get accurate results when they are fitted to the computed spectra.

\subsection {Stellar parameters from H$\gamma$}

For stars cooler than about 8000~K, the H$\gamma$ profile is a good
temperature indicator because it is almost independent of gravity, while for
hotter stars with \kt~between 8000~K and 10\,000~K it depends on both \kt~and
\kg. Above 10\,000~K, H$\gamma$ becomes a good gravity indicator, because it
is almost independent of temperature.

In order to derive the stellar parameters from the H$\gamma$ profiles given
by the KPNO spectra, we fitted the observed profiles (normalized to the
continuum level) to the  grids of profiles computed with the BALMER9 code
(Kurucz, 1993a). For each star, we used the grid computed for a microturbulent
velocity $\xi$ = 2 \ks~and the metallicity closest to that derived for the star
in a preliminary abundance analyses. We found that the fit was insensitive to
the adopted value of $\xi$.

We used an interactive routine to omit all the lines of other elements which
affect the H$\gamma$ profile, and by linear interpolation, we derived the
residual intensities R$_{\rm obs}$(i) of H$\gamma$ for each  $\lambda$(i)
sampled in the observed spectrum. We then used the same fitting procedure as
that used to derive the parameters from the energy distributions. For each
star, the parameters \kt~and \kg~are those which give the minimum $rms$
difference.

For stars cooler than 8000~K, this procedure gives \kt, but not \kg, because
the H$\gamma$ profile is not sensitive to gravity for these temperatures.
Therefore,  to derive \kt~for these stars, we adopted the average \kg~from
the Str\"omgren photometry and UV colours, since small differences in \kg~do
not change the value of \kt.

For stars with \kt~between 8000~K and 10000~K both \kt~and \kg~can be
obtained by the fitting procedure, but the situation is less satisfactory
because some ambiguity occurs in this range. For example, for [M/H] = $-$1.5,
the H$\gamma$ profile is almost the same for \kt~= 8800~K, \kg~= 3.0 as for 
\kt~= 9500~K, \kg~= 3.4. This means that very small differences in the reduction
procedure may give very different values for the parameters. Therefore, when
the parameters derived from the fitting procedure were in  reasonably
agreement with other determinations, we have given both \kt~and \kg. 
Otherwise we fixed either \kt~or \kg~and calculated the other parameter. The
stellar parameters found in this way are given on the ``H$\gamma$'' line in
Table~7. We give in parenthesis the parameters that were fixed in advance. As
for the energy distribution, the errors in the parameters were estimated from
the ranges in \kt~and \kg~for which $rms$ = $rms$(min)+50\% $rms$(min).

The main error in deriving \kt~and \kg~from H$\gamma$  comes from the
uncertainty in the normalization of the KPNO spectra; this is largely because
of a small non-linear distortion in the spectra which means that it is not a
straightforward task to decide where the wings of H$\gamma$ start. The
uncertainty in \kt~produced by the extraction procedure of the unblended
H$\gamma$ profile is of the order of 50~K.

\subsection{Stellar parameters from IUE data}

As we discussed in Sect. 5, the parameters derived from the ultraviolet fluxes
are those which lead to the most consistent values of reddening when one
compares the observed  $(18-V)$ colours and also the observed $(18-V)$ vs.
\kx~ colours with the corresponding theoretical values. The parameters found
in this way are on the ``UV" lines in Table~7.

 As a further check, we compared the whole UV-observed energy distributions,
for the stars that have both short- and long-wavelength IUE data, with model
energy distributions computed with the adopted parameters given in Table 7.
 We did not find systematic discrepancies between the models and the observed
data at 1\,600 \AA~and shorter wavelengths, as was found by Huenemoerder et al.
(1984) and Cacciari et al. (1987) using the 1979 Kurucz models.
 For more than half of these stars (HD~2857,
HD~4850, HD~13780, HD~14829, HD~31943, HD~74721 and HD~93329) the observed 
and calculated energy distributions match rather well over the entire IUE
wavelength range. For three stars (HD~78913, BD +00 0145 and HD~130201), the 
UV data suggest hotter temperatures than those adopted in Table~7. For three 
other stars (HD~8376, HD~60778 and HD~252940), the discrepancies may be caused 
by incorrect values of the adopted stellar parameters and/or uncertainties in 
the IUE and Str\"{o}mgren photometry. A detailed investigation, that compares
the observed and synthetic UV energy distributions using the latest model
atmospheres, would be of interest but is beyond the scope of this paper.
Meanwhile, we are confident that the use of the $(18-V)$ colour index gives
results for the reddening and physical parameters which are consistent with 
and give the same degree of uncertainty as those that would be derived by
using the entire IUE energy distributions.

\begin{table}
{{\bf Table 8.} Systematic differences ($\Delta$) from mean of \kt~ \\obtained by
different methods}
\begin{flushleft}
\begin{tabular}{cccc} 
\noalign{\smallskip}
\hline 
       &      &    &         \\
Method & No. of&Systematic Difference & Dispersion \\
       & Stars\kbx&from mean ($\Delta$)& ($rms$)  \\
    (1)&    (2)&   (3)&   (4) \\  
\noalign{\smallskip}
\hline 
\noalign{\smallskip}
c(\kx)\kba& 8  & +152$\pm$26 K & 69 K     \\
(a,r))\kba& 12 & +198$\pm$54 K &180 K     \\
c(\kx)\kbb&18  &$-$73$\pm$30 K & 122 K     \\
(a,r))\kbb& 8  &$-$218$\pm$64 K &169 K     \\
En~Distr & 10  & +27$\pm$60 K  &180  K     \\
UV        &25  & -10$\pm$26 K & 129 K     \\
H$\gamma$& 14 & $-$80$\pm$26 K & 95 K     \\
\kv      & 7  & $-$75$\pm$49 K  &120  K     \\

\noalign{\smallskip}
\hline 
\noalign{\smallskip}
\end{tabular}           
\\
\kbx~~Omitting HD 117880, HD 130095 \& HD139961. \\
\kba~~\keb~taken from SFD map (Table~6,~column 6).  \\
\kbb~~\keb~taken from Str\"{o}mgren colours using Moon (1985)\\~~~ UVBYLIST 
 program (Table~6,~column 8).\\

\end{flushleft}
\end{table}

\subsection{The Effective Temperatures from \kvo .}

The \kv~colours are available for nine of our candidate BHB stars (Arribas \&
Martinez Roger 1987). These \kv~colours are listed in Table~1. For seven of these
stars, \kv~colours had previously been given by Carney (1983). The mean
difference between these two sets of colours (Carney {\it minus} Arribas \&
Martinez Roger)  is 0.016$\pm$0.005; this corresponds to a temperature
difference of $\sim$50 K; presumably the systematic error in these colours is
of this order. The largest source of error in deriving temperatures in this
way is likely to come from the correction for reddening. We assumed that
\kev~= 2.72 \keb (Cohen et al. 1999) and took the \keb~to be the mean of the
\keb~ derived from the other methods given in column 3 of Table~7.  We assumed
the mean \kg~from the other determinations,  and derived \kt~by interpolation
in the \kt,~\kg~and \kvo~grid. These temperatures are given in Table~7 and
their errors are scaled from the estimated errors in \keb; they were not used
in deriving the mean \kt~but gave a useful independent check on the
temperatures obtained by other methods (see Table~7). We see that the
systematic difference between our adopted mean \kt~and the \kt~derived from
\kv~is only slightly larger than that expected from the likely systematic
errors in the \kv~colours.

\subsection{The comparison of the stellar parameters determined by the
 different  methods.}

Table~7 gives, for each star,  the straight means of \keb, \kt, and 
\kg~together with the errors of the means. In nearly all cases, the extinction
derived from the SFD maps exceeds that derived by using the Moon UVBYLIST
program (Table~6) and the use of the SFD extinctions with the $(c, (b-y))$ data
gives higher \kt~than those found by other methods. This difference is most
pronounced for low-latitude stars (HD~252940, HD~60778, HD~78913, HD~130095,
HD~130201, HD~139961, HD~161817 and HD~180903) whose computed extinction
depends upon an uncertain model of the local distribution of the interstellar
extinctions. We have therefore felt justified in rejecting the stellar
parameters that were derived using the SFD maps for these low-latitude stars. 
We also excluded the parameters determined from the Str\"{o}mgren indices
according to the Moon UVBYLIST program for the stars HD~117880 and HD~130095,
because of the excessive difference betwen the reddening derived from the Moon
code and that from the other determinations.  These excluded parameters (and 
those derived from \kv) are enclosed in square brackets in Table~7.

The differences  from these straight means were then computed for each method.
The average of these differences ($\Delta$) for \kt~are given for each method
in Table~8. The  dispersions given in column 4 of Table~8 are of the same order
as the error estimates of the \kt~given in column 4 of Table~7 but there are
significant differences. Thus the Energy Distribution method has among the
smallest errors in Table~7 but has one of the largest dispersions in Table~8. 
This, together with the undoubted presence of systematic errors associated
with each method has stopped us from using the error estimates for any attempt
at weighting the \kt~in Table~7; we have therefore adopted the straight means
for the parameters given in this Table.

\begin{table*}
{{\bf Table~9.} Abundances derived from the KPNO and CAT spectra.}
\begin{flushleft}
\begin{tabular}{llllll} 
\noalign{\smallskip}
\hline 
      & HD~2857                   & HD~4850         & BD +00 145        & HD 8376             & HD 13780 \\
      &  KPNO                     &  CAT            &    KPNO           &  KPNO               &    CAT\\
\noalign{\smallskip}
\hline 
Model &7\,550/3.0/[--1.5a]        &8\,450/3.2/[--1.5a]& 9\,700/4.0/[--1.5a] &8\,150/3.3/[--2.5a] & 7\,950/3.1/[--1.5a]\\
\km (\ks)       &3.00            & (2.00)           & (2.00)              &1.00              & (2.00) \\
\hline
\noalign{\smallskip}
\kmg   &$-$5.76 (1)&$-5.10$(1)          &$-$6.46 (1)&$-$6.86 (1)        & $-$5.58 (1)\\
\kca   &$-$7.16 (1)& \kc                &\kc        &\kc                &\kc            \\
\kti   &$-$8.19$\pm$0.23 (14)    &$-$7.65$\pm$0.04 (4)&\kc        &$-$9.12$\pm$0.14(5)& $-$7.90$\pm$0.17 (4)\\
\kcr   &$-$8.10 (1)              &\kc                 &\kc        &\kc                &\kc              \\
\kcc   &$-$7.92 (1)              &\kc                 &\kc        &\kc                &\kc               \\ 
\kfe   &$-$6.29$\pm$0.24 (6)     &$-$5.86 (1)         &\kc        &$-$7.49$\pm$0.06(2)&$-$5.97 (1)\\
\kff   &$-$6.25$\pm$0.17 (9)     &$-$5.76$\pm$0.05 (2)&\kc        &\kc                 &$-$6.01$\pm$0.1 (2) \\
\kbt &$-$11.72 (1)               &\kc                 &\kc        &\kc                 &\kc                 \\
\noalign{\smallskip}
\hline 
\\       
\hline 
\noalign{\smallskip}
            & HD~14829      &HD~16456\kba         &HD 16456\kba        & HD 31943                 &HD 31943 \\
            &   KPNO        &  KPNO               &  CAT               &    KPNO                  &   CAT     \\
\noalign{\smallskip}
\hline 
Model    &8\,900/3.2/[--2.0a] &6\,750/(2.8)/[--1.5a]&(7500)/(3.0)/[--1.5a] &7\,900/3.2/[--1.0a]       &7\,900/3.2/[--1.0a]\\
\km (\ks)& (2.00)            & 3.00               & (3.00)              &4.00                      &4.00\\
\hline
\noalign{\smallskip}
\kmg &          $-$6.47 (1)&$-$ 5.87 (1)          &$-$5.85 (1)         &$-$ 4.90 (1)        &$-$4.91 (1)\\
\kca &\kc                   &$-$7.22$\pm$0.02 (2) &\kc                 &$-$6.66$\pm$0.06 (3)&$-$6.53 (1)\\
\kti &\kc                   &$-$8.36$\pm$0.30 (14)&$-$8.31$\pm$0.17(4) &$-$7.67$\pm$0.21 (20)&$-$7.68$\pm$0.17 (4)\\
\kcr &\kc                   &\kc                  &\kc                 &$-$7.55 (1)          &\kc             \\
\kcc &\kc                   &$-$7.73 (1)          &\kc                 &$-$7.26 (1)          &\kc             \\
\kfe    &  \kc                    &$-$6.25$\pm$0.07 (6) &$-$6.19$\pm$0.06 (3)&$-$5.58$\pm$0.14 (8) &$-$5.47$\pm$0.02 (3)\\
\kff &\kc                   &$-$6.24$\pm$0.12 (12)&$-$6.19$\pm$0.03 (2)&$-$5.50$\pm$0.10 (13)&$-$5.49$\pm$0.05 (2)\\
\kbt &\kc                   &$-$11.44 (1)         &\kc                 &$-$11.15 (1)         &\kc\\

\noalign{\smallskip}
\hline 
\\
\hline 
\noalign{\smallskip}
            & HD~252940              &HD~60778            &HD 74721 &HD 78913       &HD 86986 \\
            &   KPNO                 &  KPNO              &  KPNO             &CAT            &   KPNO   \\
\noalign{\smallskip}
\hline 
Model    &7\,550/2.95/[--1.5a]         &8\,050/3.1/[--1.5a]   &8\,900/3.3/[--1.5a] &8\,500/3.25/[--1.5a] &7\,950/3.2/[--1.5a]\\
\km (\ks)       & 3.50                & 3.00               &4.00               & (2.00)            &2.50\\
\hline
\noalign{\smallskip}
\kmg &          $-$5.91 (1)          &$-$ 5.40 (1)         &$-$5.64 (1)         &$-$5.43 (1)  &$-$5.72 (1)\\
\kca &          $-$7.12 (1)          &\kc                  &\kc                 &\kc           &$-$7.07 (1)\\
\ksc &         $-$10.54 (1)          &\kc                  &$-$10.08 (1)        &\kc          &\kc         \\
\kti &          $-$8.27$\pm$0.19 (14)&$-$8.14$\pm$0.24 (15) &$-$8.11$\pm$0.13(13) &$-$8.23 (1)&$-$8.32$\pm$0.17 (4)\\
\kcc &         $-$7.85 (1)           &$-$7.79 (1)          &$-$7.64 (1)          &\kc         &$-$7.79 (1)          \\
\kfe   &       $-$6.33$\pm$0.12 (6)  &$-$6.03$\pm$0.21 (6) &$-$5.95$\pm$0.10 (5) &\kc         &$-$6.36$\pm$0.10 (4)\\
\kff &         $-$6.30$\pm$0.09 (6)  &$-$6.02$\pm$0.13 (12)&$-$5.97$\pm$0.08 (9) &\kc         &$-$6.34$\pm$0.10 (7)\\
\kbt &         $-$11.84 (1)   &$-$11.76 (1)   $-$11.61$^{S} (1)$       &\kc                  &\kc &$-$11.85 (1),$-$11.71$^{S}$ (1)\\

\noalign{\smallskip}
\hline
\\
\hline 
\noalign{\smallskip}
  & HD~87047                  & HD~87112            & HD 93329 & BD +32~2188        & HD 106304 \\
Model  &  KPNO                &  KPNO               &    KPNO          &  KPNO               &    CAT\\
\noalign{\smallskip}
\hline 
Model &7\,850/3.1/[--2.5a]      &9\,750/3.5/[--1.5a]   &8\,250/3.1/[--1.5a] & 10450/2.1/[--1.0]     & 9\,750/3.5/[--1.5a]\\
\km (\ks)   &  2.00:           & 2.00:               & 2.00               &1.00                  & (2.00) \\
\hline
\noalign{\smallskip}
\kmg   &$-$6.47 (1)              &$-5.55$(1)          &$-$5.27 (1)          &$-$5.50$\pm$0.05 (2)        & $-$5.28 (1)\\
\kca   & \kc                     &\kc                 &\kc                 &\kc                &\kc            \\
\ksc   & \kc                     &\kc                 &$-$10.25$\pm$0.03 (2)&\kc                &\kc             \\
\kti   &$-$8.89$\pm$0.12 (8)     &$-$8.19$\pm$0.11 (7)&$-$7.84$\pm$0.19 (14)&\kc                 & $-$8.29 (1)\\
\kcc   &\kc                      &$-$7.62 (1)         &$-$7.51 (1)          &$-$7.41 (1)         &\kc\\ 
\kfe   &$-$7.01$\pm$0.11 (5)     &$-$5.92 (1)         &$-$5.86$\pm$0.08 (6) &\kc                    &\kc\\
\kff   &$-$7.01$\pm$0.04 (2)     &$-$6.08$\pm$0.11 (5)&$-$5.87$\pm$0.06 (12) &$-$5.65$\pm$0.18 (10)   &\kc \\
\kbt & \kc                       &\kc                 &$-$11.27 (1)         &\kc                    &\kc  \\
\noalign{\smallskip}
\hline 

\end{tabular}           
\\
\end{flushleft}
\end{table*}
 
\begin{table*}
\addtocounter{table}{-1}

{{\bf Table~9.} Abundances derived from the KPNO and CAT spectra.(continued)}
\begin{flushleft}
\begin{tabular}{lccccc} 
\noalign{\smallskip}
\hline 
            &BD +42 2309       &HD~109995  &BD +25 2602         &HD 117880                 &HD 128801 \\
            &   KPNO           &  KPNO                 &  KPNO              &  KPNO                  &   KPNO     \\
\noalign{\smallskip}
\hline 
Model   &8\,800/3.2/[--1.5a]    &8\,500/3.1/[--1.5a]   &8\,400/3.2/[--2.0a]    &9\,300/3.3/[--1.5a]      &10\,300/3.55/[--1.5a]\\
\km (\ks)   & 2.00             & 3.00                & 4.00                 &2.00:                   &2.00\\
\hline
\noalign{\smallskip}
\kmg &      $-$5.66 (1)        &$-$5.84 (1)          &$-$6.15 (1)         &$-$5.51 (1)               &$-$5.55 (1)\\
\kca &\kc                      &\kc                  &\kc                 &\kc                       &\kc        \\
\kti &$-$8.15$\pm$0.13 (6)     &$-$8.31$\pm$0.12 (10)&$-$8.63$\pm$0.14(8) &$-$8.27$\pm$0.20 (9)      &$-$8.23$\pm$0.10 (8)\\
\kcc &\kc                      &\kc                  &$-$8.02 (1)         &$-$7.73 (1)               &$-$7.77 (1)\\
\kfe &  $-$6.13$\pm$0.22 (4)     &$-$6.24$\pm$0.12 (4) &$-$6.52$\pm$0.10 (5)&$-$6.13$\pm$0.03 (3)      &$-$6.01$\pm$0.05 (2)\\
\kff &$-$6.20$\pm$0.11 (3)     &$-$6.28$\pm$0.13 (7) &$-$6.53$\pm$0.02 (3)&$-$6.23$\pm$0.07 (6)      &$-$5.97$\pm$0.08 (6)\\

\noalign{\smallskip}
\hline 
\\
\hline
\noalign{\smallskip}
           & HD~130095  &HD~130095  &HD 130201         &HD 139961   &HD 139961 \\
            &   KPNO                 &  CAT            &  CAT                &KPNO          &CAT       \\
\noalign{\smallskip}
\hline 
Model         &9\,000/3.3/[--2.0a]  &9\,000/3.3/[--2.0a]  &8\,650/3.5/[--1.0a]  &8\,500/3.2/[--1.5a]  &8\,500/3.2/[--1.5a]\\
\km (\ks)     & 2.00                & 2.00              &(2.00)              & 3.00               &3.00\\
\hline
\noalign{\smallskip}
\kmg &   $-$6.11 (1)               &$-$6.12 (1)         &$-$4.66 (1)          &$-$5.80 (1)          &$-$5.79 (1) \\
\kti &  $-$8.72$\pm$0.18 (9)       &$-$8.86 (1)         &$-$7.82$\pm$0.16(2 ) &$-$8.29$\pm$0.16 (9) &$-$8.41 (1)\\
\kcc &  $-$8.06 (1)                &\kc                 &\kc                  &$-$7.78 (1)           &\kc          \\
\kfe &  $-$6.40 (1)                &\kc                 &\kc                  &$-$6.31$\pm$0.04 (4)  &\kc\\
\kff &     $-$6.42$\pm$0.15 (4)    &\kc                 &$-$5.54 (1)          &$-$6.20$\pm$0.17 (6) &\kc \\

\noalign{\smallskip}
\hline 
\\
\hline
\noalign{\smallskip}

           & HD~161817    &HD~167105 &HD 180903           &HD 180903              &HD 202759 \\
            &   KPNO                 &  KPNO           &  KPNO                     &CAT                    &KPNO       \\
\noalign{\smallskip}
\hline 
Model    &7\,550/3.0/[--1.5a]    &9\,050/3.3/[--1.5a]    &7\,700/3.1/[--1.5a]         &7\,700/3.1/[--1.5a]  &7\,500/3.05/[--2.0a]\\
\km (\ks)       & 3.00               & 3.00                 &3.00                    & 3.00              &2.0\\
\hline
\noalign{\smallskip}
\kmg &          $-$5.69 (1)          &$-$5.77 (1)           &$-$5.34 (1)           &$-$5.26 (1)          &$-$6.33 (1)\\
\kca &          $-$6.99 (1)          &\kc                   &$-$6.70$\pm$0.36 (3) $-$6.90$^{S}$ &\kc       &\kc         \\
\ksc &          \kc                  &\kc                   &$-$10.08$\pm$0.35 (3)  &\kc                  &$-$10.98$\pm$0.24 (2) \\
\kti &         $-$8.13$\pm$0.24 (14) &$-$8.20$\pm$0.14 (11) &$-$7.89$\pm$0.33(20)   &$-$7.84$\pm$0.01 (3) &$-$8.74$\pm$0.25 (12)\\
\kcr           &$-$8.02 (1)          &\kc                   &$-$7.99 (1)            &\kc                   &$-$8.58 (1)      \\
\kcc &         \kc                   &$-$7.80 (1)           &$-$7.69 (1)            &\kc                   &$-$8.12 (1)          \\
\kfe&          $-$6.06$\pm$0.12 (6)  &$-$6.08$\pm$0.05 (3) &$-$5.99$\pm$0.14 (8)    &$-$6.02$\pm$0.02 (2) &$-$6.67$\pm$0.05 (5)\\
\kff &         $-$6.12$\pm$0.07 (10)  &$-$6.12$\pm$0.12 (8)&$-$5.96$\pm$0.10 (9)    &$-$6.00$\pm$0.03 (2)  &$-$6.73$\pm$0.12 (8)\\
\kbt &         $-$11.60 (1)          &\kc                     &$-$11.46 (1),$-$11.31$^{S}$ (1)&\kc                  &$-$12.62 (1)$-$12.52$^{S}$ (1)\\

\noalign{\smallskip}
\hline 
\\
\hline
\noalign{\smallskip}

           & HD~202759              &HD~213468 \\
            &   CAT                 &  CAT         \\
\noalign{\smallskip}
\hline 
Model       &7\,500/3.05/[--2.0a]    &9\,150/3.3/[--1.5a]\\
\km (\ks)   &2.00                  & (2.00)             \\
\hline
\noalign{\smallskip}
\kmg &      $-$6.38 (1)            &$-$ 5.71 (1) \\
\kti &      $-$8.80$\pm$0.18 (4)   &$-$8.04$\pm$0.36 (3)\\
\kfe &      $-$6.57 (1)            &\kc         \\
\kff &      $-$6.68$\pm$0.08 (2)   &\kc           \\

\noalign{\smallskip}
\hline 
\noalign{\smallskip}
\end{tabular}           
\\
$^{S}$ Abundances derived from the synthetic spectrum analysis\\
\kba~RR Lyrae variable CS Eri at phase 0.42. A further discussion is given
  in Section 10.7 \\
\end{flushleft}
\end{table*}

\section{Abundances} 

\subsection {Abundances from KPNO and ESO-CAT spectra}

Our first estimate of the abundances (using the mean stellar parameters given
in Table~7)  was made by fitting the measured equivalent widths
(W$_{\lambda}$) of the apparently unblended lines to the computed ones. In the
case of the spectra observed at Kitt Peak\footnote{ except for BD~+00~0145 and
HD~14829 for which only the \kmg~4481~\AA~line was measured.}, we tried to
determine the microturbulent velocity ($\xi$) by assuming that, for a given
element, the abundance is independent of the equivalent widths.   The
uncertainty, however, both in the equivalent widths of the weak lines and in
the \kgf~values (especially for the lines of \kti, which are the most
numerous) severely limits this method of obtaining $\xi$. We therefore, in
addition, determined $\xi$  by comparing the observed spectra  against a
series of synthetic spectra in which $\xi$ was sampled in steps of 1.0 \ks; in
a few cases an intermediate step of 0.5 \ks~was used.

In the case of BD~+00~0145 and HD~14829 and for the stars observed at ESO we
assumed a microturbulent velocity $\xi$ of 2.0~\ks, since there were too few
lines in their spectra to allow us to derive $\xi$.

In computing the  synthetic spectra, we  used either the mean abundance
derived from the equivalent widths for species with more than one measured
line (e.g. \kfe, \kff, and \kti) or the abundance computed from a single
line if only one line of a species was available (e.g. Ba\,{\sc ii} $\lambda$4554).

The synthetic spectra were computed at a resolving power of 500\,000 and then
were degraded to 15\,000 (the nominal resolution of the Kitt Peak spectra)
using a gaussian instrumental profile.  The computed spectra were then
broadened by the rotational velocity  (\ksi) that is  given in column 2 of 
Table 15. This \ksi~was derived by fitting the observed profile of the 
\kmg~4481~\AA~to the computed profile assuming the Mg abundance that had been
derived from the measured equivalent width. No macroturbulent velocity was
considered.

The comparison of our observed spectra with the synthetic spectra showed that
some of the lines in our original list should be discarded either because they
were blended or because they were too weak. The WIDTH program was now used to
recompute new abundances from the equivalent widths (W$_{\lambda}$) of the
remaining lines using the value of $\xi$ that had been determined from the
synthetic spectra. 
 We made several iterations  using both the comparison of the observed and the 
 computed W$_{\lambda}$
 and the comparison of the observed and  synthetic
spectra until the abundances obtained by the two methods were consistent. In
the course of the successive iterations we changed the $\xi$ and the
metallicity of the models so that they were as close as possible to the values
that we derived from the abundance analysis.

The measured equivalent widths (W$_{\lambda}$), the adopted \kgf, their
sources, and the logarithmic abundances relative to the {\it total number of
atoms} are given for the individual lines for each star in Table~4 for the
KPNO spectra and in Table~5 for the ESO-CAT spectra. Table~9 lists, for each
star, the model parameters,  the microturbulent velocity  and the average
abundances derived from the measured equivalent widths of the individual lines.
We derived the barium abundance from the Ba\,{\sc ii}~4554.033~\AA~line. For a few
stars, we used only the synthetic spectra, while for some others we used the
equivalent width method in addition. Both values are given in Table~9 (that
from the synthetic spectra is identified with the superscript S).  The slight
systematic difference between the abundances obtained by the two methods may
be related to the placement of the continuum level which was fixed 
independently by Kinman for the measurement of the KPNO equivalent widths 
and by Castelli for the normalization of the whole observed spectrum.

Table~10 summarizes the abundances relative to the solar values together with
the [Mg/Fe] and [Ti/Fe] ratios.  The solar abundances, relative to the total
number of atoms, are taken from Grevesse et al.  (1996). Their
logarithmic values are  --4.46 for Mg, --5.68 for Ca, --8.87 for Sc, --7.02
for Ti,--6.37 for Cr, --4.54 for Fe, and --9.91 for Ba. A few of them are also
given in the last line of Table~10 for reference.

\begin{table*}
{{\bf Table~10.} The adopted parameters and abundances relative to the solar
 values.}
\begin{flushleft}
\begin{tabular}{lrccccccccc} 
\noalign{\smallskip}
\hline 
Star   &\kt    &\kg &$\xi$ &[Fe/H]   & [Mg/H] &[Ti/H]&[Ba/H]&[Mg/Fe] &[Ti/Fe]&[Fe/H]$^{a}$\\
       & (K)   &    &(km/s)&        &         &      &      &     &    &   \\  
    (1)& (2)   & (3) &(4) & (5)     &  (6)    & (7)  & (8)  & (9)    & (10) &(11) \\
\noalign{\smallskip}
\hline 
\noalign{\smallskip}
HD~2857$^{K}$  &7550& 3.00  & 3.0 & [--1.73]&[--1.30] & [--1.17]& [--1.84]&[+0.43]&[+0.56] & [--1.70] \\
HD~4850$^{C}$  &8450&3.20  & (2.0)& [--1.27]&[--0.64] & [--0.63]&\kc     &[+0.63]&[+0.64] & [--1.18] \\
BD~+00 0145$^{K}$  &9700&4.00  & (2.0)& \kc    &[--2.00] &\kc     &\kc     &       &\kc & [--2.45] \\ 
HD~8376$^{K}$  &8150&3.30   & 1.0 & [--2.95]&[--2.40] & [--2.10]&\kc     &[+0.55]&[+0.85] & [--2.82] \\
HD~13780$^{C}$  &7950&3.10  & (2.0)& [--1.45]&[--1.12] & [--0.88]&\kc        &[+0.33]&[+0.57] & [--1.53] \\
HD~14829$^{K}$  &8900&3.20  & (2.0)&  \kc   &[--2.01] &\kc     &\kc        &\kc    &\kc & [--2.39] \\
HD~16456$^{K}$ &6750&2.80   & 3.0 & [--1.70]&[--1.41] & [--1.34]&[--1.56]&[+0.29]&[+0.36] & [--1.82] \\
HD~16456$^{C}$ &7500&3.00   &(3.0)& [--1.65]&[--1.39] & [--1.29]&\kc    &[+0.26]&[+0.36] & [--1.80] \\
HD~31943$^{K}$  &7900&3.20   & 4.0 &[--1.04] &[--0.44] & [--0.65]&[--1.25]&[+0.60]&[+0.39] & [--0.97] \\
HD~31943$^{C}$  &  " & "     & "   &[--0.94] &[--0.45] & [--0.66]&\kc    &[+0.59]&[+0.28] & [--0.98]  \\
HD~252940$^{K}$  &7550&2.95 &  3.5  &[--1.77] &[--1.45] & [--1.25]&[--1.90]&[+0.32]&[+0.52] & [--1.80] \\
HD~60778$^{K}$  &8050&3.10   & 3.0 &[--1.49] &[--0.94] & [--1.12]&[--1.70]&[+0.55]&[+0.37] & [--1.34] \\
HD~74721$^{K}$  &8900&3.30  & 4.0  &[--1.42] &[--1.18] & [--1.09]&\kc    & [+0.24]&[+0.33] & [--1.48] \\
HD~78913$^{C}$  &8500&3.25  & (2.0)&\kc     &[--0.97] & [--1.21]&\kc    &\kc    & \kc  & [--1.43] \\
HD~86986$^{K}$  &7950&3.20   & 2.5 & [--1.81]&[--1.26] & [--1.30]&[--1.80] &[+0.55]&[+0.51] & [--1.66] \\
HD~87047$^{K}$  &7850&3.10   & 2.0:& [--2.47]&[--2.01] & [--1.87]&\kc     &[+0.46]&[+0.60] & [--2.43]  \\
HD~87112$^{K}$  &9750&3.50   & 2.0:&[--1.46] &[--1.09] & [--1.17]&\kc     & [+0.37]&[+0.29] & [--1.56] \\
HD~93329$^{K}$  &8250&3.10   & 2.0 &[--1.32] &[--0.81] & [--0.82]&[--1.39]&[+0.51]&[+0.50] & [--1.30] \\
BD~+32~2188$^{K}$&10450&2.10  &1.0  &[--1.11] &[--1.04] & \kc    &\kc    &[+0.07]&\kc  & [--1.45] \\
HD~106304$^{C}$  &9750&3.50   &(2.0)&\kc     &[--0.82] & [--1.21]&\kc    &\kc    &\kc & [--1.34] \\
BD~+42~2309$^{K}$&8800&3.20   &2.0 &[--1.63]  &[--1.20] & [--1.13]&\kc    &[+0.43]&[+0.50] & [--1.62] \\
HD~109995$^{K}$  &8500&3.10   &3.0 &[--1.72]  &[--1.38] & [--1.29]&\kc    &[+0.34]&[+0.43] & [--1.70] \\
BD~+25~2602$^{K}$&8400&3.20   &4.0 &[--1.98]  &[--1.69] & [--1.61]&\kc    &[+0.29]&[+0.37] & [--1.98]  \\
HD~117880$^{K}$  &9300&3.30   &2.0:&[--1.64]  &[--1.05] & [--1.25]&\kc    &[+0.59]&[+0.39] & [--1.51] \\
HD~128801$^{K}$  &10300&3.55   &2.0 &[--1.45] &[--1.09] & [--1.21]&\kc    &[+0.36]&[+0.24] & [--1.56] \\
HD~130095$^{K}$  &9000&3.30   &2.0 &[--1.87]  &[--1.65] & [--1.70]&\kc    &[+0.22]&[+0.17] & [--2.04] \\
HD~130095$^{C}$  &  " & "     & "  &\kc      &[--1.66] & [--1.84]&\kc    &\kc    &\kc & [--2.05]  \\
HD~130201$^{C}$  &8650&3.50   &(2.0)&[--1.00]  &[--0.20] & [--0.80]&\kc   &[+0.80]&[+0.20] & [--0.86] \\
HD~139961$^{K}$  &8500&3.20   &3.0  &[--1.71] &[--1.34] & [--1.22]&\kc    &[+0.37]&[+0.49] & [--1.68] \\
HD~139961$^{C}$  & "  & "     &3.0  &\kc     &[--1.33] & [--1.39]&\kc    &\kc    &\kc & [--1.66] \\
HD~161817$^{K}$  &7550&3.00   &3.0  &[--1.55] &[--1.23] &[--1.11]&[--1.76]&[+0.32]&[+0.44] & [--1.64] \\
HD~167105$^{K}$  &9050&3.30   &3.0  &[--1.56] &[--1.31] &[--1.18]& \kc   &[+0.25]&[+0.38] & [--1.66] \\
HD~180903$^{K}$ &7700&3.10   &3.0  &[--1.43] &[--0.88] &[--0.87]& [--1.40]&[+0.55]&[+0.56] & [--1.32] \\
HD~180903$^{C}$ & "  & "     &"    &[--1.47] &[--0.80] &[--0.82]&\kc     &[+0.67]&[+0.65] & [--1.27] \\
HD~202759$^{K}$  &7500&3.05   &2.0  &[--2.16] &[--1.87] &[--1.72]&[--2.00] & [+0.29]&[+0.44] & [--2.35] \\
HD~202759$^{C}$  & "  &"      &"    &[--2.08] &[--1.92] &[--1.78]& \kc    &[+0.16]&[+0.30] & [--2.40] \\
HD~213468$^{C}$  &9150&3.30  & (2.0)&\kc     &[--1.25] &[--1.02]&\kc     & \kc    &\kc  & [--1.67] \\
\noalign{\smallskip}
\hline
\noalign{\smallskip}
\multicolumn {2}{c}{Sun~ log(N$_{\rm elem}$/N$_{\rm tot}$)}    &       &     &Fe=--4.54  &Mg=--4.46   &Ti=--7.02  &Ba=--9.91\\         
\noalign{\smallskip}
\hline 
\noalign{\smallskip}
\end{tabular}           
\\
$^{K}$: KPNO spectra;~~ $^{C}$: CAT spectra\\
$^{a}$ Derived from \kmg~$\lambda$ 4481 (see text). \\
\end{flushleft}
\end{table*}

The ESO-CAT abundances, although based on only a few lines,  show excellent
agreement with those derived from the KPNO spectra for the non-variable stars
HD~31943, HD~130095, HD~139961 and HD~180903 and for the 
 low-amplitude variable HD~202759. The case
of the larger amplitude type-c variable HD~16456 (CS~Eri) is discussed in
Sect. 10.7.

 For the stars whose \kt~exceeds about 
 8\,500 K (or about half the stars in our sample),
 the He\,{\sc i} $\lambda$ 4471 line is visible in our spectra. Its strength agrees 
 with that predicted by the synthetic spectrum for a solar helium abundance. 
 
\subsection {The [Fe/H] abundance as a function of the equivalent width of
      \kmg~$\lambda$4481 line and the colour index \kfo} 

In most halo stars, [Mg/Fe] can be assumed either to be constant or a
slowly-varying monotonic function of [Fe/H] (see Sect. 10.5). If we have the
photometric information, we can derive the stellar parameters and then
determine [Mg/H] from the equivalent width of the \kmg~$\lambda$4481 line
even in quite low resolution spectra; [Fe/H] can then be derived by assuming
an appropriate  value for [Mg/Fe]; in this paper we assume [Mg/Fe] = 0.43.

Even if only \kfo~is available, one can estimate [Fe/H] from the the
equivalent width W of the \kmg~$\lambda$4481 doublet and the intrinsic
colour. Using the data and [Fe/H] abundances that we derived from our KPNO
spectra we found the following expression: 
\begin{eqnarray*}
 [Fe/H] = -3.350 + 0.01119W -  0.00001315W^2 \\ -0.30(B-V)_0  
\end{eqnarray*}  
where \kfo~ was obtained by using the mean extinctions given in boldface in 
column 3 of Table~7\footnote{ One may replace the \kfo~term by 
$-$0.44\kxo.}.
[Fe/H] derived from the above equation is listed in the last column of
Table~10.
The $rms$ difference between our measured [Fe/H] and those obtained from this
equation is $\pm$0.12 for the range $-$0.05$\leq$\kfo$\leq$0.17. Systematic
differences can occur between equivalent widths measured at very different spectral
resolutions. Our relation strictly applies only to spectra whose
resolution is comparable to those discussed in this paper;
it may be less accurate if used with equivalent widths
derived from lower resolution spectra.

\section {Uncertainties}

In this section we consider the effect on our abundances of uncertainties in 
\kt, \kg, and the microturbulent velocity $\xi$. We also consider errors in
log~$gf$ and NLTE effects.

\subsection{Uncertainty in the the stellar parameters
\kt, \kg~}

The quantitative dependence of the derived abundances on differences
$\Delta$\kt~= $\pm$100~K and $\Delta$\kg~= $\pm$0.1 dex in the stellar parameters
is given for the stars HD~161817, HD~139961, HD~167105, and HD~87112 in
Table~11. These stars are representative of stars having \kt~around 7500~K,
8500~K, 9000~K, and 9750~K respectively. It is seen  that the uncertainty in
\kt~affects the abundances more than the uncertainty in \kg. Furthermore, the
species most affected by uncertainties in the parameters are Cr\,{\sc i}, \kfe~and
Ba\,{\sc ii}. Their abundance changes by about 0.1~dex for  $\Delta$\kt~= $\pm$100~K.
The effect on \kmg$\lambda$ 4481 is small and decreases with increasing \kt.

\begin{table}
{{\bf Table~11.} Abundance changes produced by uncertainties \\of $\pm$100~K in
 \kt~and $\pm$0.1 in \kg}
\begin{flushleft}
\begin{tabular}{lcccc} 
\noalign{\smallskip}
\hline 
 Star     & Elem    & \multicolumn{2} {c}{$\Delta$log$\epsilon$}            & Net error\\
          &        & $\Delta$\kt &$\Delta$\kg   &            \\
          &        & $\pm$ 100   &$\pm$0.1\\
    (1)&    (2)&   (3)  &  (4)  &(5)\\
\noalign{\smallskip}
\hline 
\noalign{\smallskip}
HD~161817  &\kmg   & $\pm$0.04 &$\pm$0.04 &0.06\\
\kt~7533 K &\kca   & $\pm$0.08 &$\pm$0.02&0.08\\
\kg~3.00   &\kti   & $\pm$0.04 &$\pm$0.04 &0.06\\
           &\kcr   & $\pm$0.09 &$\pm$0.01 &0.09\\
           &\kfe   & $\pm$0.08 &$\pm$0.01 &0.08\\
           &\kff   & $\pm$0.03 &$\pm$0.03 &0.04\\
           &\kbt   & $\pm$0.09 &$\pm$0.01 &0.09\\
\\
HD~139961  &\kmg   & $\pm$0.00 &$\pm$0.02 &0.02\\
\kt~8517 K &\kti   & $\pm$0.06 &$\pm$0.02 &0.06\\
\kg~3.23   &\kcc   & $\pm$0.04 &$\pm$0.02 &0.04\\
           &\kfe   & $\pm$0.10 &$\pm$0.03 &0.10\\
           &\kff   & $\pm$0.04 &$\pm$0.02 &0.04\\
\\  
HD~167105  &\kmg   & $\pm$0.01 &$\pm$0.01 &0.01\\
\kt~9025 K &\kti   & $\pm$0.07 &$\pm$0.01 &0.07\\
\kg~3.29   &\kcc   & $\pm$0.04 &$\pm$0.02 &0.04\\
           &\kfe   & $\pm$0.10 &$\pm$0.04 &0.11\\
           &\kff   & $\pm$0.04 &$\pm$0.02 &0.04\\
\\
HD~87112  &\kmg   & $\pm$0.01 &$\pm$0.01 &0.01\\
\kt~9733 K&\kti   & $\pm$0.06 &$\pm$0.01 &0.06\\
\kg~3.48  &\kcc   & $\pm$0.03 &$\pm$0.02 &0.04\\
          &\kfe   & $\pm$0.09 &$\pm$0.04 &0.10\\
          &\kff   & $\pm$0.03 &$\pm$0.03 &0.04\\

\noalign{\smallskip}
\hline 
\noalign{\smallskip}
\end{tabular}           
\\

\end{flushleft}
\end{table}

\subsection{Uncertainty in $\xi$}  

The value of $\xi$ was assumed for the spectra of the two stars BD~00+00~145
and HD~14829 and for all the ESO-CAT spectra because there were too few lines
in these spectra to determine this quantity.   Table~12 gives the abundances
of the different species in these stars for microturbulent velocities  $\xi$ =
2~\ks, 3~\ks and 4~\ks. For a change $\Delta$$\xi$ = 1~\ks, the abundance
derived from the \kmg~$\lambda$ 4481 line changes by about 0.2 dex for the
stars observed at ESO and about 0.05 dex for the two
weaker-lined stars observed at KPNO (BD~+00~145 and HD~14829). 
 The abundance derived from the \kti~lines
is also affected by the value of $\xi$; the change varies from 0.2~dex
for HD~4850 and HD~13780 to 0.05~dex for HD~106304. The effect of $\xi$ on the
\kfe~and \kff~abundances is very small in all these stars.

\begin{table}
{{\bf Table~12.} The effect of the microturbulent velocity $\xi$ \\
on the abundances }
\begin{flushleft}
\begin{tabular}{lcccc} 
\noalign{\smallskip}
\hline 
 Star     & Elem       &\multicolumn3 {c} {$\log(N_{elem}/N_{tot}$)}\\
          &            &$\xi$=2.0 &$\xi$=3.0&$\xi$=4.0        \\
\noalign{\smallskip}
\hline 
\noalign{\smallskip}
HD~4850    &\kmg &$-$5.10&$-$5.34&$-$5.53\\
           &\kti &$-$7.65&$-$7.86&$-$7.95\\
           &\kfe &$-$5.86&$-$5.87&$-$5.87\\
           &\kff &$-$5.76&$-$5.80&$-$5.82\\
\\
BD +00 0145 &\kmg   &$-$6.46  &$-$6.51& $-$6.54\\
\\
HD~13780   &\kmg  &$-$5.48&$-$5.69&$-$5.82\\
           &\kti &$-$7.90&$-$8.11&$-$8.20\\
           &\kfe &$-$5.97&$-$5.98&$-$5.60\\
           &\kff &$-$6.01&$-$6.04&$-$6.06\\
\\
HD~14829   &\kmg   &$-$6.40  &$-$6.47&$-$6.53\\
\\
HD~78913   &\kmg   &$-$5.43  &$-$5.64&$-$5.78\\
           &\kti   &$-$8.23  &$-$8.41&$-$8.50\\
\\
HD~106304  &\kmg   &$-$5.28  &$-$5.49&$-$5.64\\
           &\kti   &$-$8.23  &$-$8.27&$-$8.29\\
\\
HD~130201  &\kmg   & $-$4.66 &$-$4.86 &$-$5.06\\
           &\kti   & $-$7.82 &$-$7.94 &$-$8.00\\
           &\kff   & $-$5.54 &$-$5.60 &$-$5.63\\
\\
HD~213468  &\kmg   & $-$5.71 &$-$5.87 &$-$6.00\\
           &\kti   & $-$7.90 &$-$7.94 &$-$7.96\\
           &\kfe   & $-$4.28 &$-$4.36 &$-$4.40\\
\\

\noalign{\smallskip}
\hline 
\noalign{\smallskip}
\end{tabular}           
\\

\end{flushleft}
\end{table}

\subsection{Errors in \kgf}

The errors in \kgf~can be a significant source of uncertainty if only a
few lines of a species are available for measurement. This can happen if the
star is very metal-poor (e.g. HD~008376) so that only the strongest lines are
measurable or, as with the ESO-CAT spectra, the observed waveband is not large.
We inferred the presence of these errors in \kgf~as follows.

Our first estimate of the abundance was made by fitting the measured
equivalent widths (W$_{\lambda}$)  of the apparently unblended lines to those 
computed by Kurucz's WIDTH program. We therefore have an abundance for  each
line and the difference between this abundance and the mean for that species
in a given star is called $\Delta$[m/H]. This quantity, when averaged over
all our program stars, ($<$$\Delta$[m/H]$>$) is shown in Fig~8 for both the
\kff~and \kti~lines. It shows little correlation with equivalent width 
(the W$_{\lambda}$ on the left of Fig. 8 are those for HD~93329 which has an 
intermediate \kt). 

$<$$\Delta$[m/H]$>$  was also computed (for the same
lines) from the BHB star data of Lambert et al. (1992) and is called
$<$$\Delta$[m/H]$>$$_{\rm LMS}$. It is seen that there is a correlation between
the values of $<$$\Delta$[m/H]$>$ determined from our data and those of Lambert
et al.; moreover the range in this quantity is markedly greater for the 
\kti~lines
 than for the \kff~ones. This scatter in $<$$\Delta$[m/H]$>$ is
greater than can be accounted for by measuring errors (the vertical error bars)
and must be caused by a factor that is intrinsic to each species and which is
common to both our calculations and those of Lambert et al. It seems most 
likely that 
it is caused by errors in the assumed \kgf.

\subsection{Non-LTE effects}

The models used to derive our abundances assume LTE conditions.  In hot stars,
however, UV  radiation can cause the \kfe~states to be underpopulated while
the \kff~lines are relatively unaffected; the effect is expected to increase
with decreasing metallicity. Lambert et al.  (1992) tried to allow for this
effect by adjusting their stellar parameters so as to make [\kfe] $-$ [\kff] =  
 $-$0.2 . Cohen \&  McCarthy (1997), however,  made no non-LTE corrections in
deriving the abundances of BHB stars in M~92. The \kt~ of their stars were in
the range 7\,500~K to 9\,375~K and were derived from their $(B-V)$ and $(V-K)$
colours. They  found  a mean value for $<$[\kfe]$-$[\kff]$>$ of only $-$0.08;
this suggests that  non-LTE effects are not significant. The abundances,
moreover,  which they found for their BHB stars were in excellent agreement
with those previously found for red giants in the same cluster. We find
$<$[\kfe]$-$[\kff]$>$ = 0.01$\pm$0.01 for the 27 spectra where we measured
both \kfe~and \kff~lines.  We therefore feel that it is unlikely that our
iron abundances are significantly compromised by non-LTE effects. Our barium
abundances (Table~9) were derived from the Ba\,{\sc ii} $\lambda$4554.03 line alone
and gave a mean LTE abundance of [Ba/Fe] from nine stars of $-$0.08$\pm$0.05;
hyperfine broadening was not taken into account and significant non-LTE
effects may be expected for this line (Mashonkina \& Bikmaev 1996, Belyakova 
et al. 1998).

\begin{figure}
\resizebox{8.3cm}{!}{\includegraphics{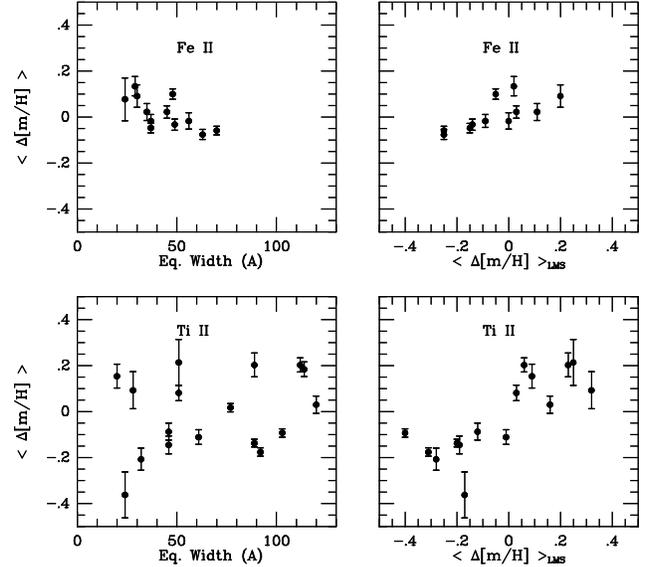}}
\hfill
\parbox[b]{88mm}{ 
\caption[]{ $\Delta$[m/H] is the difference between the abundance derived 
 from a given line and the mean abundance of that species for that star.
$<$ $\Delta$[m/H]$>$ is the mean value of this quantity for each [\kff] line 
 (above) and [\kti] line (below). This mean is plotted against the equivalent 
 width (on the left) and against the same quantity from the BHB spectra of 
 Lambert et al. (1992) (on the right) 
 .}
\label{Fig08}}
\end{figure}

\subsection{Convection}

For the coolest stars of our sample (\kt$<$8\,000~K) there may be a  problem
with the treatment of the convection in the model atmospheres. Uncertainties 
of the order of 200~K in \kt~can  be expected in the sense that \kt~is
higher for the  mixing-length parameter l/H = 1.25 that we adopted than for the
lower value l/H =0.5  suggested by  Fuhrmann, Axer \& Gehern (1993, 1994).
Also, a different convection theory, like that of Canuto \& Mazzitelli (1992)
leads to a very low convection (or no convection) in stars hotter than
7\,000~K, so that \kt~derived by adopting this theory may be lower than that
derived by us. We feel, however, that more accurate observations that allow a
more precise location of the continuum and more discussions on the theories
adopted to compute the Balmer profiles are needed in order to confirm the
superiority of other convections over that adopted by us. The effect of
convection on the colour indices and Balmer profiles, and therefore on the
\kt~derived from them, has been discussed by Smalley \& Kupka (1997), Van't
Veer-Menneret \& Megessier (1996), Castelli et al. (1997), and Gardiner et al. 
(1999).

\section{Discussion} 

\subsection{Comparison with previous observers.}

Table~13 compares model parameters and abundances found by us (KCCBHV) with
those given by other authors. Stellar abundances are relative to the solar
values from Grevesse et al. (1996), as given at the end of Table~10.
Parameters from de Boer et al. (1997) (BTS)  are only averages of previous
determinations taken from the literature. Takeda \& Sadakane (1997) 
estimated the stellar parameters of HD~161817 from the literature.
 They obtained a microturbulent velocity
$\xi$ = 4 km s$^{-1}$ from an analysis of O\~{\sc i} lines in this star and suggested that
$\xi$ is depth dependent.
 
The mean differences between our parameters and abundances and those found by
Adelman \& Philip (1990, 1994, 1996a) for the nine stars we have in common are:
\begin{description}
\item $<\Delta$[Fe/H]$>$~~=~ $+$0.08$\pm$0.05~~~($\pm$0.14)   
\item $<\Delta$[Mg/H]$>$~=~ $-$0.02$\pm$0.14~~~($\pm$0.35)   
\item $<\Delta$[Ti/H]$>$~~=~ $+$0.16$\pm$0.05~~~($\pm$0.14)   
\item $<\Delta$\kt$>$~~~~~~=~ $+$331 K$\pm$75 K~~~($\pm$212 K)   
\item $<\Delta$\kg$>$~~~~~=~ $+$0.10$\pm$0.10~~~($\pm$0.27)   
\item $<\Delta$\km$>$~~~~~~~~~=~ +1.0$\pm$0.4~~~($\pm$1.0)  
\end{description}
where the numbers in parentheses are the $rms$ differences between the
individual determinations. The most significant difference is in \kt~and this
may well be traceable to different assumptions for interstellar reddening. The
largest of these is the 800~K difference in \kt~for HD~130095 for which there
is a large range in the different estimates of \keb. In spite of this, the
differences between the abundance estimates for this star are quite small. For
further comments on HD 130095 see Sect. 10.6.

 Gray et al. (1996) give stellar parameters for BHB stars that were 
determined from Philip's Str\"{o}mgren photometry, classification-dispersion 
spectra and spectral synthesis. The mean differences between our parameters and
 theirs for the ten stars in common are:  
\begin{description}
\item $<\Delta$[Fe/H]$>$~~=~ $-$0.26$\pm$0.05~~~($\pm$0.16)   
\item $<\Delta$\kt$>$~~~~~~=~ $+$40 K$\pm$91 K~~~($\pm$272 K)   
\item $<\Delta$\kg$>$~~~~~=~ $-$0.09$\pm$0.02~~~($\pm$0.06)   
\end{description}
The systematic difference between our \kt~and those of Gray et al. are much
smaller than for those given by Adelman \& Philip. The abundance estimates of
Gray et al. from their low resolution spectra, however, average 0.2 to 0.3 
dex more metal rich than ours. 

\begin{table*}
{{\bf Table~13.} Comparison of stellar parameters and abundances 
from different authors}
\begin{flushleft}
\begin{tabular}{lrccccccccc} 
\noalign{\smallskip}
\hline 
Star   &\kt    &\kg &$\xi$ &[Fe/H]   & [Mg/H] &[Ti/H]&[Ba/H]&Source\\
       & (K)   &    &\ks \\  
    (1)& (2)   & (3) &(4) & (5)     &  (6)    & (7)  & (8)  & (9)  \\
\noalign{\smallskip}
\hline 
\noalign{\smallskip}

HD~2857  &7550&3.00  & 3.0  &[--1.73] &[--1.30] & [--1.17]&[--1.84] &KCCBHV\\
           &7700&3.10  & \kc  & [--1.5]  & \kc     & \kc     & \kc     &GCP \\
HD~14829  &8900&3.20  & \kc  &[--2.39] &[--2.01] & \kc     &\kc      &KCCBHV\\
           &8700&3.30  & \kc  & [--2.0]  & \kc     & \kc     & \kc     &GCP \\
HD~60778  &8050&3.10  & 3.0  &[--1.49] &[--0.94] & [--1.12]&[--1.70] &KCCBHV\\
           &8600&3.30  & \kc  & [--1.0]  & \kc     & \kc     & \kc     &GCP \\
HD~74721  &8900&3.30  & 4.0  &[--1.42] &[--1.18] & [--1.09]& \kc  &KCCBHV\\
           &8600&3.30  &1.4  &[--1.40] &[--0.96] & [--1.00]& \kc   &AP96\\
           &8600&3.30  & \kc  & [--1.5]  & \kc     & \kc     & \kc     &GCP \\
HD~86986  &7950&3.20  & 2.5 & [--1.81]&[--1.26] & [--1.30]&[--1.80]&KCCBHV\\
           &7800&3.10   & 2.2 & [--1.80]&[--1.21] & [--1.35]&[-2.15]&AP96\\     
           &8050&3.20  & \kc  & [--1.5]  & \kc     & \kc     & \kc     &GCP \\
HD~93329  &8250&3.10  & 2.0 & [--1.32] &[--0.81] & [--0.82]&[-1.39]&KCCBHV\\
           &8150&3.10   & 2.4 & [--1.40] &[--0.96] & [--0.98]&[-1.65]&AP96\\ 
BD~+42 2309 &8800&3.20  & 2.0  &[--1.63] &[--1.20] & [--1.13]&\kc      &KCCBHV\\
           &8400&3.30  & \kc  & [--1.5]  & \kc     & \kc     & \kc     &GCP \\
HD~109995  &8500&3.10  &3.0  & [--1.72] &[--1.38] & [--1.29]&\kc    &KCCBHV\\
           &8150&3.25  &1.7  & [--1.89] &[--1.28] & [--1.39]& \kc   &AP94,AP96\\
           &8300&3.20  & \kc  & [--1.5]  & \kc     & \kc     & \kc     &GCP \\
           &8300&3.15  &\kc  &\kc       &\kc      &\kc      & \kc   &BTS\\  
HD~128801  &10300&3.55 &2.0 &[--1.45] &[--1.09] & [--1.21]&\kc    &KCCBHV\\
           &10250&3.40 &0.0 &[--1.26] &[--0.79] &[--1.33] &  \kc  &AP94,AP96\\
HD~117880  &9300&3.30  & 2.0: &[--1.64] &[--1.05] & [--1.25]&\kc      &KCCBHV\\
           &9200&3.40  & \kc  & [--1.5]  & \kc     & \kc     & \kc     &GCP \\
HD~130095  &9000&3.30  &2.0 &[--1.87] &[--1.65] & [--1.70]&\kc    &KCCBHV\\
           &8300&3.45  &2.0 &[--2.03] &[--1.55] & [--2.09]& \kc   &AP94,AP96 \\
           &8950&3.40  & \kc  & [--1.5]  & \kc     & \kc     & \kc     &GCP \\
           &8800&3.40  &\kc &\kc      &\kc      &\kc      & \kc   &BTS\\        
HD~139961  &8500&3.20  &3.0 &[--1.71] &[--1.34] & [--1.22]& \kc      &KCCBHV\\
           &8750&3.30  &\kc &\kc      &\kc      &\kc      &\kc       &BTS\\ 
HD~161817  &7550&3.00   &3.0 &[--1.55] &[--1.23] &[--1.11]&[--1.76]&KCCBHV\\
           &7225&2.80   &2.3& [--1.66]&[--1.98] &[--1.43] &[--2.01]&AP94,AP96\\
           &7600&3.10  & \kc  & [--1.2]  & \kc     & \kc     & \kc     &GCP \\
           &7500&2.95  &\kc &\kc      &\kc      &\kc      &\kc        &BTS\\
           &7500&3.00  &4.0 & [$\sim$--1.5]&\kc      &\kc      &\kc        &TS\\
HD~167105  &9050&3.30   &3.0  &[--1.56] &[--1.31] &[--1.18]& \kc   &KCCBHV\\
           &8550&3.30    &2.0  &[--1.80] &\kc     &[--1.42]&\kc    &AP94,AP96\\
HD~202759  &7500&3.05   &2.0  &[--2.16] &[--1.87] &[--1.72]&[--2.00]&KCCBHV\\
           &7000&2.30    &0.6  &[--2.36] & \kc    &[--1.85]&  \kc   &AP90\\
           &7400&3.10    &\kc  &\kc      &\kc     &\kc     &\kc      &PB\\ 
\noalign{\smallskip}
\hline 
\noalign{\smallskip}
\end{tabular}           
\\
KCCBHV:this paper; AP90, AP94, AP96: Adelman \& Philip (1990; 1994; 1996a);
GCP: Gray et al. (1996) \\
BTS: de Boer at al. (1997); PB: Przybylski \& Bessell (1974); TS: Takeda \&
Sadakane (1997)

\end{flushleft}
\end{table*}

\subsection{Comparison of BHB abundances with those of other types of halo stars.}

Excluding BD~+32~2188,  BD~+00~0145 and  HD~16456, we have 28 stars that from
their stellar parameters, abundances, \ksi~and kinematics have a very high
probability of being BHB stars
  HD~202759 has been classified as
a type~c RR Lyrae star, but its $V$-amplitude is so low ($<$ 0.1 mag), and its
\kt~is so high (7500~K), that it has been included with the BHB stars. The
[Fe/H] of these 28 stars lie in the range $-$0.99 (for HD~31943) to $-$2.95
(for HD~8376) with a mean value of $-$1.67$\pm$0.08 and an $rms$ dispersion
($\sigma$) about this mean of $\pm$0.42\footnote{The 24 BHB stars for which
we derived abundances from high resolution spectra have a mean [Fe/H] of $-$1.66$\pm$0.09. The four BHB stars for which an abundance was estimated from the
\kmg~($\lambda$4481) line have a mean [Fe/H] of $-$1.71$\pm$0.27}. 
 We compare these parameters with
those of other types of halo stars in Table~14. The small group of nearby red
horizontal branch stars are taken from Pilachowski et al. (1996).
The nearby RR Lyrae stars include those with abundances by Clementini et al.
(1995) and by Lambert et al.  (1996). The red giants are those within 600 pc
from the sample given by Chiba \& Yoshi (1998). The halo globular clusters are
those listed by Armandroff (1989). The first sample is a subset of 21 of these
clusters whose [Fe/H] has been given by Carretta \& Gratton (1997). The second
sample contains all those in Armandroff's list, using Carretta \& Gratton's
abundances for 21 of the clusters while for the remainder, the abundances
given by Armandroff (which are on the Zinn \& West (1984) scale) were
converted to the system of Carretta \& Gratton using the  quadratic relation
given in their paper\footnote{
The most metal-poor cluster, NGC 5053, lies outside the range  of this
relation. This makes the metal-poor limit of this cluster sample  uncertain
but scarcely affects its mean value.}. 
The halo clusters, on the  average, appear to be 0.1 or 0.2 dex more
metal-rich than the field halo stars.  On the Zinn \& West scale, they would
have had more comparable metallicities.                               The red
giant sample contains a greater fraction of very metal-poor stars than the
other groups. Thus,  30\% of the red giants have [Fe/H]$\leq$$-$2.00 while
only between 5 and 10\%  of the globular clusters are this metal-poor; this
difference is significant  at better than  the 1\% level. This is possibly
because  many of the red giants were discovered in the objective-prism 
surveys of Bond (1970,~1980) which, while being  kinematically unbiased,
tended to accentuate the discovery of the most  metal-poor stars. The large
subdwarf samples of Ryan \& Norris (1991),  although they contain stars in the
range +0.01$\geq$[Fe/H]$\geq$$-$3.70  and presumably include thick disk stars,
have a maximum frequency in [Fe/H]  at $-$1.65. This is similar to what we
find for the field halo stars but  not for the halo globular clusters where
the maximum frequency is $\sim$0.3  dex more metal-rich.  Thus, although the
[Fe/H] abundances which we have derived for the BHB stars  is in general
agreement with those found for other local halo stars, they are  appreciably 
more metal-poor than those of the halo globular  clusters. This discrepancy
requires further investigation\footnote{
Current  abundance estimates of late-type halo stars are generally based on
LTE  analyses. The problem of NLTE effects in these stars is discussed by
Gratton et  al. (1999) and by  Th\'{e}venin \& Idiart (1999).}.

\subsection{Comparison with ZAHB models.}
   
\begin{table}
{{\bf Table 14.} Comparison of the distribution of [Fe/H] in our BHB stars 
 with that of other samples of halo stars.}
\begin{flushleft}
\begin{tabular}{cccc}
\hline\noalign{\smallskip}
Type \& No. &\multicolumn{3}{c}{[Fe/H]}              \\
\cline{2-4} 
of stars      & Range                             & Mean & $\sigma$         \\
\noalign{\smallskip}
\hline\noalign{\smallskip}
                             &       &        &                        \\
BHB stars\kba~  (28)  &$-$0.99 to $-$2.95 &$-$1.67$\pm$0.08&$\pm$0.42    \\
RHB stars\kbb~  (14)  &$-$1.17 to $-$2.26 &$-$1.62$\pm$0.11&$\pm$0.40    \\
RR Lyrae\kbc~   (39)  &$-$1.11 to $-$2.49 &$-$1.61$\pm$0.06&$\pm$0.35  \\
R. Giants\kbd~ (46)  &$-$0.92 to $-$2.82 &$-$1.78$\pm$0.07&$\pm$0.50   \\
Halo~$\oplus$~\kbe~  (21)  &$-$0.96 to $-$2.16 &$-$1.50$\pm$0.08&$\pm$0.35  \\
Halo~$\oplus$~\kbf~  (76)  &$-$0.79 to $-$2.71 &$-$1.40$\pm$0.04&$\pm$0.35  \\
                             &       &        &                        \\
\hline\noalign{\smallskip}
\end{tabular}
\\
\kba~~BHB stars (this paper).                    \\
\kbb~~Pilachowski et al. (1996).  \\
\kbc~~Halo RR Lyraes (see text).                            \\
\kbd~~Chiba \& Yoshii (1998) (Red Giants within 600 pc). \\  
\kbe~~Halo globular clusters~(see text).  \\
\kbf~~Halo globular clusters~(see text).  \\
\end{flushleft}
\end{table}

\begin{figure}
\resizebox{8.3cm}{!}{\includegraphics{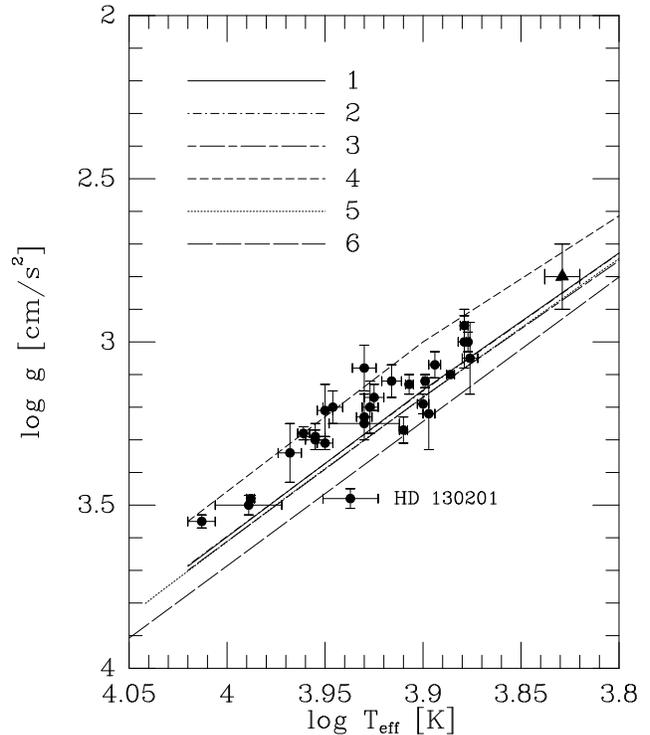}}
\hfill
\parbox[b]{88mm}{ 
\caption[]{The 28 BHB stars (filled circles) and the RR Lyrae star HD~016456 
(filled triangle) in the \klt~ $-$ \kg~ plane using the mean values given in 
Table~11. 
The lines show the ZAHB O-enhanced models of Dorman et al.(1993)~(1) 
and the models of Straniero et al. (priv. comm.) (2) 
and the He-enhanced models of
Sweigert (1997, 1999) for $\Delta$X$_{\rm mix}$~=~0.00 (3) and $\Delta$X$_{\rm mix}$~=~0.10 (4).
The models of Bono \& Cassisi for [Fe/H]~=~\kl~2.5 and \kl~1.7 are 
shown by (5) and (6) respectively.} 
\label{Fig09}}
\end{figure}

The \kt~and \kg~that we adopted for the analysis of the Kitt Peak and 
ESO-CAT spectra 
(Table~10) are plotted in Fig~9.   The 28 stars that  have a high probability
of being BHB stars are plotted as filled circles and  the c-type RR Lyrae star
HD~16456 as a filled triangle. For comparison we show the ZAHB models of
Dorman et al. (1993) with [m/H]~=~$-$1.48 and [O/Fe]~=~0.6, the
models of Straniero et al. (1998, priv. comm.) with [m/H]~= $-$1.3
(equivalent to [m/H] = --1.6 with $\alpha$-enhancement +0.4,
see Salaris et al. 1993) and the He-enhanced models of Sweigert
(1997, 1999) ($\Delta$X$_{\rm mix}$~=~0.0 and 0.10\footnote{$\Delta$X$_{\rm
mix}$ is a measure of the amount of helium that is mixed into the envelope of
the red giant precursor of the HB star.} with [m/H]~= $-$1.56).  We also show
models by Bono \& Cassisi (1999, priv. comm.) for [Fe/H] ~= $-$1.7 and $-$2.5; 
these illustrate the small metallicity dependence that is present. The agreement
is generally satisfactory except for HD~130201 whose \kt~is not very well 
determined.   A similar plot for the BHB stars in globular clusters (both
metal-poor and the metal-rich NGC 6388, NGC 6441, NGC 362, and 47 Tuc) has
been given in Fig. 8 of the recent review by Moehler (1999). At \klt~=~3.95,
the metal-poor globular cluster BHB have \kg~in the range 2.90 to 3.44 and
are mostly concentrated in the range 3.10 to 3.40. We have eleven BHB with
\klt~in the range 3.93 to 3.97 and  and their mean \kg~is 3.27, so there is
good agreement between the field and cluster BHB stars in the \kt~ $vs$ \kg~
plot. Both field and cluster BHB stars tend to lie slightly above the ZAHB,
suggesting either that some evolution is present or that some He-enhancement
is required. The difference is, however, comparable with the errors in the
computed gravities so that no definitive conclusion is possible.

\begin{figure}
\resizebox{8.3cm}{!}{\includegraphics{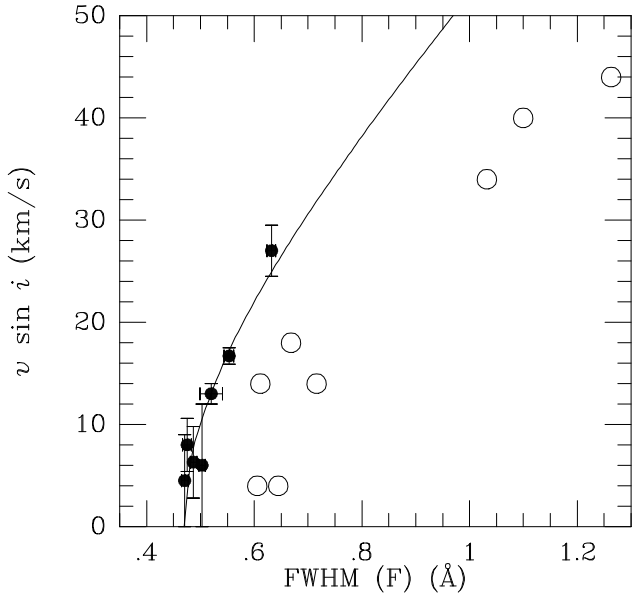}}
\hfill
\parbox[b]{88mm}{ 
\caption[]{A plot of \ksi~against the FWHM of the \kmg~$\lambda$4481 doublet
for BHB stars observed by Peterson, Tarbell \& Carney 
(1983)~(filled circles)
and IAU standards~(open circles). The adopted calibration is shown by the 
curve. }
\label{Fig10}}
\end{figure}

\begin{table}
{{\bf Table~15.} The rotational velocity (\ksi) for BHB star
 candidates\kq~(determined  from the \kmg~$\lambda$4481 line).  The
 Str\"{o}mgren index $\beta$ for the same stars determined both  
 photometrically and from our spectra.}
\begin{flushleft}
\begin{tabular}{lcccc} 
\noalign{\smallskip}
\hline 
      &        &     &    \\
      &\ksi\kba &\ksi\kbb  &\kr\ka & \kr\kw      \\
 Star &(\ks )&(\ks)&  &    \\
    (1)&    (2)&   (3)  &  (4) & (5)\\
\noalign{\smallskip}
\hline 
\noalign{\smallskip}
HD~2857 &28  &30  & 2.787 & 2.779\\
HD~4850 &\kc &14  & 2.846 &\kc  \\
HD~8376 &20  &10  & 2.835 & 2.837 \\
HD~13780 &\kc &14  & 2.816 &\kc   \\
HD~14829 &\kc &07  & 2.858 &\kc  \\
HD~16456 &15  &14  &  \kc  &\kc  \\
HD~31943 &13  &16  & 2.814 & 2.806  \\
HD~252940 &25  &24  &2.768&2.776  \\
HD~60778 &15  &11  &2.834&2.835   \\
HD~74721 &10  &02  &2.859&2.856   \\
HD~78913 &\kc &14  &2.842&\kc     \\
HD~86986 &15  &04  &2.825&2.827  \\
HD~87047 &12  &00  &2.797&2.810  \\
HD~87112 &10  &03  &2.840&2.823   \\
HD~93329 &15  &07  &2.825&2.832   \\
BD~+32~2188&05 &00  &2.633&2.592   \\
HD~106304 &\kc &10  &2.845&\kc    \\
BD~+42~2309&35 &35  &2.844&2.856  \\
HD~109995 &27  &25  &2.848&2.852   \\
BD~+25~2602&20 &12  &2.850&2.855   \\
HD~117880 &15  &13  &2.855&\kc     \\
HD~128801 &14  &04  &2.816&2.800   \\
HD~130095 &12  &07  &2.855&2.847   \\
HD~130201 &\kc &16  &2.860&\kc    \\
HD~139961 &35  &39  &2.858&2.852  \\
HD~161817 &17  &17  &2.746&2.777   \\
HD~167105 &22  &21  &2.849&2.856  \\
HD~180903 &20  &17  &2.800&2.789 \\
HD~202759 &15  &07  &2.770&2.759 \\
HD~213468 &\kc &12  &2.849&\kc   \\

\noalign{\smallskip}
\hline 
\noalign{\smallskip}
\end{tabular}           
\\
\kba~Determined from synthetic spectra. \\
\kbb~Determined from FWHM of line (Sect. 9.4). \\
\kq~Omitting BD~+00~0145 because of the poor quality of the spectrum. \\
\ka~~Adopted  photometric value (Table~1)  \\
\kw~~Determined from H$\gamma$ (Sect. 3).    \\

\end{flushleft}
\end{table}

\subsection{Projected rotational velocities (\ksi )}

Peterson et al. (1983) measured the projected rotational
velocities (\ksi~) of eight of the brighter field BHB stars from echelle
spectra (resolution of 24\,000) and found rotations of up to 30 \ks~. 
Peterson (1983, 1985a and 1985b) also measured the  \ksi~of HB stars in the
globular clusters M3, M5, M13, M4 and NGC 288. More recently, the \ksi~of 67
HB stars in M3, M5, M13 and NGC 288 have been measured by Peterson et al.
(1995, hereafter PRC). Also, Cohen \& McCarthy (1997) have determined \ksi~
for 5 HB stars in M92  from HIRES Keck spectra. Behr et al. (2000) have also
measured \ksi~for stars in M13. Rotations of up to 40~\ks~were found in both
M13 and M92 for HB stars whose \kt~were less than 11\,000 K. PRC  could find
no correlation between \kf~and \ksi. Cohen \& McCarthy  suspected a possible
trend of \ksi~with abundance.

\begin{figure}
\resizebox{8.3cm}{!}{\includegraphics{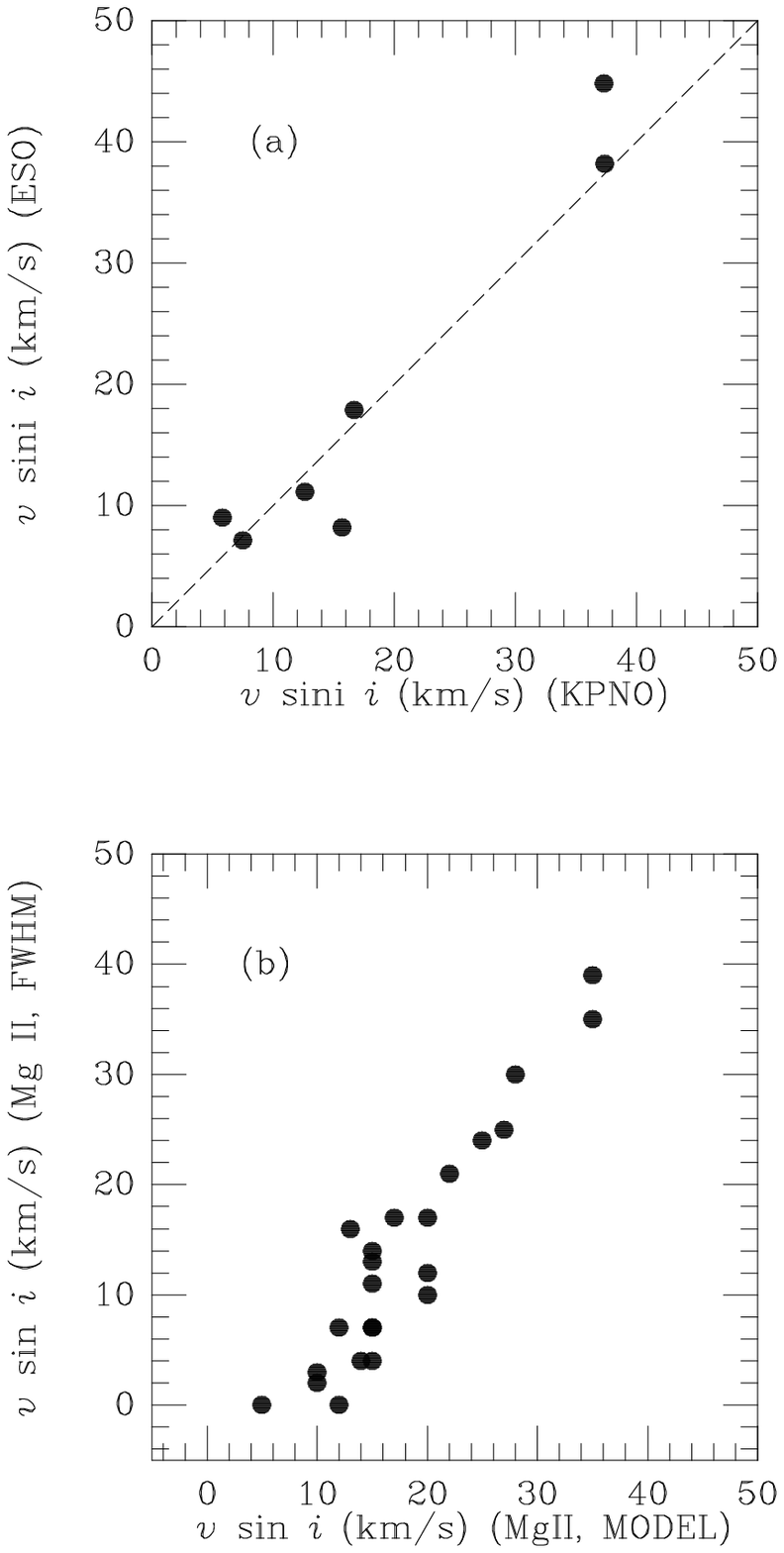}}
\hfill
\parbox[b]{88mm}{ 
\caption[]{{\bf a} A comparison of the \ksi~obtained (using the FWHM of
the \kmg~doublet) from spectra taken with the ESO-CAT (ordinate) with those
obtained from spectra of the same stars taken with the Kitt Peak
coud\'{e} feed (abscissa).
{\bf b}      A comparison of the \ksi~obtained (using the FWHM of the 
\kmg~doublet) for the Kitt Peak coud\'{e} feed spectra (ordinate) with that
determined from the same line using the synthetic spectra. (abscissa).
 }
\label{Fig11}}
\end{figure}

\begin{figure}
\resizebox{8.3cm}{!}{\includegraphics{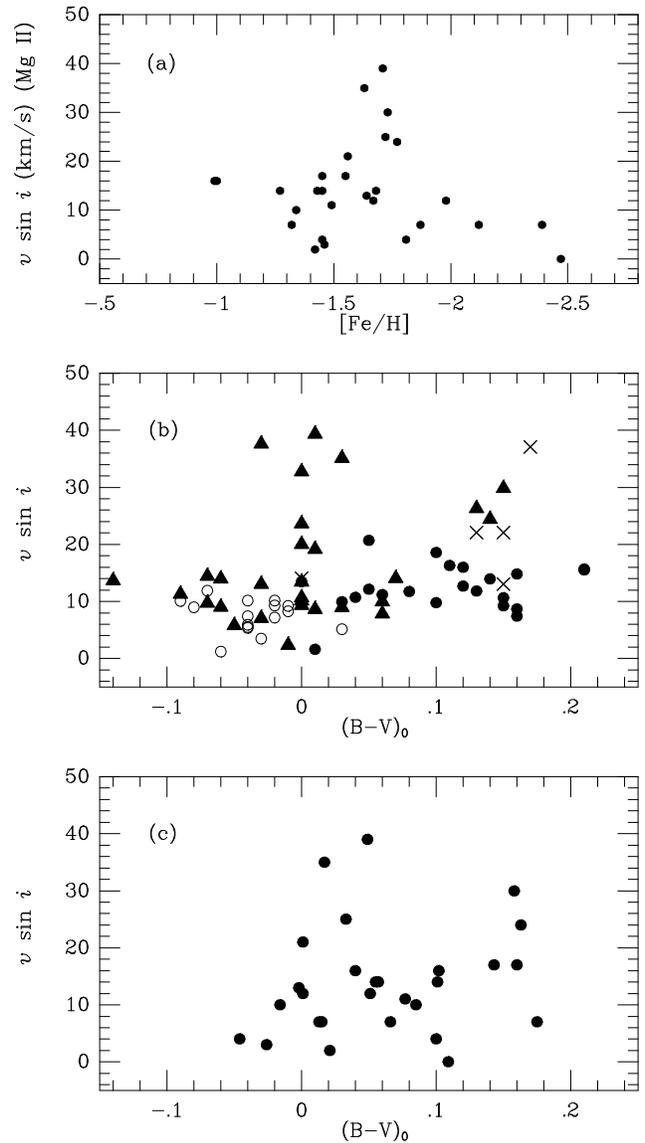}}
\hfill
\parbox[b]{88mm}{ 
\caption[]{{\bf a} A plot of \ksi~(ordinate) against [Fe/H] (abscissa)
 for our field BHB stars. {\bf b} A plot of \ksi~(ordinate) against 
 \kfo~(absicissa) for BHB stars in the globular clusters NGC 288 (open
 circles),  M3 (filled circles) and M13 (filled triangles) from Peterson et al.
 (1995) and also M92 (crosses) from Cohen \& McCarthy  (1997). 
 {\bf c} A plot of \ksi~(ordinate) against \kfo~(abscissa) for our field BHB
 stars.
 }
\label{Fig12}}
\end{figure}

The resolution of most of our spectra (15\,000) is not enough for us to make
definitive measurements of \ksi, {\it but we can distinguish quite easily
between stars with a \ksi~of less than 15 \ks~and those with a \ksi~$\sim$
30~\ks}. We chose to use the \kmg~($\lambda$ 4481) line\footnote{This close
doublet was used by Slettebak (1954) for his study of the rotational
velocities of stars of spectral types B8 to A2. The line was also used by
Glaspey et al. (1989) to derive \ksi~for two HB stars in NGC 6752. The line
is strong over a wide range of spectral types and is essentially unblended.}
 and measured its FWHM (F) with the IRAF routine that employs a
simple gaussian fit. The \ksi~of seven field BHB stars observed by Peterson
et al. (1983) were used to convert the FWHM  to \ksi~with the relation:
\begin{displaymath}
     v \sin i  = 59.0 \times \surd~(F^{\rm 2} - K)
\end{displaymath}
where F is in \AA~and  the constant K is 0.221 for the Kitt Peak spectra and
0.151 for the ESO CAT spectra. The fit for the Kitt Peak spectra is shown in
Fig~10. A number of early-type stars whose \ksi~are given in the IAU
Transactions (1991) were also observed and they are shown by open circles.
Their \ksi~follow the same trend with F as the calibrating BHB stars (filled
circles) but their \ksi~are systematically lower for a given F. The reason
for this discrepancy is not understood but we have chosen to follow the
calibration defined by the observations of Peterson et al. (1983) because our
main interest is to compare our \ksi~with those obtained by PRC for the BHB
stars in globular clusters. We point out, however, that the use of our
relation for \ksi~$>$ 30~\ks~does involve a small  extrapolation beyond the
range of  the calibration. Had  we used a calibration based on the IAU
standards, our computed \ksi~would have been about 60\% of those given in
Table~15.
\begin{table*}
{{\bf Table~16.} Mean \ksi~and deprojected rotational velocities in field and
cluster stars}
\begin{flushleft}
\begin{tabular}{ccccc} 
\noalign{\smallskip}
\hline 
       &      &    &    &     \\
System & No. of&~~~~~\ksa~~~~~~ &~~~~~$\bar{v}$~~~~~&~~~~~$\surd~(\overline{v-\bar{v})^{\rm 2}}$~~~~~   \\
       & Stars  &\multicolumn{3}{c}{\ks~}    \\
    (1)&    (2)&   (3)&   (4)&    (5) \\  
\noalign{\smallskip}
\hline 
\noalign{\smallskip}
Field BHB stars (this paper) &  28 & 14.0  &  17.8 &  10.3    \\
Clusters M3\ka~, M13\ka~, M92\kw~ \& NGC 288\ka~ & 72 & 13.6 & 17.3 & 9.0   \\
Clusters M3\ka~, \& NGC 288\ka~ & 38 & 10.3 & 13.1 & 3.8   \\
Clusters M13\ka~ \& M92\kw~ & 34 & 17.3 & 22.0 & 10.7  \\
\noalign{\smallskip}
\hline 
\noalign{\smallskip}
\end{tabular}           
\\
\ka~~Data from Peterson et al. (1995).   \\
\kw~~Data from Cohen \& McCarthy (1997).    \\

\end{flushleft}
\end{table*}

In Fig~11(b), we compare the \ksi~that  were determined from the FWHM of the
\kmg~line 
with the estimates of the rotational broadening that were obtained in fitting
the observed and computed \kmg~line profiles. 
   There is a good correlation between the
two for \ksi~$\goa$ 15 \ks~; for smaller \ksi, the fractional errors in the 
estimates are greater and so there is a poorer correlation. In any case, we 
should not 
expect the two quantities to be identical since the \ksi~determined from the
\kmg~line have been forced onto the system of another observer, whereas the
rotational broadenings deduced from the model involve different assumptions.
In Fig 11 (a) we compare the \ksi~that were obtained
from the ESO-CAT spectra with those obtained for the same seven stars 
(six BHB stars and HD~140194) with the 
KPNO coud\'{e} feed. The good agreement between these independent estimates of
\ksi~fully supports the conclusion that our data can be used to distinguish
between  stars with a \ksi~$\sim$~30 \ks~and those of lower rotational
velocity.
 
The dependence of \ksi~on the metallicity is shown in panel (a) of  Fig~12.
Although the six most metal-weak stars have lower than average \ksi, it is not
thought that these data show any {\it significant} trend of \ksi~with
metallicity. The middle and lower panels of Fig~12 show plots of \ksi~against
 \kfo~for the HB stars in globular clusters (middle) and for our field
BHB (below). The distributions in the clusters and in the field are similar
and in neither case is there a trend seen between \ksi~and colour.

The interpretation of the observed distribution of \ksi~in terms of a
randomly oriented population has been discussed by Chandrasekhar \& M\"{u}nch
(1950) and by Brown (1950). Brown, in particular, points out that the true
distribution of rotational velocities can only be determined from relatively
large samples. It is possible to put some constraints on the true distribution
using the expressions    given by Chandrasekhar \& M\"{u}nch for the mean and
mean square deviation of this distribution (their equation (20)). Table~16  
gives the mean projected rotational velocity (\ksa~), the mean true rotational
velocity ($\bar{v}$) and the root mean square deviation of this true
rotational velocity (\ksb~) in \ks~for our sample of field BHB stars and for
various samples of globular cluster HB stars. Bearing in mind that our
measured \ksi~ undoubtedly have somewhat larger observational errors than
those  of the globular cluster HB stars, the \ksa~, $\bar{v}$ and \ksb~of our
sample well match the whole sample of globular cluster HB stars. This suggests
that the two subgroups of globular clusters with low $\bar{v}$ and \ksb~ (M3
\& NGC 288) and high $\bar{v}$ and \ksb~ (M13 \& M92) are fairly equally
represented in the field. None of these samples show significant evidence for
skewness so the characterization of the true velocity distribution in terms of
$\bar{v}$ and \ksb~is sufficient. It is to be noted that the \ksb~of the low
velocity group must be very largely produced by observational error so that
the intrinsic dispersion in this subgroup must be very low.
   
\subsection{Abundances of the $\alpha$-elements.}

It is well known that the $\alpha$-elements are more abundant relative to iron
in metal-poor halo stars than in disk stars with solar abundances ( Wheeler et
al. 1989). The exact form of this enhancement may differ
somewhat from element to element. Thus Boesgaard et al. (1999) have found a
linear relation between [O/H] and [Fe/H] in the range 0.0$>$[Fe/H]$>$$-$3.0,
but the relation is less well-defined for other $\alpha$-elements such as Mg
and Ti. The mean abundances of these two elements (relative to iron) are given
in Table~17 for the BHB stars in our sample and for a number of other samples
of metal-poor stars of similar metallicity. All of these other samples are
late-type halo stars except for the old metal-poor selection taken from the
thick-disk stars of Edvardsson et al. (1993) and the halo RR Lyrae sample of
Clementini et al. (1995). Some systematic differences may be expected between
the abundance ratios found for the different samples because they are derived
from different lines of these elements and also different ionization states
and undoubtedly systematic errors are present in their assumed  \kgf. Also,
the abundance determined from the \kmg~$\lambda$4481 line can be quite
sensitive to the assumed microturbulent velocity (Table~12). Under these
circumstances, we consider that the $\alpha$-element enhancement in our BHB
sample is in reasonable agreement with other recent determinations for halo
stars.

\begin{table}
{{\bf Table~17.} Mean values of [Mg/Fe] and [Ti/Fe] \\ from various sources.}
\begin{flushleft}
\begin{tabular}{ccccc}
\hline\noalign{\smallskip}
 $<$[Mg/Fe]$>$ &$<$[Ti/Fe]$>$ & $<$[Fe/H]$>$ & No. & Ref.  \\ 
\noalign{\smallskip}
\hline\noalign{\smallskip}
+0.43$\pm$0.03 & +0.44$\pm$0.03 & $-$1.66 & 24  &  (1)   \\
+0.33$\pm$0.02 & +0.23$\pm$0.02 & $-$0.68 & 19  &  (2)   \\
+0.37$\pm$0.05 & +0.32$\pm$0.02 & $-$1.60 &  8  &  (3)   \\
 \kc           & +0.28$\pm$0.02 & $-$1.62 & 11  &  (4)   \\
+0.30$\pm$0.01 & \kc            & $-$1.86 & 60  &  (5)   \\
+0.48$\pm$0.02 & +0.42$\pm$0.02 & $-$2.13 & 20  &  (6)   \\
+0.42$\pm$0.03 & +0.27$\pm$0.02 & $-$2.15 &  9  &  (7)   \\
+0.23$\pm$0.03 & +0.20$\pm$0.03 & $-$1.60 & 11  &  (8)   \\
+0.37$\pm$0.02 & +0.30$\pm$0.02 & $-$1.09 & 16  &  (9)   \\
\hline\noalign{\smallskip}
\end{tabular}
\\
(1)~~BHB stars (this paper).                    \\
(2)~~Edvardsson et al. (1993) (Thick disk:  \\
~~~~~Age$\geq$~10~Gyr;~Orbital Ecc.~$\geq$~0.35;~[Fe/H]$\leq$$-$0.50 ).  \\
(3)~~Clementini et al. (1995) (halo RR Lyraes).  \\
(4)~~Gratton \& Sneden (1991) (metal-poor dwarfs and giants).  \\
(5)~~Pilachowski et al. (1996) (halo giants).  \\
(6)~~Magain (1989) (halo dwarfs).  \\
(7)~~Nissen et al. (1994) (metal-poor dwarfs and subgiants). \\
(8)~~Stephens (1999) (halo dwarfs: eccentric orbit). \\
(9)~~Clementini et al. (1999) (Hipparcos stars:   \\
~~~~~~~~~~~~~~~~~~~~~~~~~~~~~~~~~~~~~~~  [Fe/H]$\leq$$-$0.50 ).  \\
\end{flushleft}
\end{table}

\subsection{BHB Binaries and HD 130095.}

\begin{table}
{{\bf Table~18.} Radial velocities of HD~130095.  }
\begin{flushleft}
\begin{tabular}{lll}
\hline\noalign{\smallskip}
 Date (UT)  & Radial velocity (\ks~)  & Source   \\
\noalign{\smallskip}
\hline\noalign{\smallskip}
   1960~Apr~23   & +46.0    &   (1)     \\ 
   1963~Feb~13   & +73      &   (2)     \\
   1963~Mar~01   & +64      &   (2)     \\
   1963~May~19   & +55      &   (2)     \\
   1964~Apr~28   & +61.0    &   (1)     \\ 
   1964~May~28   & +42.9    &   (1)     \\ 
   1969~May~23   & +63      &   (2)     \\
   \kc           & +65$\pm$2.4 & (3)     \\
   1980~Jul      & +83$\pm$6  &  (4)    \\
   1982~Apr~08   & +64.3$\pm$1.0 &  (5)  \\
   1982~Apr~17   & +65.0$\pm$1.0 &  (5)  \\
   \kc           & +96$\pm$2     &  (6)  \\
   1995~Apr~29   & +65.6$\pm$0.8 &  (7)   \\
   1995~May~03   & +66.0$\pm$0.7 &  (8)   \\
   
\hline\noalign{\smallskip}
\end{tabular}
\\
(1)~~Przybylski \& Kennedy (1965b)   \\
(2)~~Hill (1971)   \\
(3)~~Greenstein \& Sargent (1974)   \\
(4)~~Kodaira \& Philip (1984)   \\
(5)~~Peterson et al. (1983) \\
(6)~~Adelman \& Philip (1990)  \\
(7)~~This paper (ESO-CAT)     \\
(8)~~This paper (KPNO coud\'{e} feed)  \\
\end{flushleft}
\end{table}

Binaries may be expected among halo stars and a discussion of their possible
effect on the abundances has been given by Edvardsson et al. (1993) and
Clementini et al. (1999). We have no direct evidence from the spectra that
there are any binaries in our sample except that HD~130095 may have a 
variable radial velocity although it does not appear to vary in light (ESA
Hipparcos Catalogue 1997, Stetson 1991). Although the published radial
velocities of this star (Table 18) show a spread of over 50 \ks, more than
half of these velocities lie in a 5~\ks~range centered on +63~\ks. It does
not seem entirely impossible that HD 130095 has a constant velocity of
+63~\ks~and that the errors of the velocities that are outside this range
have been greatly underestimated. If, however, the spread is real, then a
period of about seven months seems to be possible, although far from certain. 
Now if $P$ is the period in years, $a$ is the semi-major axis of the orbit (in
A.U.) and $m_{1}$ and $m_{2}$ are the masses of the two components (in
M$_{\sun}$), then
\begin{displaymath}   
    a^{3}  =  P^{2} \times (m_{1} + m_{2}) 
\end{displaymath}   
If we assume equal components with a combined mass of 1.2 M$_{\sun}$, then the
semi-major axis will be 0.74 A.U.; this is somewhat larger than the radius of
the red giant progenitor of the HB star ($\sim$100R$_{\sun}$). The other
component might possibly be an equally metal-poor subdwarf ([Fe/H]~=~$-$2.0)
whose lines would not be easily detectable in the spectrum of HD~130095. Such
a star would be much less luminous than but of comparable mass to the HB star.
Such a companion would not be particularly bright in the infrared and so would
not have been discovered in the survey for infrared-bright companions of halo
stars by Carney (1983).

It is known (Smart 1931) that 
\begin{displaymath}   
    A \sin i = 6875 P (\alpha + \beta ) \surd (1 - e^{2})
\end{displaymath}   
where A is the semi-axis major (in km), T is the period (in days), $e$ is the
orbital eccentricity and ($\alpha$ + $\beta$) is the velocity amplitude. If  
we assume a velocity amplitude of 50 \ks, then  we find 
\begin{displaymath}   
    \sin i  = 1.5 \times \surd (1 - e^{2})                 
\end{displaymath}   
which requires that $e > $ 0.75. Thus  the published radial velocities are not
incompatible with HD 130095 being a binary, but it does seem highly desirable
to make new velocity measurements over a period of several months so that the
reality of the variability can be confirmed and a period established. The star
is relatively bright ($V$ = 8.15) and at declination $-$27$\degr$; the
observations would most easily be made in the southern hemisphere.

\subsection{The RR Lyrae variable CS Eri (HD~16456)} 

\begin{table*}
{{\bf Table~19.} Spectroscopic observations of CS Eri (HD~16456).}
\begin{flushleft}
\begin{tabular}{ccccccc}
\hline\noalign{\smallskip}
\noalign{\smallskip}
Source & \multicolumn{2}{c}{Phase}&$T_{\rm eff}$&[Fe/H]\ka&[Fe/H]\kw& Rad. vel.  \\
       & Gen. Catalogue & Hipparcos&                   &   &   & (\ks)       \\
\hline\noalign{\smallskip}
                       &      &            &      &        &     &       \\
Solano et al.          & 0.12 & 0.20 & 6928 &$-$1.36 &\kc  & \kc       \\
Solano et al.          & 0.31 & 0.39 & 6679 &$-$1.45 &\kc  & \kc       \\
This Paper (Kitt Peak) & \kc  & 0.42 & 6750\kj &$-$1.70 &$-$1.69 & $-$139.8   \\
This Paper (ESO-CAT)   & \kc  & 0.94 &(7500)\kj&$-$1.65 &$-$1.65 &$-$158.9    \\
                       &      &            &      &        &     &       \\
\hline\noalign{\smallskip}
\end{tabular}
\\
\ka~~from the equivalent widths of both the \kfe~and \kff~lines. \\
\kw~~from the equivalent widths of the \kff~lines only.             \\
\kj~~\kg~=~3.0                     \\
\end{flushleft}
\end{table*}
 
Solano et al. (1997) observed CS Eri (HD~16456) with an Image Tube
spectrograph (resolving power 19\,000) on the SAAO 1.9-m telescope at
Sutherland in July, 1995. They determined abundances by assuming a
microturbulent velocity (\km) of 3.6 \ks~and a \kg~of 2.75. A summary of
their observations and ours is given in Table~19. Solano et al. found the
phases of their observations from the ephemeris of CS Eri given in the General
Catalogue of Variable Stars (Kholopov et al. 1985) (column 2 of Table 19). We
have calculated phases for all the observations using the more recent ephemeris
given in the Hipparcos Catalogue (1997) (column 3 of Table 19)\footnote{The
radial velocities indicate that these phases are reasonably correct. DH Peg is
a c-type RR Lyrae star that has a $V$-amplitude of 0.51 mag that is only
slightly smaller than the $V$-amplitude (0.55 mag) of CS Eri. Jones, Carney \&
Latham (1988) have determined a precise radial velocity curve for DH Peg so
that the difference between  the radial velocity and the $\gamma$-velocity at
each phase is known and this may be scaled by the $V$-amplitudes to predict the
corresponding differences for CS Eri. From these differences we derive
$\gamma$-velocities of $-$145.2 and $-$150.3 \ks~for CS Eri from the Kitt
Peak and ESO-CAT spectra respectively. These agree well with the
$\gamma$-velocity of $-$147 \ks~given by Solano et al. (1997).}. The effective
temperatures  which are given by Solano et al. and also the one which we 
derived from the Kitt Peak spectrum are given in column 4. CS Eri 
is intermediate in metallicity and amplitude to the two c-type
variables T Sex ($\Delta$$V$ = 0.42 mag) and TV Boo ($\Delta$$V$ = 0.60 mag) and
has a similar period.
Using the \kt~given for these stars by Liu \& Janes (1990), we deduce that
the maximum and minimum \kt~for CS Eri should be 7475~K and 6725~K
respectively. This minimum \kt~is in good agreement with the \kt~determined
from the Kitt Peak spectrum which was taken near minimum (phase 0.42). The
abundance deduced from the ESO-CAT spectrum (phase 0.92, near maximum)
assuming \kt~= 7500~K agrees well with that deduced from the Kitt Peak
spectrum; their mean is [Fe/H] = $-$1.67. Table~20  also gives the [Fe/H] that
was derived for the \kff~lines alone since, at the \kt~of RR Lyrae stars,
the strengths of these lines are less sensitive both to \kt~and NLTE effects
than those of \kfe~(Fernley \& Barnes, 1997). Our abundances for [Fe/H] are
therefore $\sim$0.2 dex lower than those found by Solano et al.(1997).

\section{Summary and Conclusions}

The purpose of this paper is to determine stellar parameters (e.g. \ksi,
~\kt~\&~\kg) and chemical abundances that will allow us to isolate a local
sample of BHB stars by their physical properties. All of our sample of thirty
one candidate stars appear to belong to the halo,   but BD~+32~2188 (a
post-AGB star), BD~+00~0145 (a possible cool sdB star) and  HD~16456 (the
RR~Lyrae star CS Eri) are not BHB stars. HD~202759, although classified as an
RR Lyrae star (AW Mic), has such a low $V$-amplitude ($<$ 0.1 mag) and high \kt~
(7\,500~K) that it has been included with the BHB stars. Our spectra of
HD~14829, HD~78913, HD~106304 and HD~213468 were not of sufficient quality
for a complete abundance analysis although we were able to estimate [Fe/H]
from their \kmg~($\lambda$4481) lines. 
 %
%
 %
 %

Of the twenty eight stars which we classify as BHB stars, the most doubtful is
HD~139961 because 
it has the largest \ksi~and also an unusually low orbital eccentricity
(0.22)\footnote{
to be discussed in forthcoming paper with Christine Allen.}. 
It is also NSV~7204 in the New Catalogue of Suspected Variable stars,
 Kukarkin et al. (1982).
Corben et al. (1972) found a range of 0.08 magnitudes in $V$ over six
observations. The 85 observations of this star in the ESA Hipparcos catalogue,
however, show a range of only 0.05 magnitudes; this corresponds to an 
$rms$ deviation of only 0.01 magnitudes.
Its colour, moreover,  does not put it near the edge of the
instability strip, so that its variablity seems questionable. 
 {\it The existence of stars such as
HD~139961 shows how difficult the classification of BHB stars can be and how
necessary it is to use all available criteria}. When large numbers of stars
are to be surveyed, simpler methods may have to suffice but one must then
expect to get more  misclassifications. 
 Thus,   Wilhelm et al. (1999) classify BHB stars  with broad band
$UBV$ colours, Balmer-line widths and the Ca\,{\sc ii} (K-line) equivalent widths.
Among the 18 stars in common with our sample, they  classify 
the broad-lined A-star \object{HD 203563} as an FHB
star and their [Fe/H] average 0.32$\pm$0.08 more metal-poor than ours with 
individual stars differing from our [Fe/H] by as much as 0.8 and 0.9 dex.

Projected rotational velocities (\ksi) were determined for each star by
calibrating the FWHM of the \kmg~($\lambda$4481) line against the \ksi~of
seven of the stars in our sample that had previously been determined from
echelle spectra by Peterson et al. (1983). No obvious trend of \ksi~ was found
with either \kfo~or abundance. A simple analysis of the \ksi~(following
Chandrasekhar \& Munch 1950) shows that the deprojected distributions of these
rotational velocities are similar to those found in globular clusters. Both
have a $\overline{v}$ of $\sim$17 \ks~that is intermediate between that of
the high rotational velocity clusters (M13 and M92) and the low rotational
clusters (M3 and NGC 288).

BD~+00~0145, HD~14829, HD~78913, HD~106304 and HD~213468  should be
reobserved since we did not obtain spectra of sufficient quality for a
complete analysis. Improved equivalent widths and \ksi~could be obtained for
all our BHB stars by using a higher resolution and a larger waveband (e.g. by
using an echelle spectrograph) so that more lines would be available. Improved
abundances, however, require a better understanding of the physical conditions
in the stellar atmospheres and more accurate $gf$ values as well as more
certain determinations of the interstellar extinctions. In this latter
connection, more reliable determinations of the extinction would be possible
if \kv~colours were available for our entire sample. It is possible that
HD~130095 is a binary. Its reported velocity variations should be checked so
that (if these are real) a period can be derived.
 
As we noted earlier, many of our BHB stars were selected
from the early type stars that were found in surveys for high proper motion;
our sample may therefore be expected to have a kinematic bias. This bias
(inter alia) will be examined in a following paper, where we shall compare the
galactic orbits of these BHB stars with those of other nearby halo stars.

\begin{acknowledgements}

  We thank Saul Adelman for making his spectrum of HD~161817 available to us 
  and also Giuseppe Bono and Santino Cassisi for providing their ZAHB models
  before publication. We also thank Allen Sweigart for sending us his 
  He-enhanced models in electronic form. We are grateful to 
  John Glaspey for helpful comments on a provisional draft of this paper 
  and the referee (Klaas de Boer) for questioning the validity of models
  for representing far-UV spectra of BHB stars and for many suggestions for
  improving the style and readability of the paper.
  We are pleased to acknowledge 
  the use of the IUE Final Archive which is sponsored
  and operated by NASA/ESA. 
  This research has made use of the Simbad database, operated at CDS,
  Strasbourg, France.  
      
\end{acknowledgements}

\appendix

\section{Comments on other possible BHB star candidates.} 

Philip \& Adelman (1993) found 19 BHB star candidates by searching the Hauck
\& Mermilliod photometric catalogue (1980) for stars with the appropriate
Str\"{o}mgren indices (e.g. one of their criteria was that the c$_{\rm 1}$
index should exceed 1.15). Bragaglia et al. (1996) made preliminary
measurements of the \ksi~of fourteen of these stars and noted that their
rotations were mostly too large for them to be BHB stars.  Adelman \& Philip
(1996b) obtained high resolution spectra of seven of these stars
(\object{HD~15042}, 
\object{HD~42999},
\object{HD~47706},
\object{HD~48567},
\object{HD~49224},
\object{HD~67426}
\& \object{HD~79566}) 
and also concluded that their rotational velocities were too high  for them to
be BHB stars. Of the remaining seven stars observed by Bragaglia et al., five
(\object{HD~53042}, 
\object{HD~67542}, 
\object{HD~128855}, 
\object{HD~181119} 
\& \object{HD~185174}) 
have \ksi~greater than 60 \ks~. Two, however, (\object{HD~83751} and~ 
\object{HD~140194}) have \ksi~$\sim$ 30 \ks~which is within the range of
rotations observed for BHB stars; both stars have Population I
kinematics\footnote{
The radial velocities of HD~83751 and  HD~140194 are +13.5 and  +1.2 
\ks~respectively.} 
and roughly solar abundances; thus in spite of their low \ksi~, they are
unlikely to be BHB stars. The remaining five of the nineteen candidates listed
by Philip \& Adelman were not observed by us but some comments can be made on
the probability that they are BHB stars.  \object{HD 100548} was classified as
G8 III by Upgren (1962) from its objective prism spectrum. The photometry of
this star listed in the Hauck \& Mermilliod catalogue (1980) appears to be
spurious because the star is not found among those in the listed reference
(Drilling \& Pesch 1973). Three of the remaining stars (\object{HD~94509},
\object{HD~120401} \& \object{HD~304325}) have very low galactic latitudes (b
$\leq$ 3\degr) while \object{HD~123664} is likely to be a member of the
Scorpio-Centaurus Association (Glaspey 1972, Slawson et al. 1992).
 It therefore seems unlikely that any of Philip \& Adelman's  nineteen
BHB star candidates have a high probability of being BHB stars. Their work was
valuable, however, because it has shown the need to use criteria in addition
to Str\"{o}mgren photometry in the identification of these stars.

Listed below are a number of other stars that have sometimes been suggested to
 be BHB stars; this list is not intended to be exhaustive. Spectra of one of
them (BD~+33~2171) should be obtained since its classification is doubtful from
the available data. The others are almost certainly not BHB stars.
\begin{description}
\item[\object{HD~52057}] Stetson (1991). Kilkenny \& Hill  (1975) classified
 the star as B6 and almost certainly subluminous.
\item[\object{HD~57336}] FHB~24 in Philip (1984). Huenemoerder et  al. (1984)
 noted that the star has Population I metal-line characteristics. It is
 broad-lined.
\item[\object{BD~+33~2171}] FHB~2 in Philip (1984). Its colour \kf~=~+0.276 is
 too red for it to be a BHB star if the reddening given by the STD maps (1998)
 (\keb~=~0.021) is correct. The \ksi~of 42~\ks~is also somewhat high for a BHB
 star.
\item[\object{HD~176387}] Stetson (1991) is the RR Lyrae star MT Tel. 
\item[\object{HD~203563}] Stetson (1991) is broad-lined. 
\item[\object{HD~214539}] Stetson (1991). Feast et al. (1955) discovered 
 its very high radial velocity (+333 \ks) 
 and Przybylski (1969) found it to be metal-poor and considered it
 to be an HB star. A two-sigma upper limit  to its Hipparcos parallax (ESA
 1997), however, means that it cannot be closer than 735 pc which would give it
 an $M_V$ of $-$2.1 or brighter so that it cannot be a HB star.

\end{description}

\end{document}